%% file: main_SI_sep.tex
\documentclass[
    aps,
    amssymb,
    amsmath,
    prx,
    twocolumn,
    showpacs,
    superscriptaddress,
    longbibliography,
]{revtex4-2}

\usepackage{lmodern}

\usepackage{graphicx}
\graphicspath{{figs_main/}{figs_SI_theory/}{figs_SI_experiment/}}

\usepackage{dcolumn}
\usepackage{bm}
\usepackage{subfigure}
\usepackage{color}
\usepackage{soul}

\usepackage{amsfonts,verbatim}

\usepackage[margin=0.6in]{geometry}

\usepackage{tikz}

\usepackage[english]{babel}
\usepackage{times}
\usepackage{dsfont}
\usepackage{latexsym}
\usepackage{float}
\usepackage{afterpage}
\usepackage{enumitem}
\usepackage{mleftright}
\usepackage[normalem]{ulem}
\usepackage{enumitem}

\definecolor{lR}{rgb}{1, 0.8, 0.79}

\usepackage{eso-pic, graphicx}

\usepackage{listings}
\usepackage{multirow}
\usepackage{xcolor,colortbl}

\usepackage{bbm}
\usepackage{newtxtext,newtxmath}

\usepackage{upgreek}
\input{Commands3.tex}

\renewcommand{\alpha}{\upalpha}
\renewcommand{\beta}{\upbeta}
\renewcommand{\gamma}{\upgamma}
\renewcommand{\delta}{\updelta}

\renewcommand{\epsilon}{\upepsilon}
\renewcommand{\varepsilon}{\upvarepsilon}
\renewcommand{\phi}{\upphi}
\renewcommand{\varphi}{\upvarphi}
\renewcommand{\kappa}{\upkappa}
\renewcommand{\xi}{\upxi}
\renewcommand{\lambda}{\uplambda}
\renewcommand{\psi}{\uppsi}

\renewcommand{\mu}{\upmu}
\renewcommand{\pi}{\uppi}
\renewcommand{\zeta}{\upzeta}

\renewcommand{\vartheta}{\uptheta}
\renewcommand{\theta}{\upvartheta}

\renewcommand{\eta}{\upeta}
\renewcommand{\tau}{\uptau}

\renewcommand{\rho}{\uprho}
\renewcommand{\sigma}{\upsigma}

\renewcommand{\Omega}{\Upomega}
\renewcommand{\omega}{\upomega}

\newcommand{\change}[1]{{\color{black}{#1}}}

\newcommand{\nocontentsline}[3]{}
\newcommand{\tocless}[3]{
\vspace{1em}
\bgroup\let\addcontentsline=\nocontentsline#1{#2\label{#3}}\egroup
}

\definecolor{Ablue}{rgb}{0.96,0.24,0.00}

\definecolor{Abluetitle}{rgb}{0.,0.24,0.51}

\definecolor{orange}{rgb}{0.96,0.24,0.00}

\definecolor{darkred}{rgb}{0.55, 0.0, 0.0}

\definecolor{darksalmon}{rgb}{0.91, 0.59, 0.48}
\definecolor{maroon}{cmyk}{0,0.87,0.68,0.32}

\setcitestyle{numbers,round,citesep={,\kern-.24em}}

\definecolor{mustard}{rgb}{1.0, 0.86, 0.35}

\addto\captionsenglish{\renewcommand{\figurename}{Fig.}}

\definecolor{Gray}{gray}{0.85}
\definecolor{LightCyan}{rgb}{0.88,1,1}
\newcolumntype{a}{$>${\columncolor{Gray}}c}
\newcolumntype{b}{$>${\columncolor{White}}c}

\usepackage{array}
\newcolumntype{L}[1]{$>${\raggedright\let\newline\\\arraybackslash\hspace{0pt}}m{#1}}
\newcolumntype{C}[1]{$>${\centering\let\newline\\\arraybackslash\hspace{0pt}}m{#1}}
\newcolumntype{R}[1]{$>${\raggedleft\let\newline\\\arraybackslash\hspace{0pt}}m{#1}}

\newcolumntype{P}[1]{>{\centering\arraybackslash}p{#1}}
\newcolumntype{M}[1]{>{\centering\arraybackslash}m{#1}}

\usepackage[
    colorlinks=true,
    citecolor=black,
    urlcolor=black,
    linkcolor=black,
    filecolor=black
]{hyperref}

\mleftright
\makeatletter
\@addtoreset{section}{part}
\makeatother

\newcommand{\beginsupplement}{%
        \setcounter{table}{0}
        \renewcommand{\tablename}{Supplementary Table}
        \renewcommand{\thetable}{\arabic{table}}%
        \setcounter{figure}{0}
        \renewcommand{\thefigure}{S\arabic{figure}} %
        \renewcommand{\theHfigure}{S\arabic{figure}} 
		\setcounter{page}{1}
		\renewcommand{\figurename}{Fig.} 
		\renewcommand{\thesection}{\:S\arabic{section}}
		\setcounter{section}{0}
        \setcounter{equation}{0}
        \renewcommand{\theequation}{S\,\arabic{equation}}
     }

\newcommand{\affA}{Department of Chemistry, University of California, Berkeley, Berkeley, CA 94720, USA.}
\newcommand{\affB}{Department of Physics, KTH Royal Institute of Technology, SE-106 91 Stockholm, Sweden.}

\newcommand{\affD}{Max Planck Institute for the Physics of Complex Systems, N\"othnitzer Str.~38, 01187 Dresden, Germany.}
\newcommand{\affE}{Chemical Sciences Division,  Lawrence Berkeley National Laboratory,  Berkeley, CA 94720, USA.}
\newcommand{\affF}{CIFAR Azrieli Global Scholars Program, 661 University Ave, Toronto, ON M5G 1M1, Canada.}
\newcommand{\affG}{Division of Physics, Mathematics, and Astronomy, California  Institute of Technology, 1200 E  California Blvd, Pasadena, CA 91125, USA.}

\begin{document}
\include{main}

\include{SI}

\end{document}

%% file: Commands3.tex
\newcommand{\norm}[1]{ \left| \left| {#1} \right| \right| }
\newcommand{\exptval}[1]{\left< {#1} \right>}

\newcommand{\identity}{\mathds{1}}

\newcommand{\nvfield}{\eta}
\newcommand{\potential}{\varphi}
\newcommand{\Hfloquet}{\mathcal{H}_F}
\newcommand{\Hdd}{\mathcal{H}_\mathrm{dd}} 
\newcommand{\Hfield}{\mathcal{H}_\mathrm{pot}} 
\newcommand{\Hfull}{\mathcal{H}} 
\newcommand{\Hkick}{\mathcal{H}_\mathrm{SL}} 
\newcommand{\Hsp}{\mathcal{H}_\mathrm{SP}} 
\newcommand{\tdd}{t_\mathrm{acq}} 
\newcommand{\tkick}{t_\mathrm{p}} 
\newcommand{\period}{\tau}

\newcommand{\Heffdd}{\mathcal{H}^{(0)}_{F, \mathrm{dd}}}
\newcommand{\Heff}{\mathcal{H}^{(0)}_F} 
\newcommand{\Heffpi}{\mathcal{H}^{\vartheta{\approx}\pi}_{\mathrm{eff}}} 
\newcommand{\Heffnopi}{\mathcal{H}^{\vartheta{\napprox}\pi}_{\mathrm{eff}}}

\newcommand{\Rabi}{\Omega} 
\newcommand{\detuning}{\delta\omega} 

\newcommand{\id}{\mathds{1}}

\newcommand{\tm}{{\text -}}

\newcommand{\tacq}{t_{\R{acq}}}
\newcommand{\xg}{\gamma}
\newcommand{\tzc}{t_{\R{zc}}}
\newcommand{\tw}{t_{\R{wait}}}
\newcommand{\Sint}{S_{\R{int}}}
\newcommand{\xt}{\vartheta}

\newcommand{\xl}{\lambda}
\newcommand{\xr}{\rho}
\newcommand{\xo}{\omega}
\newcommand{\xph}{\phi}

\newcommand{\app}{\approx}

\newcommand{\Bpol}{\textbf{B}_{\R{pol}}}
\newcommand{\tpol}{t_{\R{pol}}}

\newcommand{\Cs}{{}^{13}\R{C}}

\newcommand{\Hs}{{}^{1}\R{H}}

\newcommand{\mO}[0]{\mathcal{O}}

\newcommand{\mHdd}[0]{\mH_{\R{dd}}}

\newcommand{\mHeff}[0]{\mH_{\R{eff}}}

\newcommand{\xy}[0]{\xhat\tm\yhat}

\newcommand{\yz}[0]{\yhat\tm\zhat}

\newcommand{\xO}{\Omega}

\newcommand{\C}{$^\mathrm{13}C$}

\newcommand{\mH}[0]{\mathcal{H}}

\newcommand{\rt}{\rightarrow}

\newcommand{\beq}{\begin{equation}}
\newcommand{\eeq}{\end{equation}}
                  
\newcommand{\benum}{\begin{enumerate}}
\newcommand{\eenum}{\end{enumerate}}
                    
\newcommand{\bit}{\begin{itemize}}
\newcommand{\eit}{\end{itemize}}
\newcommand{\xhat}{\hat{\T{x}}}
\newcommand{\yhat}{\hat{\T{y}}}
\newcommand{\zhat}{\hat{\T{z}}}
\newcommand{\nhat}{\hat{\T{n}}}

\newcommand{\bea}{\begin{eqnarray}}
\newcommand{\eea}{\end{eqnarray}}

\newcommand{\bev}{\begin{verbatim}}
\newcommand{\eev}{\end{verbatim}}

\newcommand{\zt}{\times}

\newcommand{\qt}{\tau}

\newcommand{\T}[1]{\textbf{#1}}
\newcommand{\I}[1]{\textit{#1}}
\newcommand{\R}[1]{\textrm{#1}}

\newcommand{\zfl}[1]{\protect\label{fig:#1}}
\newcommand{\zfr}[1]{\figurename\,\ref{fig:#1}}

\newcommand{\zsl}[1]{\label{sec:#1}}
\newcommand{\zsr}[1]{Sec. \!\!\!\ref{sec:#1}}

\newcommand{\ket}[1]{\left\vert{#1}\right\rangle}

\newcommand{\expec}[1]{\left\langle #1\right\rangle}

\newcommand{\ba}{\left\{ \begin{array}{lr}}
\newcommand{\ea}{\end{array}\right.}

\newcommand{\Tr}[1]{\textrm{Tr}\left\{{#1}\right\}}

\newcommand{\blist}[1]{
 \begin{list}{#1}
 \begin{align}
	 arrow
 \end{align}
 $\checkmark\star
  { \setlength{\itemsep}{3pt}
     \setlength{\parsep}{2pt}
     \setlength{\topsep}{3pt}
     \setlength{\partopsep}{0pt}
     \setlength{\leftmargin}{1em}
     \setlength{\labelwidth}{1em}
     \setlength{\labelsep}{0.5em} } }
\newcommand{\elist}{
  \end{list}  }

\newcommand{\bef}
{
\begin{figure}[htbp]
\centering
}

\newcommand{\eef}{\end{figure}}


%% file: main.tex
\title{Nanoscale engineering and dynamical stabilization of mesoscopic spin textures}
\author{Kieren Harkins}\thanks{These authors contributed equally to this work}\affiliation{\affA}
\author{Christoph Fleckenstein}\thanks{These authors contributed equally to this work}\affiliation{\affB}
\author{Noella D'Souza}\thanks{These authors contributed equally to this work}\affiliation{\affA}
\author{Paul M. Schindler}\thanks{These authors contributed equally to this work}\affiliation{\affD}
\author{David Marchiori}\thanks{These authors contributed equally to this work}\affiliation{\affA}
\author{Claudia Artiaco}\affiliation{\affB}
\author{Quentin Reynard-Feytis}\affiliation{\affA}
\author{Ushoshi Basumallick}\affiliation{\affA}
\author{William Beatrez}\affiliation{\affA}
\author{Arjun Pillai}\affiliation{\affA}
\author{Matthias Hagn}\affiliation{\affA}
\author{Aniruddha Nayak}\affiliation{\affA}
\author{Samantha Breuer}\affiliation{\affA}
\author{Xudong Lv}\affiliation{\affA}\affiliation{\affG}
\author{Maxwell McAllister}\affiliation{\affA}
\author{Paul Reshetikhin}\affiliation{\affA}
\author{Emanuel Druga}\affiliation{\affA}
\author{Marin Bukov}\affiliation{\affD}
\author{Ashok Ajoy}\email{ashokaj@berkeley.edu}\affiliation{\affA}\affiliation{\affE}\affiliation{\affF}

\begin{abstract}
\textbf{Abstract:}
Thermalization, while ubiquitous in physics, has traditionally been viewed as an obstacle to be mitigated. \change{In contrast, we demonstrate here the use of thermalization in} the generation, control, and read-out of “shell-like” spin textures with interacting $\Cs$ nuclear spins in diamond, wherein spins are polarized oppositely on either side of a critical radius. The textures span several nanometers and encompass many hundred spins; they are created and interrogated without manipulating the nuclear spins individually. Long-time stabilization is achieved via prethermalization to a Floquet-engineered Hamiltonian under the electronic gradient field: the texture is, therefore, meta-stable \change{and} robust against spin diffusion. \change{This enables the state to endure} over multiple minutes before it decays. Our work \change{on spin-state engineering} paves the way for applications in quantum simulation and nanoscale imaging.
\end{abstract}

\maketitle
\pagebreak

\textbf{Teaser:}
Stable nanoscale spin textures emerge by harnessing thermalisation principles in \change{non-equilibrium} systems of coupled \C-nuclei.

\vspace{.5cm}
\textbf{Main Text:}
\vspace{-1cm}
\tocless\section{Introduction}{sec:Intro}
The quest to elucidate how isolated quantum systems approach equilibrium, has spurred a thriving field at the intersection of contemporary theoretical and experimental research. This has been marked by ongoing developments in nonequilibrium dynamics, such as the Eigenstate Thermalization Hypothesis for closed quantum systems~\cite{Srednicki1995,rigol2008thermalization,Deutsch2018}, anomalous transport and emergent hydrodynamics~\cite{zu2021emergent, wei2022quantum,wienand2023emergence}, or the creation of prethermal ordered states of matter~\cite{Else17,Machado2020}.

Thermalization in quantum systems proceeds analogously to its classical counterpart: entropy grows with time, erasing traces of the system’s prior history \change{and gradually reducing accessible information}. In most cases, such loss of information is irreversible, impeding the capacity to harness quantum systems. Consequently, a broad endeavor has been underway to retard, even preclude, these thermalization processes. Techniques range from physically isolating quantum systems (e.g. in vacuum) or through quantum control~\cite{viola00}, cooling them to near absolute zero temperatures~\cite{Bloch2008_UltracoldGasesReview}, or, in a complementary manner, exploiting theoretical paradigms of many-body localization to inhibit the onset of thermalization~\cite{Abanin2019_MBLReview}.

Though the thermalization process is often depicted as leading to mundane, featureless states, in this work, we demonstrate its utility in preparing structured quantum states. Specifically, in a system of interacting nuclear spins at high temperature (${>}100\,$K), we exploit out-of-equilibrium thermalizing dynamics to controllably engineer and stabilize \I{shell-like} nuclear spin polarization textures that span several nanometers and envelop hundreds of nuclear spins (\zfr{fig1}A). Simultaneously, we continuously observe the formation and stabilization of these textures with high temporal resolution over prolonged, multiple-minute-long periods. This spatiotemporal control, facilitated by thermalization, \change{removes} the need for local spin \change{control} and bypasses technical challenges of differentiating spins with near-identical resonance frequencies. Given these methodological advantages, our approach has direct implications for quantum memories~\cite{king_optically_2012,Meriles_longterm_2016}, spintronics~\cite{Awschalom_ferromagtex_2001,Awschalom_doughnuts_2004}, and nanoscale magnetic resonance imaging~\cite{Abobeih_atomicimaging_2019,Budakian_NanoMRI_2022}.

Our experiments are in a model system of Nitrogen Vacancy~(NV) center electrons in single crystal diamond surrounded by $\Cs$ nuclear spins (\zfr{fig1}A)~\cite{Jelezko01}.
The sparsely distributed NVs encompass ${\sim}10^4$ $\Cs$ nuclei, spanning a radius of ${\app}12\,$nm~\cite{Ajoy19relax}.
Data collected is an \change{average} over all NV center orientations with respect to the magnetic field.
Our strategy is built upon three elements.
Firstly, the optically polarizable NV electron serves as a ``polarization injector” and a nanoscale ``antenna”.
In its role as an antenna, it generates a nanoscale magnetic gradient field through the hyperfine interaction; this induces local displacements of the nuclear spin resonance frequencies as a function of distance from the NV center~\cite{Ajoy15}.
Secondly, under this gradient field, a time-periodic drive stabilizes the nuclei into a metastable state~\cite{ho2023quantum}, characterized by a shell-like polarization texture (\zfr{fig1}B).
Finally, spatially dependent dissipation from the NV is exploited to observe these textures: $\Cs$ nuclei in the electronic proximity relax faster such that signal at later times is dominated by farther away spins; this provides a means to \I{serially} probe spins farther away from the electron, without locally measuring them (\zfr{fig1}B).

To demonstrate this new toolbox, we first exhibit a simple means to produce spin textures. Here the NV is exploited to successively inject ``hyperpolarization'' -- polarization that is orders of magnitude greater than Boltzmann levels -- of alternate sign into the $\Cs$ nuclei. However, the resulting \T{State-Engineered} spin texture lacks stability: its domain boundary melts due to nuclear spin diffusion~\cite{dalessio2016quantum}.

Exploiting thermalization, however, offers a solution — creating polarization textures with domain boundaries protected against spin diffusion. We create an effective inhomogeneous parent Hamiltonian with spatial characteristics derived from the NV antenna field (\zfr{fig1}C). Irrespective of the initial state~(see methods), the spin polarization naturally \I{prethermalizes}~\cite{abadal2020floquet,fleckenstein2021thermalization,peng2021floquet} to its drive-induced quasi-equilibrium state, forming spin shells with constant-in-time domain boundaries. We observe the formation of this generated \T{Hamiltonian-Engineered}~\cite{Bukov15,goldman2014periodically,eckardt2017colloquium} texture continuously with unprecedented resolution over a lifetime spanning several minutes. Numerical simulations corroborate these observations and illustrate a stable spin texture spanning several nanometers.

\begin{figure}[t]
  \centering
 {\includegraphics[width=0.49\textwidth]{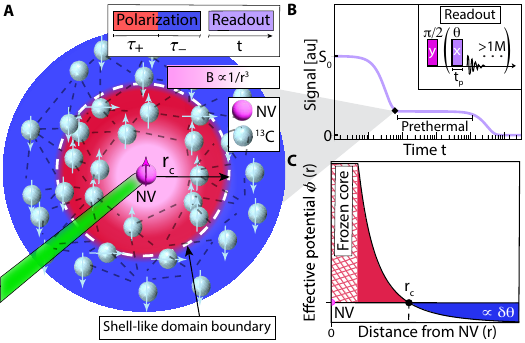}}
    \caption{\T{System and readout.} (A) \I{Spin-textures.} Controllable ``shell-like” spin texture of positive and negative polarization (shaded red and blue, respectively) is generated in a nanoscale ensemble of $\Cs$ nuclear spins surrounding a central NV electron (manipulated by an external laser, shown in green). The domains are separated by the critical radius $r_c$ at the domain boundary (dashed white line). The texture encompasses ${\sim}\mO(100)$ spins within the critical radius and remains stable for minutes. The optically-pumped electron serves as a spin injector, and produces a nanoscale magnetic field gradient $B$ (shaded light pink) that stabilizes the spin texture in a prethermal state. 
  (B) \I{Experimental schematic} showing prethermalization caused by \change{periodic} driving at high-magnetic field (HF) $B_0{\geq}7$ T, with a train of spin-locking $\xt$-pulses of length $t_p$. $\xy$ spin polarization is interrogated in windows between the pulses for total readout times $t{>}60$s (${>}0.5$M pulses). (C) $\Cs$ spin texture is generated and stabilized by a spatially-varying potential $\varphi(r)$ created upon driving with $\vartheta{\approx}\pi$ in the presence of the NV gradient (cf.~Sec.~\ref{sec:HamiltonianEngineering}).
  }
\zfl{fig1}
\end{figure}

\begin{figure*}[t]
  \centering
  {\includegraphics[width=1\textwidth]{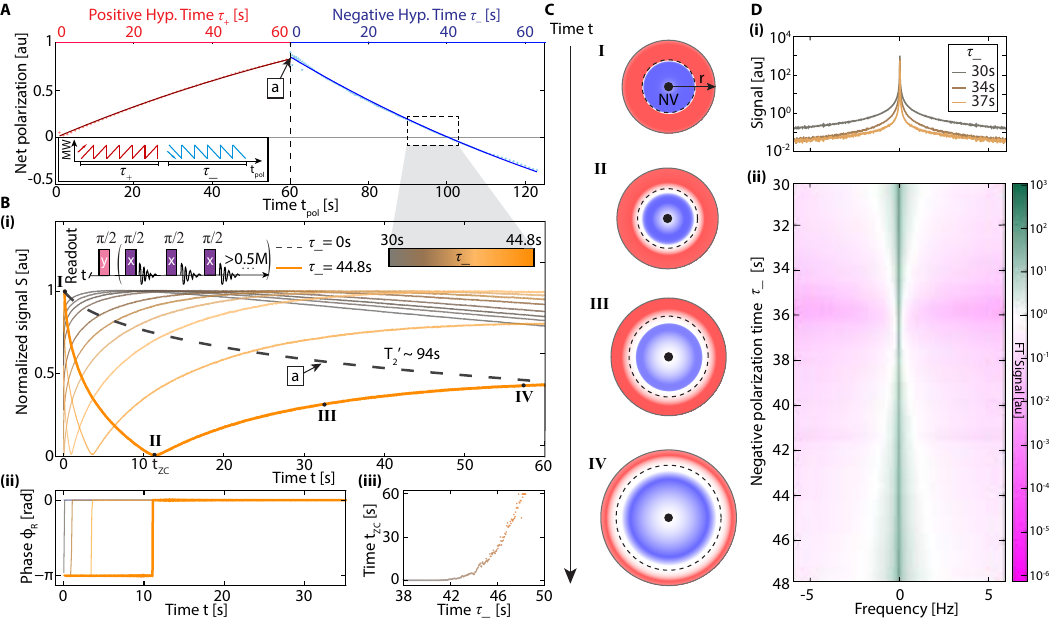}}
     \caption{\T{Spin textures via hyperpolarization injection (State Engineering).} 
    (A) Net $\Cs$ spin polarization under positive ($\qt_+$, red points), \change{followed by negative $\qt_+{=}60$s ($\qt_-$, blue points) hyperpolarization injection} at room temperature. Solid lines represent a biexponential fit. $\tpol{\app}97$s box corresponds to \change{${\app}$0} net polarization.
    (B) Normalized spin-lock decays under \change{periodic} drive with $\xt{\app}\pi/2$, showing (i) signal $S$ and (ii) rotating-frame phase $\xph_R$.
\I{Dashed black line:} spin-lock decay corresponding to $\qt_-{=}0$ (\I{a} in A), display long lifetimes $T_2’{\app}$93.4s. \I{Colored lines:} Data for different $\qt_-$ values. Decay profiles for $\qt_-$ inset from (A) exhibit a sharp zero-crossing at $t{=}\tzc$ (II) and associated phase inversion (see ii). \I{Dark orange line:} $\qt_-{=}44.8\,$s emphasized for clarity. \change{Full dataset shown in movie S1.}
(iii) Zero-crossing $\tzc$ movement with $\qt_-$ demonstrates control of spin texture.
    (C) \I{Schematic representation} of spin textures with increasing experiment progression time $t$. Representative points (I-IV) are marked on $\qt_-{=}44.8$s in (B). NV is at $r{=}0$. Positive (negative) polarization is shaded red (blue). Electron-driven dissipation decreases polarization (white) around $r{=}0$. Growing texture size and increasing white region at \change{+/- boundary} indicate spin diffusion; dashed line represents polarization regime.
    (D) (i) \change{\textit{Fourier Transformed $\Cs$ NMR spectrum}} of full $\qt_-{=}30, 34,$ and $37$s data in (B). (ii) FT data for varying $\qt_-$ is plotted on a logarithmic scale comprising $91$ $\qt_-$ slices, separated by $0.2\,$s with a ${>}10^9$ variation in intensity. 
    Spectral narrowing with increasing $\tau_-$ indicates that fast-relaxing \C-nuclear spins near the NV are first depolarized until inversion of total signal with $\qt_-{\app}37$s (boxed region in (A)).
}
\zfl{fig2}
\end{figure*}

\begin{figure*}[t]
  \centering
  {\includegraphics[width=1\textwidth]{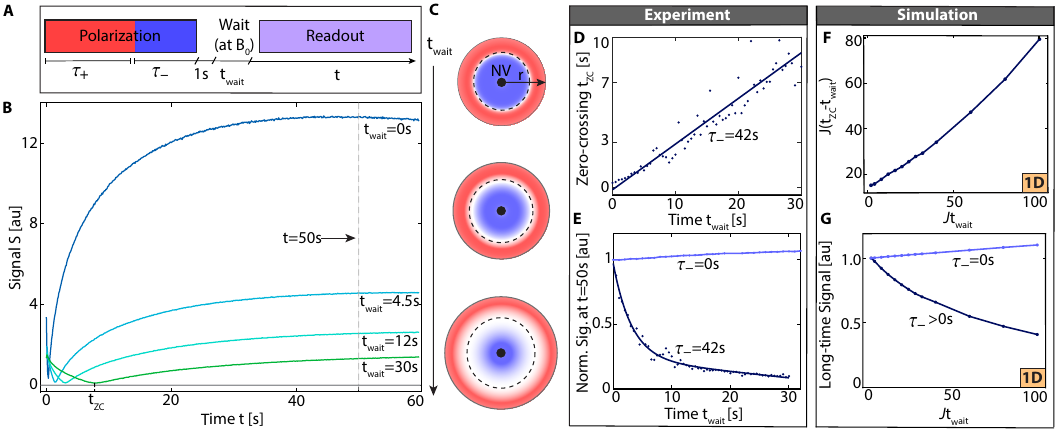}}
       \caption{
\T{Melting of spin-texture due to spin diffusion (State Engineering).}
(A) \I{Experiment schematic.} Waiting period $\tw$ at high-field is introduced after successive $\{\qt_+,\qt_-\}$ spin injection, and prior to application of the \change{periodic} drive in \zfr{fig2}.
(B) Representative signal traces showing changes in decay profiles with variable $\tw$ for texture generated with $\qt_+{=}60\,$s and $\qt_-{=}42\,$s (bolded trace in \zfr{fig2}B). 
Decrease in signal amplitude between $\tw{=}0$ and $\tw{=}30\,$s evidences the instability of state-engineered spin texture. A full movie of this data set, with phase information, can be found in movie S2.
(C) \I{Schematic representation} showing the melting of spin texture during $\tw$ due to spin diffusion at the boundary between polarization layers, dashed line serves as guide to the eye for the domain wall boundary.
(D) Points show movement of the zero-crossing $\tzc$ position with $\tw$ for data in (B). Solid line is a linear fit.
(E) Signal intensity at $t{=}50$s in (B) plotted for different values of $\tw$ normalized to its value corresponding to the $\tw{=}0$ case. Melting of the spin texture due to diffusion manifests as a decrease in signal. Solid line is a biexponential fit. Light blue line corresponds to case without spin texture ($\qt_+{=}60s$, $\qt_-{=}0$).
(F-G) \I{Simulations} corresponding to (D-E) showing zero-crossing times and $\vert \mathrm{min}(\mathcal{I}_x)\vert$ extracted from numerical time-evolution in a one-dimensional short-range model using a similar shell-like initial state (see \zfr{fig5} and Methods). Simulations show qualitative agreement with experimental results.
}
\zfl{fig3}
\end{figure*}

\begin{figure*}[t]
  \centering
  {\includegraphics[width=1\textwidth]{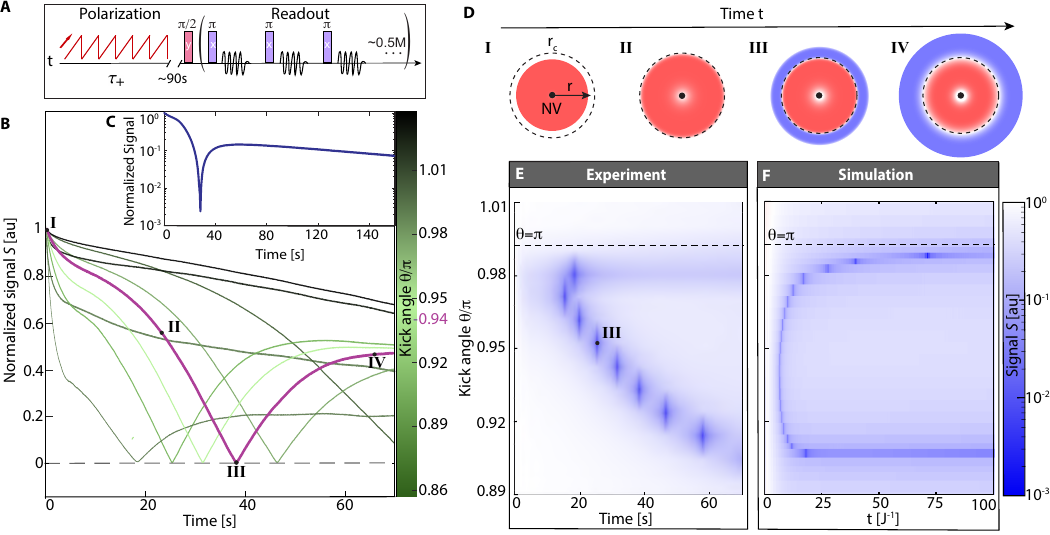}}
 \caption{\T{Robust spin textures by Hamiltonian engineering.}
(A) \I{Experiment schematic.} Spins are hyperpolarized for $\qt_+{=}90$s and subject to a spin-locking train with $\xt{\app}\pi$.
(B) Measured signal $S$ undergoes sharp zero-crossing and associated sign inversion (phase signal is analogous to \zfr{fig2}B(ii) and not shown). Colored lines show variation with $\xt$ (see colorbar). Each trace has ${\app}0.5$M points. Dark purple trace highlights representative data at $\xt/\pi{=}0.94$ for clarity, with sharp zero-crossing at time $\tzc$.
(C) Analogous signal to bolded purple trace in (B), with small frequency offset (see SI~\ref{sec:exp param}). Signal is plotted on a logarithmic scale and extends for $t{>}150$s. Dramatic signal zero-crossing is evident.
(D) \I{Schematic representation} of formed spin texture for key points of the bolded trace in (B) (marked I-IV). Signal zero-crossing arises due to thermalization to an effective Hamiltonian $\mHeff$ bearing spatial texture arising from the NV-imposed gradient (\zfr{fig1}C). Dashed black line indicates the domain boundary at $r_c$. Spin texture remains robust against spin diffusion with $t$ (see \zfr{fig5}A).
(E) \I{Movement of zero-crossing with $\xt$}. 2D color plot showing logarithmic scale visualization of data in (B) plotted with respect to $\xt$ (horizontal slices). Zero crossing appears as an abrupt decrease in signal (colored blue). $\xt{=}\pi$ slice is marked and corresponds to a rapid signal decay. $\tzc$ occurs at later times for smaller $\xt$. Point III corresponding to zero-crossing in (B) is marked.
(F) \I{Numerical simulations} performed with LITE for the simplified short-range model with the experimental data (see ~\zfr{fig5} and SI~\ref{sec:approximate_dynamics}).
}
\zfl{fig4}
\end{figure*}

 \begin{figure*}
	\centering
	{\includegraphics[width=1\textwidth]{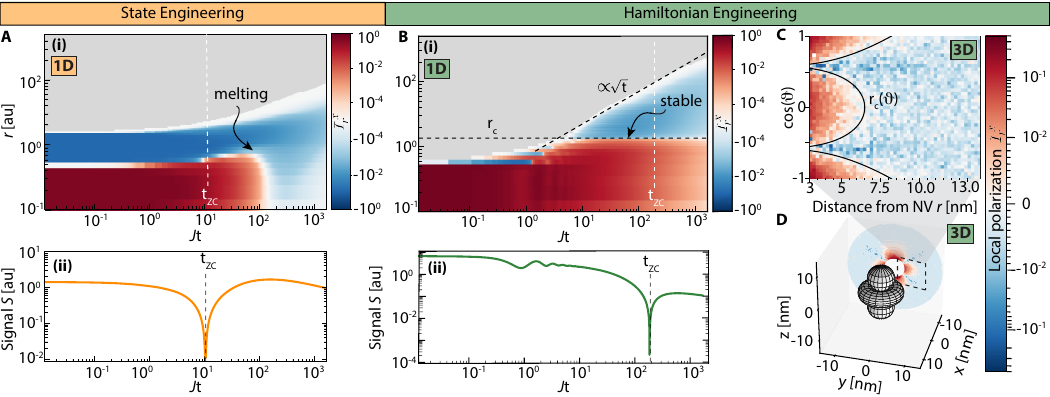}}
		\caption{
    \T{Simulations of spin texture formation and evolution} using the effective model Hamiltonians for (A) \T{State Engineering} and  (B-D) \T{Hamiltonian Engineering}. Panels (A,B) are obtained from \textit{1D short-range simulations} for infinitely extended systems including effects of dissipation using the LITE algorithm~(see SI~\ref{sec:LITE}). System extends in both directions, but only positive values are shown. Horizontal axis displays interrogation time $t$ in $J^{-1}$ units, vertical axis displays distance $r$ from NV in arbitrary units~(a.u.). 
    (A) \I{Spin texture via State Engineering.} We imprint the spin texture in the initial state using 11 positively polarized spins close to the NV center at $r=0$ followed by 18 negatively polarized spins. 
    (i) Colors display polarization $\mathcal{I}_r^x(t)$ (colorbar) at different sites at time $t$ (vertical slices). Panel displays spreading dynamics with $t$. Formed shells melt under diffusion (see SI~\ref{sec:one_d_pi_half}). 
    (ii) Integrated polarization over the spin ensemble, corresponding to signal $S$ measured in \zfr{fig2}. Simulations reveal formation of a sharp zero-crossing in either case ($\tzc$ marked), occurring at the instance of zero total net polarization. 
    (B) \I{Spin texture via Hamiltonian engineering}. The initial state at $t{=}0$ contains $11$ positively polarized spins within a background of spins in a fully mixed state. 
    (i) Polarization $\mathcal{I}_r^x(t)$ (colorbar) showing that spin texture forms via thermalization. Late time behavior (dashed line) follows energy diffusion ${\propto}\sqrt{t}$. Shell critical radius $r_c$ (horizontal dashed line) is stabilized.
    (ii) Integrated ensemble polarization showing formation of zero-crossing analogous to \zfr{fig4}.
    (C, D) \I{Formed shells in three-dimensions} via Hamiltonian Engineering, obtained from late-time dynamics of classical three-dimensional long-range simulations (see SI fig.~\ref{fig:classical_simulation} for full time traces). For the classical simulation dissipation is not considered and spins within the frozen core~(white region in D) are not simulated.
    }
    \zfl{fig5}
\end{figure*}

\tocless\section{Results}{sec:Results}

\tocless\subsection{Spin texturing via hyperpolarization injection}{sec:StateEngineering}
\T{State-Engineered} shells are created by alternately injecting positive and negative hyperpolarization~\cite{Ajoy17,Ajoy_DNPCombs_2018} for periods $\tau_+$ and $\tau_{-}$, respectively (\zfr{fig2}A inset). Direct $\Cs$ hyperpolarization occurs rapidly over a short range, with distant nuclei polarized more slowly via spin diffusion. Consequently, $\Cs$ nuclei close to the NV center are negatively polarized, while more distant nuclei are left positively polarized, yielding the shell-like texture. Varying $\tau_{-}$ at a fixed $\tau_+$ provides a means to tune the shell size.

Spin diffusion is driven by internuclear dipolar interactions with the Hamiltonian
\begin{equation}
\label{eq:Hdd}
    \mHdd = \sum_{k<\ell}b_{k\ell}\left(3 I_{k}^zI_{\ell}^z-\mathbf{I}_k\cdot\mathbf{I}_\ell\right), 
\end{equation}
where $b_{k\ell}{=}J_\mathrm{exp} (3\cos^2(\theta_{k\ell})-1)/{r}_{k\ell}^3$, $J_\mathrm{exp}{=}\mu_0 \hbar \gamma_n^2/4\pi$, $\cos(\theta_{k\ell}) {=} \mathbf{B}_0\cdot\mathbf{r}_{k\ell}/(\vert \mathbf{B}_0\vert r_{k\ell} )$, $\mathbf{r}_{k\ell}$ is the inter-spin vector, $\mathbf{B}_0=B_0\zhat$ is the external field. The median $\Cs$ coupling strength is $\expec{b_{kl}}{=}J{\app}0.6\,$kHz. 

\zfr{fig2}A depicts the \I{net} $\Cs$ polarization during spin injection, for $\tau_+{=}60\,$s, after which the sign of the injected polarization is reversed (dashed vertical line). In the boxed region around $\tpol{\approx}100\,$s ($\tau_-{=}40\,$s,) the total net polarization approaches zero. Although there is limited net polarization, the system can still exhibit locally large spin expectation at different sites. To observe this, the $\Cs$ nuclei are subject to a \change{periodic} protocol at $B_0{=}7\,$T (\zfr{fig2}B(i) inset) involving a series of spin-locking $\xt$-pulses of duration $t_p$, at Rabi frequency $\Omega$, and separated by time interval $\tau$. Spin-locking with $\xt{\napprox}\pi$ yields the leading order effective average Hamiltonian~\cite{eckardt2017colloquium}~(see SI~\ref{subsec:SpinLocking}),
\begin{equation}
\label{eq:Heff_SL}
    \mathcal{H}_{\mathrm{SL}} = - \frac{1}{2} \sum_{k,\ell} b_{k\ell} \left( 3 I^x_k I^x_\ell - \mathbf{I}_k\cdot \mathbf{I}_\ell \right)\, ,
\end{equation}
\change{which conserves $\xhat$-polarization (see SI~\ref{subsec:SpinLocking}); higher-order contributions break the conservation-law leading to a slow decay of $\xhat$-polarization~(quasi-) conservation.}
Here ${>}560,000$ pulses are applied over a period of $t{\gtrsim}60\,$s. \change{The quasi-conserved $\xhat$-polarization results in long transverse lifetimes $T_2^{\prime}$~\cite{Beatrez21}.}


This process can be quasi-continuously tracked in real-time. Between the pulses, rapid and non-destructive (inductive) interrogation of $\Cs$ Larmor precession occurs at a rate of $\tau^{-1}{\sim}10\,$kHz. We record the nuclear polarization amplitude $S$ and phase $\phi_R$ in the $\xy$ plane in the rotating frame, where $\phi_R{=}0(\pi)$ refers to a vector along $\xhat (-\xhat)$. 

Each point in \zfr{fig2}A can therefore be expanded into a secondary dimension $(S,\phi_R)$, as shown in \zfr{fig2}B, considerably increasing the information content compared to optical detection approaches~\cite{Chekhovich_spindiffusion_2023}. Consider first the point \I{(a)} at the top of the polarization buildup curve in \zfr{fig2}A ($\qt_-{=}0$). Considering normalized signal $S$ in \zfr{fig2}B (black dashed line), a long-lived decay bearing lifetime $T_2^{\prime}{\app}93.4\,$s is evident, reflective of prethermalization to $\mathcal{H}_{\mathrm{SL}}$ and the resulting $\xhat$ (quasi-)conservation~\cite{Beatrez21}. In comparison to this slow, featureless, decay, normalized signal profiles, the zero total net polarization region (boxed in \zfr{fig2}A) exhibits distinct differences. For clarity, we consider first the representative trace (bold orange) in \zfr{fig2}B(i) at $\qt_{-}{=}44.8\,$s. Normalized $S$ here features a sharp zero-crossing at $t{=}\tzc$(${\app}11.34\,\mathrm{s}$), accompanied by a simultaneous reversal in total net polarization sign from $-\xhat$ to $+\xhat$ (\zfr{fig2}B(ii)). The region in the zero-crossing vicinity (1\% variation) itself encompasses ${\sim}10^4$ points, showcasing the rapid and non-invasive dynamics sampling — $\qt^{-1}{>}14J$ in these experiments. Additionally, the data captures dynamics up to very long times, $Jt{>}1.2{\zt}10^6$, surpassing comparable, local Hamiltonian engineering experiments on different platforms by several orders of magnitude~\cite{Roos_emergenthydro_2022,Cappellaro_disorder_2023,zu2021emergent}.

The other colored traces in \zfr{fig2}B(i) show corresponding signals $S$ for different $\qt_-$ (see colorbar). The zero-crossing point $\tzc$ shifts to the right with increasing $\tau_{-}$. The extracted $\tzc$ values are elaborated in \zfr{fig2}B(iii). A movie showcasing data for 151 changing values of $\tau_{-}$ is accessible in movie S1.

Quasi-conservation of the net $\xhat$-polarization naively suggests signal curves that appear almost constant on time-scales $\ll T_2'$, regardless of the spatial $\xhat$-polarization structure imposed on the initial state. However, this would contradict our observations; to rationalize the data in \zfr{fig2}, recall that the NV electron is strongly coupled to a phonon bath while the $\Cs$ spins are only weakly coupled to it. Consequently, the NV acts as a dominant local relaxation source for the $\Cs$ nuclei and the dynamics of the couple $\Cs$ spins has to be described within the framework of open systems. Thereby, proximal nuclei dissipate polarization (${\sim}1/r^6$) faster than more distant ones, as can be derived with the Lindblad master equation in the strong coupling limit (see Methods and Sec.~\ref{sec:SI-open-system}). Measurements at \I{longer} times $t_m$, therefore, serially probe nuclei \I{further} away from the NV as $\Cs$ closer to the NV (with relaxation times $<t_m$ do not contribute to the commutative signal.

\zfr{fig2}C schematically represents the polarization distribution during the interrogation period $t$, focusing on specific points (I-IV) along the bold orange trace ($\qt_{-}{=}44.8$ s) in \zfr{fig2}B. For simplicity, we depict the generated texture as spherical shells, although in reality, it possesses an angular dependence inherited from the hyperfine interaction (see \zsr{numerics}). Starting from the initial texture ($t{=}0$), proximal $\Cs$ polarization undergoes dissipation from the NV center (shown white), gradually revealing polarization at greater distances as $t$ increases. Zero-crossing at $t{=}\tzc$ corresponds to an equal distribution of positive and negative polarization (\zfr{fig2}C II). Further dissipation leads to a polarization sign inversion (\zfr{fig2}C III-IV). Overall, \zfr{fig2} illustrates the ability to discriminate spins without relying on the electronic “frozen core”, constituting a departure from previous work ~\cite{Jannin_spindiffusion_2021,Chekhovich_spindiffusion_2023}.

The different stages in \zfr{fig2}C exhibit distinct signatures in the $\Cs$ NMR spectrum. In \zfr{fig2}D we report  the $\Cs$ spectrum for 91 $\qt_-$ values, separated by $0.2\,$s intervals, obtained by applying a Fourier transform (FT) to the sign-corrected data in \zfr{fig2}B. The Fourier intensity is shown on a logarithmic scale spanning over nine orders of magnitude. The wide dynamic range reflects the high signal-to-noise ratio in our experiments. $\Cs$ spins closer to the NV produce broader spectral lines because they experience faster relaxation, manifesting as stronger contributions to the spectral wings around 0Hz. Conversely, more distant $\Cs$ nuclei are centrally located in the spectrum due to longer relaxation times. With increasing $\qt_{-}$, depolarization initially affects the spectral wings, resulting in an apparent narrowing of the spectrum at $\qt_{-}{\app}37\,$s. Subsequently, the inversion of the central feature, corresponding to bulk $\Cs$ nuclei, follows suit. 

Spin texture generated via \T{State Engineering} (\zfr{fig2}) is not intrinsically stable. Over time, any imprinted domain boundaries begin to dissolve due to spin diffusion. Such “melting” of spin texture can be observed in experiments where a delay period $\tw$ (${<}30\,$s) is introduced before the (shell-like) $\zhat$-polarized state is rotated into the $\xhat$-$\yhat$ plane and the driving is started (see \zfr{fig3}A). During $\tw$, NV-driven dissipation of the $\zhat$-polarized texture is negligible. This is due to the increased energy gap ($\gamma_n B_0$, as opposed to $\Omega$, see SI~\zsr{relax}), and is reflected in the extremely long nuclear lifetimes $T_{1n}{\app}1\,\mathrm{h}{\gg}T_2^{\prime}$ under these conditions.  Consequently, spin dynamics in the $\tw$ period is primarily affected by spin diffusion. Simulations based on a Lindblad master equation confirm this picture (see Methods).

\zfr{fig3}B shows the signal $S$ dependence on the delay period $\tw$, starting from a spin texture produced with $\qt_+{=}60\,$s and $\qt_-{=}42\,$s. Spin diffusion gradually homogenizes (flattens) the polarization distribution as $\tw$ increases (\zfr{fig3}B). This leads to a rightward shift in the zero-crossing point $\tzc$ and a decrease in signal amplitude, as evident in \zfr{fig3}B, and as schematically depicted in \zfr{fig3}C. 

The spin diffusion-mediated flattening of the polarization distribution can be directly observed 
by considering the signal decreasing with $\tw$ for a fixed value of $t$. 
\zfr{fig3}E shows $S(\tw)$ at $t{=}50\,$s, normalized against the value at $\tw{=}0$ to highlight changes with $\tw$. For a homogeneously polarized state ($\qt_-{=}0$), the thus normalized signal increases slightly with increasing $\tw$ due to polarization diffusing away from the NV, and hence being subjective to lower effective relaxation from it.
These observations are supported by numerical simulations (\zfr{fig3}F-G) obtained by solving the Lindblad equation for a simplified toy model (see \zsr{numerics} and SI sec.~\ref{sec:LITE}). All qualitative features and intuition from the microscopic dynamics are well reproduced.

\tocless\subsection{Robust spin shells by Hamiltonian engineering}{sec:HamiltonianEngineering}

To stabilize the generated spin texture, we introduce a second approach of \T{Hamiltonian Engineering}.
This stems from a surprising observation: when deploying the pulse train with flip angle $\xt {\approx} \pi$ (\zfr{fig4}A) and starting with $\Cs$ spins positively polarized ($\qt_{+}{=}60\,$s, $\qt_{-}{=}0\,$s), we observe that over time, the signal $S$ exhibits a sharp zero-crossing and subsequent sign inversion that can persist longer than $180\,$s.
This is highlighted in representative traces in \zfr{fig4}B-C, on linear (B) and logarithmic (C) scales, respectively, both with ${>}10^5$ points. The emergence of spin shells here is counterintuitive. 
Since for $\vartheta{\approx}\pi$ no conservation law shields the initial state from rapid heat-death, one would expect the interacting nuclear spin system to quickly relax to a featureless infinite-temperature state~\cite{abanin2015heating,dalessio2016quantum}.
Despite extensive use of $\pi$-trains (CPMG experiments~\cite{Carr_CPMG_1954,Meiboom_CPMG_1958}) in various contexts including dynamical decoupling and quantum sensing, to the best of our knowledge, this phenomenon has hitherto not been previously reported. 

We attribute the emergence of the long-lived signal to Hamiltonian engineering facilitated by the simultaneous action of the nanoscale electronic field gradient dressed by the $\pi$-train drive.
In particular, the NV electron spin, thermally polarized (${\approx}12.6\,\%$ polarized at $9.4\,$T and $100\,$K), induces a \change{hyperfine magnetic gradient field $\nvfield({\mathbf{r}})$} on the nuclear spins.
As a consequence, the direction and magnitude of the Rabi field $\Omega$ experienced by the nuclei depends on their proximity to the electron (see \zfr{fig1}A).
When $\xt{\neq}\pi$ (cf.~\T{State Engineering}, \zfr{fig2}-\zfr{fig3}), this merely alters the \change{(quasi-)conservation} axis $\xhat{\rt}\xhat'(r)$ (see SI~\ref{subsec:SL_notpi}).
On the other hand, when $\xt{=} \pi$ exactly and we ignore the NV-induced gradient field, the absence of $\xhat$-polarization conservation leads to a rapid, yet trivial, decay of the spins within $T_2^*$ ~\cite{Beatrez21}.

However, in the presence of the NV gradient field $\nvfield({\mathbf{r}})$, for $\xt{=}\pi + \epsilon$ (with $\left|\epsilon\right| {\ll} \pi)$, spin dynamics is governed by the effective average Hamiltonian (see SI~\ref{subsec:SL_pi})
\begin{equation}
\label{eq:Heff}
    \mHeff = \mHdd + \sum_\mathbf{r} \potential_{\mathbf{r}} I_{\mathbf{r}}^x \, ,
\end{equation}
where $\potential_{\mathbf{r}} = \nvfield({\mathbf{r}}) + \epsilon/\period$ denotes the effective spatially-varying on-site potential (\zfr{fig1}C).
We operate in the regime of a weak electronic field gradient ($\potential_{\mathbf{r}} {<} 1\,$kHz), far from the frozen-core limit — a significant departure from previous experiments~\cite{Abobeih_atomicimaging_2019}.
\change{Notably, for $\epsilon<0$, with increasing distance ($r{=}\norm{\bm{r}}$) from the NV, the spatial inhomogeneity in $\potential_{\bm{r}}$ flips sign at a critical radius $r_c$ from positive~($\potential_{\bm{r}}{>}0$) in proximity to the NV~($r{<}r_c$) to negative~($\potential_{\bm{r}}{<}0$) on-site potentials for far-away nuclear spins~($r{>}r_c$); the exact position of $r_c$ is determined by the $\Cs$ slice for which $\potential_{\bm{r}} {=} 0$~(\zfr{fig1}C).}
Note that here we ignore the angular $\theta$ dependence for simplicity and return to it in Sec.~\ref{sec:numerics}.

Crucially, the sign change of $\potential_{\mathbf{r}}$ on either side of $r_c$ means that $\mHeff$ internally encodes spatial structure, and $r_c$ serves as the domain boundary.
Then, irrespective of the initial state, applying the Eigenstate Thermalization Hypothesis~\cite{Srednicki1995,rigol2008thermalization,Deutsch2018}, one predicts that the spins thermalize to the quasi-equilibrium \change{state} $\rho_{\R{eq}} {\propto} \exp(-\beta \mH_{\R{eff}}) {\approx}  \mathds{1}-\beta \mH_{\R{eff}}$, with the \change{energy of the initial state} determining \change{by the free parameter} $\beta$.
The \change{spatially resolved} $\xhat$-polarization in the thermal state, $\mathcal{I}_\mathbf{r}^x{=}\mathrm{Tr}(I_\mathbf{r}^x \rho){\app} -\beta\potential_{\mathbf{r}}/2$, then attains opposite signs on either side of $r_c$\change{; thus, the spatially inhomogeneous field $\potential_{\bm{r}}$ imprints the} spatial structure of a quasi-equilibrium shell stabilized within the prethermal plateau (\zfr{fig1}B). 

The highlighted traces in \zfr{fig4}B-C track such shell formation.
\change{The counterintuitive formation of negative polarization during thermalization is a direct result of the lack of polarization conservation, and hence the absence of polarization diffusion; instead, the thermalization dynamics can be understood in terms of energy diffusion.}
The initial polarization profile $\xr_{\R{i}}$\change{, being positively polarized and localized in the proximity of the NV,} (\zfr{fig4}D I) is far from the prethermal \change{shell-like} equilibrium $\xr_{\R{eq}}$ (\zfr{fig4}D IV): it possesses a finite amount of energy with respect to $\mHeff$, which is initially localized within the polarized region.
Over time this energy diffuses through the system due to energy (quasi-)conservation~(see SI~\ref{sec:energy_diff}).
\change{When energy diffuses beyond the critical radius $r_c$, spins in the region $r{>}r_c$ start aligning with the on-site potential $\potential_{\bm{r}}$;} consequently, spins in the region $r{>}r_c$ begin to sequentially flip, driving the system, towards a quasi-equilibrium, $\xr_{\R{i}}{\rt}\xr_{\R{eq}}$ (see \zfr{fig4}D III).
As the negative polarization in the outer region ($r{>} r_c$) surpasses the positive polarization in the inner region ($r{<} r_c$) in magnitude, the total polarization along the $\xhat$ direction undergoes a sign inversion, resulting in a zero-crossing
Importantly, since the spin texture \change{corresponds to the quasi-equilibrium state} and total polarization is not conserved, the domain boundary $r_c$ remains stable on prethermal timescales \change{($t<T_2^\prime$)}.
This stands in contrast to \T{State Engineering} (see \zfr{fig5}). Furthermore, the signal drop observed here is orders of magnitude larger compared to \T{State Engineering}, since we do not operate in a regime of nearly zero total net polarization, indicating a larger spin texture gradient~(Methods).

Experiments in \zfr{fig4}B-C, therefore, uniquely enable real-time observation of spin flipping dynamics during the prethermalization process. This capability, combined with the large accessible values of $Jt\,{>}10^6$, allows us to investigate thermalizing spin dynamics with exceptional resolution over extended time periods, surpassing the capabilities of previous experiments by many orders of magnitude ~\cite{Roos_emergenthydro_2022,Cappellaro_disorder_2023,zu2021emergent}. 

Colored traces in \zfr{fig4}B show similar experiments for different values of $\xt$ in the vicinity of $\xt{=}\pi$ (see colorbar). The zero-crossing position is observed to change with $\xt$. A movie of this data is accessible in movie S3. \zfr{fig4}E recasts this movie for 120 different values of $\xt$ in a 2D plot on a logarithmic scale in intensity, where the darker blue colors highlight the zero crossing. This visualization clearly elucidates the movement of $\tzc$ with $\xt$. Additionally, a distinct slice in the plot displays rapid signal decay (indicated by the dashed line). This is attributed to bulk $\Cs$ nuclei, which are far removed from the NV influence, corresponding for which $\xt{=} \pi$.
The qualitative behaviour can be reproduced in numerical simulations (see Methods and Refs.~\cite{Kvorning2022,Artiaco2024}). While the exact shape of the zero-crossing arc depends on the details of the model (such as $\potential_{\mathbf{r}}$), the simulations capture the diverging behaviour of $\tzc$ found for $\vartheta \to \pi^-$.

Based on data in \zfr{fig4}, we estimate that the domain spans ${r_c{\app}2.7\,\mathrm{nm} \,\cdot\, \sqrt[3]{|3\cos^2\theta-1|}}$ encompassing ${\approx}150$ spins for the representative line in \zfr{fig4}B~($\xt{=}0.94\,\pi$)~(see Methods and SI Sec.~\ref{subsec:crossing_radius}).
We are able to exert control over the $r_c$ domain boundary by \change{engineering the effective Hamiltonian}~\eqref{eq:Heff}. By adjusting the kick-angle $\xt$ we can modify the effective on-site potential $\potential_{\bm{r}}$, and hence $r_c$, allowing us to tune the shell size between tens~(${\app}50$) up to mesoscopic numbers~(${\gtrsim}300$) of spins~(see SI Sec.~\ref{subsec:crossing_radius})~\cite{Terblanche_C13_1997}. We observe that the movement of $\tzc$ matches the theoretical prediction (\zfr{fig4}E,F).
Varying Rabi frequency $\Omega$ or the frequency offset can modify the Hamiltonian and allow further control over $r_c$ (SI \zsr{offset}).
Furthermore, thermalization to shell-like texture is found to be robust to lattice orientation (SI \zsr{orient}) or initial states employed (SI \zsr{hyp_time}).

\tocless\subsection{Numerical Results}{sec:numerics}
Quantum simulations here use Local-Information Time Evolution (LITE)~\cite{Kvorning2022,Artiaco2024}, a recently developed algorithm suited to investigate diffusive and dissipative quantum dynamics. Numerical tractability constrains us to work with a simplified one-dimensional toy-model Hamiltonian, but which captures all relevant features of $\mathcal{H}_{\mathrm{eff}}$~(Methods).

Figure \ref{fig:fig5}A(i) shows simulations corresponding to \T{State Engineering} experiments. Time is represented on the horizontal axis in units of \change{in units of the inverse median coupling} ($J^{-1}$), the vertical axis represents the distance from the NV (both axes are on a logarithmic scale), and the colorbar shows the local $\xhat$-polarization $\mathcal{I}_r^x$. Considering an initial shell-like texture produced by spin injection, we observe that the subsequent dynamics are characterized by melting domain walls, causing the magnetization gradient to diminish over time. Figure~\ref{fig:fig5}A(ii) shows the total net polarization $S{=}|{\sum_r \mathcal{I}_r^x}|$ — analogous to that measured in experiment. We observe a zero-crossing similar to \zfr{fig2}B. Notably, in \T{State Engineering}, we rely on dissipation to enable the observation of the sign-inversion of the total polarization: in the absence of dissipation,  total $\xhat$ polarization is (quasi)-conserved and thus will appear constant on prethermal timescales.

In contrast, \zfr{fig5}B considers \T{Hamiltonian Engineering}, and polarization initially confined in the vicinity of the NV. Over time, by virtue of energy diffusion, the spins prethermalize to a polarization gradient under the NV-induced potential. Peripheral spins become endowed with negative polarization (blue region), and ultimately, over time, the total polarization inverts as the majority of spins feature negative polarization. Importantly, the domain $r_c$ (dashed horizontal line) separating the local positive and negative polarization regions is solely set by the on-site potential $\potential_\mathbf{r}$ present in $\mathcal{H}_\mathrm{eff}$ and is, therefore, stationary. As the spins thermalize, the (non-conserved) polarization adapts to the \I{stable} gradient profile imposed by $\mathcal{H}_\mathrm{eff}$. Irrespective of the initial state, the diffusion-propelled dynamics therefore induce a polarization gradient with a time-invariant domain boundary.

Considering the net polarization (\zfr{fig5}B(ii)), we observe a steep zero-crossing in qualitative agreement with experiments. We note that the zero-crossing $\tzc$ arises even in the absence of dissipation (see SI~\ref{sec:one_d_pi}), as observed in the long-time slices in \zfr{fig5}B(i). Dissipation only serves to extinguish polarization at distances close to the NV (i.e., accelerating the inevitable diffusion-induced polarization inversion).

To demonstrate that these results hold beyond one-dimensional short-range models, we additionally perform classical simulations based on a three-dimensional long-range model comprising ${\approx}10^3$ dipolar-interacting spins on a diamond lattice (Methods). In contrast to the one-dimensional quantum simulations, here the system is finite and free of dissipation so that equilibrium is reached in finite time. Notably, here, the system thermalizes to a three-dimensional spin texture which exhibits the analytically predicted $\theta$-dependence of the NV gradient field with $r_c$ spanning several nanometers (cf. Sec.~\ref{sec:HamiltonianEngineering}).

\tocless\section{Discussion}{sec:discussion}
Our experiments introduce several novel features. The creation, stabilization, and observation of spin texture are achieved through \I{global} spin control and readout.  Despite this, a good degree of local nuclear discrimination is shown to be attainable by utilizing an electron as a \I{controllable} spin injector, gradient source, and dissipator.  Notably, unlike other experiments~\cite{Abobeih_atomicimaging_2019}, our approach is not confined to the constraints presented by diffusion barrier limits,  enabling examination of a mesoscopically large number of nuclear spins around each electron. The injected hyperpolarization exceeds the Boltzmann polarization levels by orders of magnitude compared to previous studies~\cite{Jannin_spindiffusion_2021}. 

Our method of continuous interrogation in the rotating frame, as opposed to point-by-point probes in the lab-frame~\cite{Jannin_spindiffusion_2021}, offer a distinct methodological advantage.
It facilitates a rapidly sampled ($J\qt{<}0.07$) visualization of polarization dynamics while reaching into the very-long-time regime ($Jt{>}1.2{\zt}10^6$).
The latter is orders of magnitude beyond previous related  experiments~\cite{Roos_emergenthydro_2022,Cappellaro_disorder_2023,zu2021emergent}, permitting direct observation of emergent stabilization. The dynamics observed complement recent experiments with cold atoms~\cite{gring2012relaxation,kaufman2016quantum,polkovnikov2016thermalization,wei2022quantum,morningstar2022anomalous}, but occur in a distinct and novel context: at the nanoscale, in the solid-state, and in the limit of strongly interacting dipolar-coupled spins (for which $JT_{2}^{\ast}{\app}1$).

Our dynamical stabilization protocol operates distinctly out of equilibrium.
The dynamics is governed by an emergent quasi-conservation of energy and explicitly breaks polarization conservation, allowing us to inhibit \change{polarization} diffusion.
In conjunction with the electronic hyperfine field, this protocol stabilizes localized, shell-like spin textures, even in the absence of local control.
Note that no obvious static protocol could yield similar outcomes in our system. 
The underlying physics of this dynamical stabilization deviates from traditional state engineering approaches, is versatile, and can encompass a variety of models (short- or long-range, clean or weakly-disordered interacting systems, quantum or classical models), across different dimensions.

The elucidated \T{Hamiltonian Engineering} protocol, therefore, highlights the untapped capabilities of nonequilibrium control techniques in manipulating physical systems. For example, we envision our approach to be utilized within quantum simulators to induce on-demand magnetic domain wall states, which can potentially be useful for investigations of spin dynamics \cite{ljubotina2017spin,wei2022quantum,rosenberg2023dynamics}.
Depending on the structure of the effective Hamiltonian and the initial state, one can also stabilize and investigate spin textures at negative-temperature~\cite{braun2013negative}. 
Our work, therefore, presents a first instance of a novel idea that demonstrates the applicability of concepts from the emerging field of nonequilibrium control, to engineer stable many-body states with tailored attributes~\cite{boyers2019floquet}. 

Lastly, the manipulation and control of the orientation of spin polarization at sub-nanometer length scales itself, may open up several promising future directions.
The use of ubiquitously occurring nuclear spins, as elucidated in this work, can broaden the application of spin texturing to diverse systems, moving beyond previously considered magnetic materials~\cite{Awschalom_ferromagtex_2001,Awschalom_spinpatterns_2003,Awschalom_doughnuts_2004,Kensuke_VortexBeam_2023,king_optically_2012,West_GaAsWells_1995}.
\change{
This may enable applications in quantum memories~\cite{king_optically_2012,Meriles_longterm_2016} and solid-state systems at low temperatures, such as phosphorus-in-silicon systems~\cite{Madzik2020,Morello2020}, where spin texturing within nuclear spins (e.g., silicon-29) around phosphorus donors could create a memory resource that can be used in conjunction with the donor electronic qubit, similar to Refs.~\cite{Sullivan2022,Muhonen2014,Boehme2012}. 
While our experiments have been performed on nuclear spins, our theoretical results apply to any spin system, e.g., electronic spins, opening possible applications in spintronics~\cite{Awschalom_ferromagtex_2001,Awschalom_doughnuts_2004} through nanoscale electronic spinbath texturing~\cite{Reimer10,Doherty16,Lozovoi2022,Gulka2021,Lozovoi2023}.
Moreover, by harnessing hyperpolarized nuclei as sensors~\cite{Sahin21} the results in this work may be applied to spatiotemporal quantum sensing.}
Controllable spin textures may also be applied to non-invasive nanoscale chemical imaging in materials science and biology. We envision employing targeted electron spin labels to map the radial spin densities of different nuclei (e.g., $\Hs$ and $\Cs$) within molecules.


\tocless\section{Materials and Methods}{sec:methods}

\begin{center}
    \textbf{EXPERIMENT}
\end{center}
\vspace{-3em}

\tocless\subsection{Sample}{sec:sample}
Experiments here employ a single-crystal diamond sample (3x3x0.3mm) from Element6. It contains $\Cs$ spins at natural abundance $(1.1\,\%)$, and ${\app}1\,$ppm NV$^{-}$ concentration, corresponding to inter-NV spacing of ${\sim}25\,$nm, and each NV center has  ~$10^4 \Cs$ surrounding it. The sample also hosts ${\gtrsim}20\,$ppm of substitutional nitrogen impurities (P1 centers). For experiments in \zfr{fig2}-\zfr{fig4}, the sample is oriented such that its [100] face is approximately parallel to $B_0$. However, as elucidated in  SI, \zsr{orient}, these results are qualitatively independent of the sample orientation.

\tocless\subsection{Experimental Apparatus}{sec:expt}


\begin{figure}[t]
  \centering
 {\includegraphics[width=0.49\textwidth]{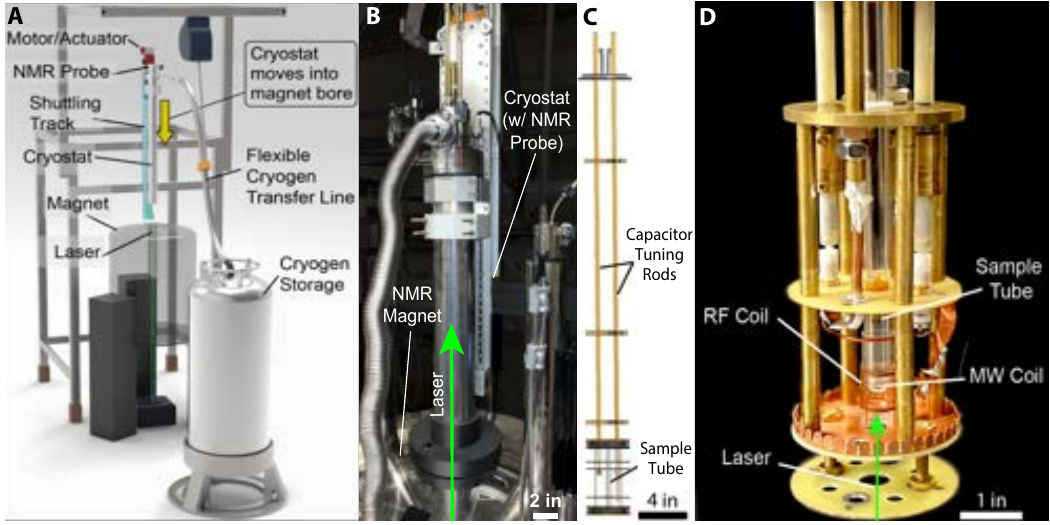} }
     \caption{\T{Novel instrumentation for cryogenic DNP.} (A) \I{Device construction.} CAD model showing instrument and highlighting main components. It consists of a $9.4\,$T superconducting magnet, surrounded by an aluminum frame to which a 4K cryostat is mounted on a belt-driven actuator, bearing a high-torque motor. Laser illuminates sample from the bottom. (B) Photograph showing key features of the instrument. Cryostat is shown at low-field position $\Bpol$; actuator truck and mounts are visible. (C) CAD model showing the schematic of the NMR/DNP probe. It fits snugly within the cryostat, providing additional RF shielding. ${>}3$ft long tuning rods extend through the top of the probe and allow for RF cavity tuning and impedance matching, even under vacuum and cryogenic conditions. (D) \I{Photograph} of constructed probe. A close-up of coil arrangement is shown, highlighting the RF saddle coil and centrally placed microwave loop employed for hyperpolarization. Aperture at probe base enables optical access to the sample.}
\zfl{method_instrument}
\end{figure}

State Engineering experiments (\zfr{fig2}-\zfr{fig3}), carried out at room temperature, employ an apparatus described before in~\cite{Ajoy_widecycle_2019}. For Hamiltonian Engineering (\zfr{fig4}), on the other hand, we introduce a novel instrument (\zfr{method_instrument}) for $\Cs$ hyperpolarization and interrogation at cryogenic temperatures. Low temperatures allow access to higher Boltzmann electronic polarization, consequently stronger electronic gradient fields, and slower electronic relaxation rates. The instrument supplies for the first time (to our knowledge) a cryogenic field cycling capability for optical dynamic nuclear polarization (DNP) experiments, allowing simultaneous operation at variable fields (1mT-9.4T), and controllable cryogenic temperatures down to 4K (although we restrict ourselves to 77K in this work).

 The device utilizes an Oxford SpectrostatNMR cryostat under continuous flow cryogenic cooling, which is mechanically moved (``shuttled") from lower (few mT) fields into a $B_0{=}9.4\,$T NMR magnet (Oxford). The low-field position situated 640mm above the magnet center ($\Bpol{=}36\,$mT) is employed for hyperpolarization; the cryostat is then shuttled to high-field ($9.4\,$T) where the $\Cs$ nuclei are interrogated. Shuttling occurs via a belt-driven actuator (Parker) powered by a motor (ACS) fitted with a high-torque gearbox for enhanced load-bearing capacity for the heavy (25lb) cryostat. The actuator carries a movable stage to which two custom-designed clamps secure the cryostat. A 1.6 m flexible transfer line allows for continuous cooling during shuttling and operation over ${>}$1 week. Shuttling occurs at 7 mm/s and takes ${\app}90\,$s. Long $\Cs$  $T_1$ lifetimes (${>}1$hr) at fields exceeding 0.1T, mean that there is minimal loss of polarization during shuttling. We note that room temperature experiments (\zfr{fig2}-\zfr{fig3})), in contrast, involve shuttling in ${<}1$s.

The diamond sample is secured at the bottom portion of the cryostat in a custom-built NMR/DNP probe (\zfr{method_instrument}). The probe is top-mounted in the cryostat and includes two coils: a loop through which microwaves (MWs) are applied for DNP, and a saddle coil for NMR detection. A custom arrangement employing O-rings at the probe top allows the NMR coil to be frequency-tuned and impedance matched without breaking vacuum. 

For optical DNP at $\Bpol$, an optical window at the cryostat bottom permits illumination by a laser ($\xl{=}532$nm, Coherent) directed into the bore using a 45$^\circ$ mirror. A pair of piezo-driven Zaber mirrors ensure optimal alignment into the cryostat center, aided by a camera mounted with a 637nm long-pass filter at the top of the cryostat. A TTL-triggered mechanical shutter controls illumination timing to within 1$\mu$s.

Hyperpolarization employs MW chirps applied across the NV EPR spectrum. MWs are generated by a Tabor Proteus arbitrary waveform transceiver (AWT) and gated by a Mini-Circuits ZASWA-2-50DR+ switch. MWs are amplified in two stages by ZHL 2W-63-S+ and  ZHL-100W-63+ amplifiers. A Varian VNMRS console generates the RF pulses at the $\Cs$ Larmor frequency (${\app}$100MHz), while the NMR signal is detected in windows between the pulses, filtered, and amplified by a Varian preamplifier, and digitized by the AWT. We apply ${>}0.5$M pulses, yielding a continuously interrogated NMR signal for up to $t{=}180$s that is non-destructive since spins are only weakly coupled to the RF coil. Timing of all events — lasers, MW application, mechanical shuttling, triggering NMR detection, and signal digitization — is synchronized by the Swabian pattern generator, controlled by MATLAB.

\tocless\subsection{Hyperpolarization Methodology}{sec:hyp}
Hyperpolarization follows ~\cite{Ajoy_DNPCombs_2018}. Application of a continuous-wave 532nm green laser (5W) induces polarization in the NV electrons to the $m_s{=}0$ state at $\Bpol$. This is transferred to $\Cs$ nuclei through successive traversals of rotating frame Landau-Zener level anti-crossings. Practically, this is accomplished by utilizing MW chirps generated by a Tabor Proteus AWT, sweeping across the NV center EPR spectral bandwidth (25 MHz) at a rate of 200 Hz. The laser is turned off before the sample is shuttled into the magnet. For State Engineering experiments, bulk polarization typically reaches levels of 0.6\%~\cite{Beatrez2022}.

\tocless\subsection{Data Acquisition and Processing}{sec:acq}
Induced $\Cs$ Larmor precession signal into the RF saddle coil is captured by a Tabor Proteus AWT, and digitized every 1ns in $\tacq$ windows between the pulses. It is decimated to preserve memory and improve acquisition and processing speed (here 64-fold). A Fast Fourier Transform (FFT) is applied to extract signal amplitude $S$ and phase $\xph$ in each $\tacq$ window. Phase signals in each window are linearly offset by the phase accrued during $t_p$ pulse periods, a phase unwrapping algorithm is used to unfold this trivial phase development, yielding phase $\xph_R$ of the spins in the rotating frame. $(S,\xph_R)$ forms the basis for the experimental data in this work.

\begin{figure}[t]
  \centering
 {\includegraphics[width=0.49\textwidth]{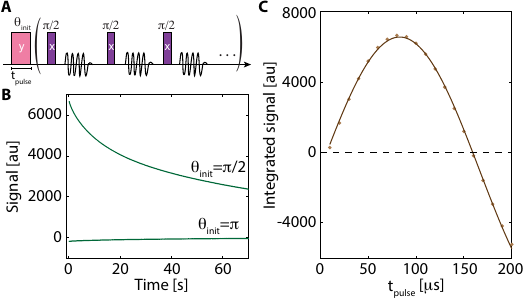}}
  \caption{\T{Estimation of flip angle $\xt$}. (A) Schematic of spin-locking Rabi experiment, consisting of a train of $\pi/2$ pulses, where the angle of the first pulse $\xt_{\R{init}}$, is varied. (B) Typical signals in case of $\xt_{\R{init}}{=}\pi/2$ and $\xt_{\R{init}}{=}\pi$, with total measurement time of $t{=}70$s, corresponding to $>0.5$M pulses. (C) Variation of integrated signal $\Sint$ with length of the first pulse. Solid line represents sinusoidal fit. Slight off-resonance in the pulses leads to a finite signal even at $\xt_{\R{init}}{=}0$.}
\zfl{method_rabi}
\end{figure}

\begin{figure}[t]
  \centering
 {\includegraphics[width=0.49\textwidth]{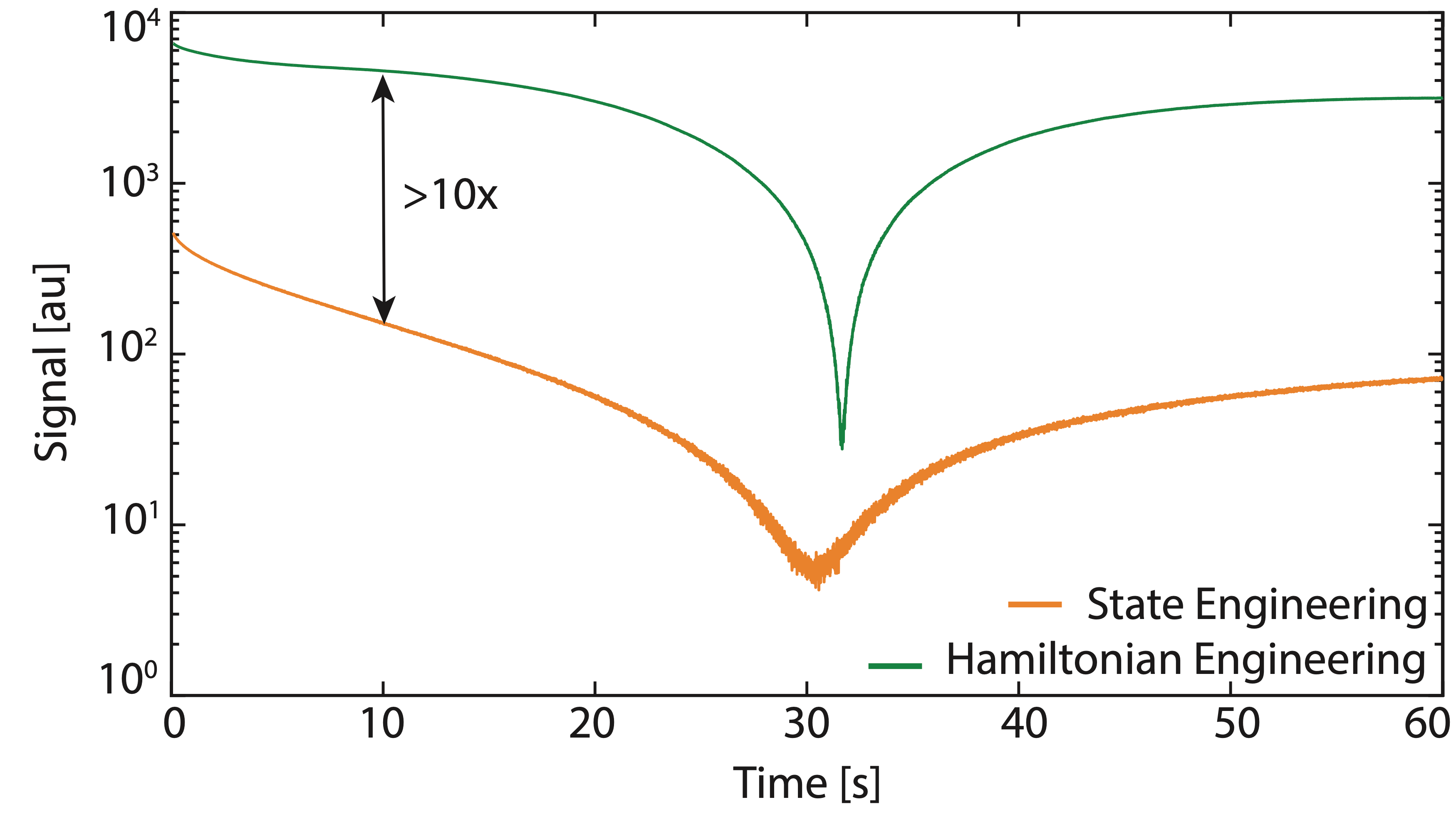}}
  \caption{\T{Comparison of State and Hamiltonian Engineering signals}. Comparison of signal $S$ for State Engineering data in main text \zfr{fig2} (taken at room temperature) and Hamiltonian Engineering data from main text \zfr{fig4} (taken at $100\,$K) for comparable zero crossing times $\tzc{\app}31\,$s. Data is shown on a logarithmic scale for clarity. Measured signal is over an order of magnitude greater in the Hamiltonian Engineering method.}
\zfl{method_compare}
\end{figure}

\tocless\subsection{Estimation of Kick Angle $\xt$}{sec:rabi}
To estimate kick angle $\xt$, important in \zfr{fig4}, we conduct $\Cs$ Rabi experiments and fit the obtained nutation to a sinusoidal curve. A spin-locking train of ${\sim}\pi/2$ pulses and continuous readout (see \zfr{method_rabi}B) for $t{=}70\,$s is employed to enhance measurement SNR. The pulse length of the initial pulse $\xt_{\R{init}}$ is varied, and the integrated signal is measured, as shown in \zfr{method_rabi}C. High SNR yields a high-precision estimate of $\xt$, with an error of within $1\,\%$. 

\tocless\subsection{Comparison of Signals in State and Hamiltonian Engineering}{sec:Compare}
For clarity, we contrast in \zfr{method_compare} the measured signal $S$ amplitude for experiments of State engineering (\zfr{fig2}-\zfr{fig3}) and Hamiltonian engineering (\zfr{fig4}). Spin texture created by State Engineering starts from a regime of \I{low} net polarization (\zfr{fig2}A), and sign inversion (characterized by the $\tzc$ zero-crossing), occurs due to electronic dissipation. On the other hand, Hamiltonian engineering starts with large net polarization, and sign inversion occurs on account of thermalization into the imposed potential. As a result, the measured signal, and the magnitude of the zero-crossing drop in the Hamiltonian engineering method is over one order of magnitude larger than State engineering (see \zfr{method_compare}). The strong signal and robustness of the domain boundaries highlight the advantages of Hamiltonian engineering for stable spin texture generation and readout.

\tocless\subsection{Additional Results on Hamiltonian Engineering}{sec:additional}
To further validate the physical picture in \zfr{fig4} and \zfr{fig5}B for Hamiltonian engineering, additional studies are conducted to probe the thermalization process. A summary of these findings is provided here, with further details available in the SI. First, to confirm that the zero-crossing observed in \zfr{fig4}B does not arise from spins tipping towards the $\zhat$-axis where they are unobservable, we interrupted the spin-locking drive with a $\pi/2$ pulse at $t {=} \tzc$. The data (SI \zsr{interrupt}) reveal no generation of $\zhat$ during $t$, supporting the thermalization model above. 
Next, we investigated the relative roles of diffusion and dissipation leading up to $\tzc$ by studying profiles similar to \zfr{fig4}B with varying temperatures (\zsr{temp}). Temperature serves as a control parameter for relaxation, as it strongly lengthens (${\gtrsim}$50-fold) the electronic $T_{1e}$ while only linearly changing the strength of the electronic gradient. Decreasing the temperature results in a rightward shift of $\tzc$ due to the slower rate of electron dissipation, consistent with theoretical expectations (\zfr{fig5}B).

\vspace{1em}
\begin{center}
    \textbf{THEORY}
\end{center}
\vspace{-3em}

\tocless\subsection{Effective Hamiltonians for State and Hamiltonian Engineering protocols}{sec:Heff_methods}

As explained in the main text, the experimental system is subject to a drive which consists of a periodic train of $\xhat$-pulses. When the period of switching is small compared to the energy scales of the physical system, this driving induces prethermalization wherein the dynamics is governed by an effective Hamiltonian~\cite{ho2023quantum}, before the system eventually heats up to a featureless infinite-temperature state. 

In the SI~(Sec.~\ref{sec:effective_Hamiltonian}), we provide a detailed derivation of the approximate effective Hamiltonians $\mathcal{H}_{\mathrm{SL}}$, Eq.~\eqref{eq:Heff_SL}, and $\mHeff$, Eq.~\eqref{eq:Heff}, that govern the dynamics of the nuclear spin system for the {State} and {Hamiltonian Engineering} approaches, respectively. In particular, using exact simulations on system sizes up to $N{=}16$ quantum spins, we obtain an excellent agreement between the dynamics generated by the effective Hamiltonian and the exact periodically driven system in all temporal regimes of interest (see SI, Sec.~\ref{sec:effective_Hamiltonian}). 
Therefore, for the theoretical analysis in the main text, we work with static effective Hamiltonians. We emphasize that their applicability is limited to the duration of the prethermal plateau.

\tocless\subsection{Dissipation induced by the NV center}{sec:dissipation_methods}

In the strong coupling limit between the NV center and $\Cs$ spins, the dissipation induced by the NV-center can be rigorously derived from the Hamiltonian of the full system in three dimensions (see Sec.~\ref{sec:SI-open-system}). Unlike previous works~\cite{Coish04}, we integrate out the phonon bath and the NV electron obtaining a Markovian master equation for the $\Cs$ spins only. Such a Markovian master equation involves on-site Lindblad jump operators with coupling constants decaying as $\sim 1/r^6$ as a function of the distance $r$ from the NV center (which is effectively short-ranged). Thus, as a simplification, in our approximate one-dimensional quantum simulations, we approximately solve the Lindblad equation with local jump operators $L$ acting only on the site with index $r=0$ (representing the location of the NV) with isotropic coupling constants,
\begin{equation}
    L_+ = \frac{1}{2}\left(I_0^x + i I_0^y\right)\,, ~~ 
    L_- = \frac{1}{2}\left(I_0^x - i I_0^y\right)\,,~~ 
    L_z = I_0^z.
\end{equation}
Such operators generate both dephasing and dissipation in the system. Both effects are produced by the jump operators derived in Sec.~\ref{sec:SI-open-system} for the Hamiltonian of the full system in three dimensions. 

\tocless\subsection{Numerical simulations}{sec:numeric_methods}

\tocless\subsubsection{Quantum simulations}{subsec:quantum_numerics}

We use the novel algorithm LITE (local-information time evolution, see Refs.~\cite{Kvorning2022, Artiaco2024}) to simulate the dynamics of the $\Cs$-spins with respect to the effective Hamiltonians subject to dissipation. LITE is designed to investigate the out-of-equilibrium dynamics of quantum systems, including open systems governed by the Lindblad equation. Its adaptive system size allows us to effectively simulate infinite systems. 
Motivated by the experimental setup, the system is initialized in a spatially inhomogeneous partially polarized state, where a relatively small number of spins in proximity to the NV center ($r{=}0$) carry a finite partial $\xhat$-polarization, while spins located far from $r{=}0$ are fully mixed $\rho_{r \gg a} {=} \id_2/2$. 

To use the LITE toolbox effectively, we investigate numerically tractable one-dimensional short-range toy models akin to the effective (three-dimensional) Hamiltonians of the coupled $\Cs$ system 
\begin{eqnarray}
    \label{eq:toy_H}
    \mathcal{H}_{\pi/2} &=& - \frac{1}{2} \sum_r J_r \left(3 I^x_r I^x_{r+a}  -\mathbf{I}_r \cdot \mathbf{I}_{r+a}\right), \nonumber \\
    \mathcal{H}_{\pi} &=& \sum_r J_r(3I^z_r I_{r+a}^z -\mathbf{I}_r \cdot \mathbf{I}_{r+a})+ \potential_r I_r^x.
\end{eqnarray}
The subscripts indicate the corresponding partner Hamiltonian in the actual system: $\pi/2 $ refers to the system kicked with $\vartheta=\pi/2$ resulting in the effective Hamiltonian $\mathcal{H}_\mathrm{SL}$; likewise, subscript $\pi$ refers to a kick angle $\vartheta\approx \pi$ with the associated Hamiltonian $\mathcal{H}_\mathrm{eff}$.
In the simulations, $J_r{=}J_0+W$ is taken weakly disordered with $W$ drawn uniformly at random from the interval $[-0.3\vert J_0\vert,0.3\vert J_0 \vert]$ and with lattice constant $a$. The on-site potential follows $\potential_r {\sim} \frac{1}{r^3}- \delta\xt$ with $r=0$ corresponding to the location of the NV center, and $\delta\vartheta$ is the deviation in the kick angle from $\pi$ (see SI~\ref{subsec:SL_pi}).
For the simulations in \zfr{fig5} we use $a=0.2/\sqrt[3]{\pi}$, $\delta\xt=0.05\pi$,  $J_0=-0.025$, and $\gamma{=}0.1\vert J_0\vert $

While these toy models may appear simplistic compared to the actual experimental system, as they lack the complexity of higher dimensions and long-range spin-spin couplings, they capture the essential physics of diffusion (and dissipation), since they obey the same conservation laws compared to their three-dimensional counterparts. Therefore, the conclusions drawn from our numerical results can be qualitatively extended to the experimental system.

\tocless\subsubsection{Classical simulations}{subsec:classical_numerics}

In addition, we also performed classical simulations analyzing the dynamics of the system with hundreds of dipolar-interacting spins of the full three-dimensional long-range interacting model. Spins are placed randomly on the vertices of a 3D diamond lattice of a finite extent, with lattice constant $a{=}0.356\,\mathrm{nm}$, and the classical evolution, generated under the effective Hamiltonian $\mHeff$~\eqref{eq:Heff}, is studied. In the classical limit, the evolution of the spins $\mathcal{I}_k$ is described by~(see SI, Sec.~\ref{subsec:classical_algorithm})
\begin{equation}
\label{eq:classicalEOM_methods}
        \frac{\mathrm{d}}{\mathrm{d}t} \boldsymbol{\mathcal{I}}_k(t) =  \boldsymbol{\mathcal{I}}_k(t) \times \boldsymbol{\nabla}_{\boldsymbol{\mathcal{I}}_k} \mHeff (\boldsymbol{\mathcal{I}}_k(t)) \, .
\end{equation}
In particular, for the classical simulation we are only considering closed system dynamics, i.e. we do not include dissipation as is done in the LITE simulations. However, as we also comment on later dissipation is not an essential ingredient for {Hamiltonian Engineering}.

As in the experiment, we average the simulation data over many~(here $100$) different lattice configurations.
For each lattice configuration, we consider an ensemble of $150$ partially polarized initial states drawn from a spatially inhomogeneously polarized distribution, where all spins within a radius $r{<}7\,\mathrm{nm}$ are polarized and all others unpolarized, without correlations between the spins. We then evolve this ensemble using the classical Hamilton equations (Eq.~\eqref{eq:classicalEOM_methods}), and compute the ensemble-averaged polarization, see Sec.~\ref{sec:classical_simulation}.

\tocless\subsection{Details on the Theoretical Results for State Engineering}{subsec:state}

\begin{figure}
    \centering
   {\includegraphics[width=0.5\textwidth]{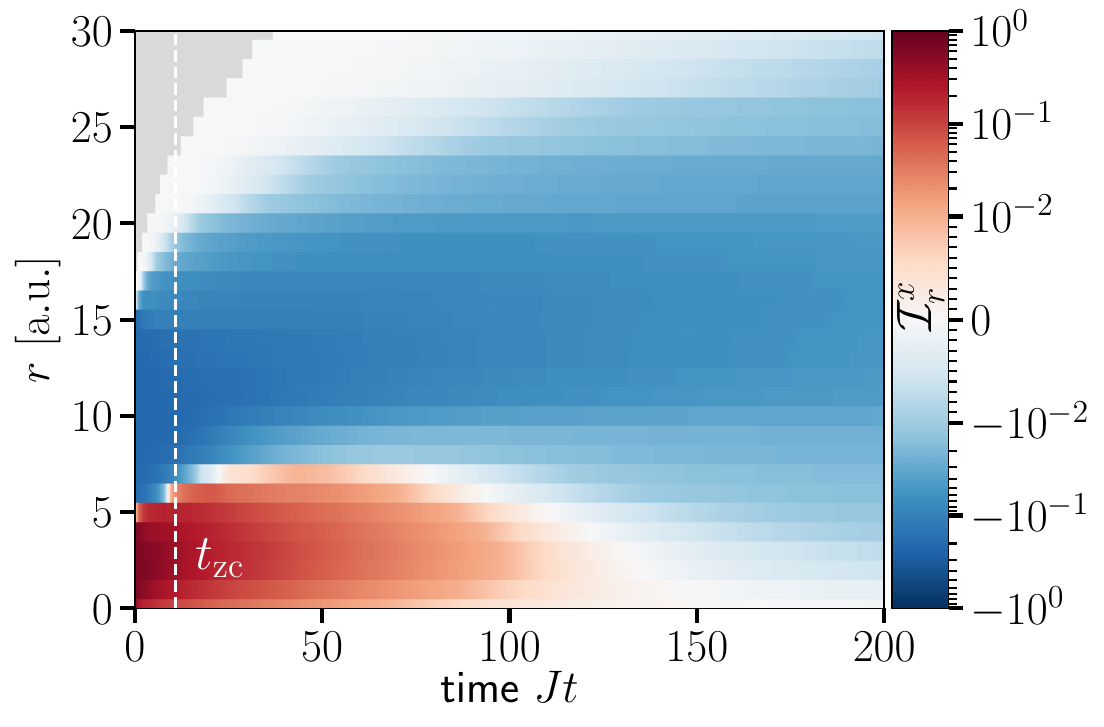}}
    \caption{
    \textbf{Zero crossing radius for State Engineering.} 
    Zoom-in to $Jt<200$ region in \zfr{fig5}A(i) with adjusted color map scale to reveal polarization dynamics close to NV~($r=0$). The dynamics of positive polarization close to NV are dominated by dissipation such that it decays to zero before the crossing radius can move outwards due to diffusion.
    }
    \label{fig:zerocrossing_pihalf}
\end{figure}

\tocless\subsubsection{Zero crossing radius for {State Engineering}}{subsubsec:rc}

In \zfr{fig5}A(i) the polarization dynamics close to the NV center seem to show an abrupt sign inversion from positive to negative polarization. However, this is only an artefact originating from the logarithmic time scale. In fig.~\ref{fig:zerocrossing_pihalf} we show the region close to $r{=}0$ and around the potential sign inversion on a linear scale. Here, it becomes clear that the strong dissipation close to the NV leads to a quick decay of positive polarization. Therefore, the positive polarization cannot diffuse outwards. In fact, the negative polarization diffuses symmetrically in both directions filling the polarization hole left by the decayed positive polarization. 
Thus, the positive polarization does not abruptly invert; rather it is absorbed by the NV and the negative polarization can freely diffuse into what initially was a positively polarized regime.
In the absence of dissipation, both positive and negative polarization would diffuse outwards leading to an increase in crossing radius (see SI fig.~\ref{fig:one_d_comparison}).

\tocless\subsubsection{Waiting time analysis}{subsubsec:twait}

Typically, the dissipation strength is indirectly controlled by tuning the temperature of the sample. Here, the peculiar form of the dissipation obtained in the singular coupling limit (see SI, Sec.~\ref{sec:SI-open-system}), where jump operators are only the spin operators along the $\zhat$-axis, opens up another possibility. While the strength of the dissipation itself remains immutable, the effect of the Lindblad jump operators on the system strongly depends on the direction of polarization: spins polarized along $\zhat$ experience only dephasing, whereas spins polarized in any other direction undergo both dephasing and dissipation. Experimentally, this is reflected in the extremely long $T_1$ lifetimes of $\zhat$-polarized states; during this period, the system still evolves under diffusion and any initial domain wall starts to melt over time (cf.~SI, fig.~\ref{fig:polarization_profile}). Since dissipation does not affect the spins uniformly, different signals are expected to emerge depending on the waiting time $t_\mathrm{wait}$ after which the system is finally rotated out of the $\zhat$-axis.

\begin{figure}
    \centering
   {\includegraphics[width=0.5\textwidth]{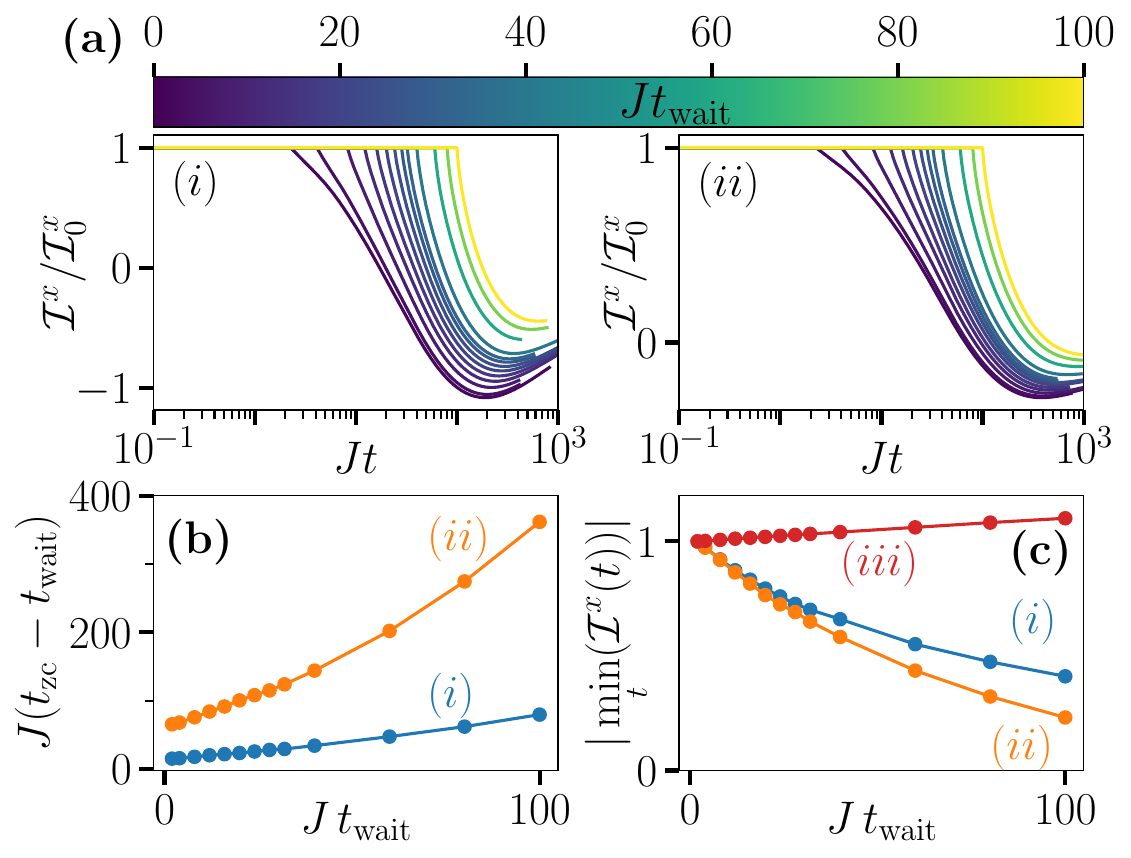}}
    \caption{ 
    \textbf{Waiting time simulations} 
    \textbf{(a)} Time-evolution curves of the total $\xhat$-polarization for different waiting times. After $t_{\mathrm{wait}}$ the dissipation acting on the site with $r=r_{\mathrm{NV}}$ is switched on. For increased waiting time, the corresponding zero-crossing time increases since polarization spreads diffusively during $t_\mathrm{wait}$ diminishing the effect of dissipation which acts most strongly at the position of the NV ($r_\mathrm{NV}=0$).
    We use an initial domain wall state with a total of $N_++N_-=31$ initially polarized spins where the $N_+=13$~(panel $(i)$, blue) and $N_+=11$~(panel $(ii)$, orange) spins closest to $r_\mathrm{NV}$ have $p_r=0.6$ and the remaining $N_-=31-N_+$ spins have $p_r=-0.2$ (here, $\rho_n\propto 1 + p_r I^x_r$). The dissipation parameters are $\gamma_z = \gamma_+ = \gamma_- = 0.5 J$, and the other simulation parameters as in SI fig.~\ref{fig:free_diffusion}.
    \textbf{(b)} The zero-crossing time $t_{\mathrm{zc}}$ of the curves in (a) measured from $t_\mathrm{wait}$ as a function of $t_\mathrm{wait}$. 
    \textbf{(c)} The absolute value of the minimum attained value of the total $\xhat$-polarization of the curves in (a) as a function of the waiting time, normalized with respect to $t_\mathrm{wait}=0$. In addition to the domain wall initial states, we also perform a simulation with a uniformly polarized initial state with $N_-=0$ ($N_+=31$)~(red curve, (iii)).
    }
    \label{fig:method_waiting_time}
\end{figure}

To simulate this effect we initialize the system in domain wall states akin to fig.~\ref{fig:polarization_profile} and evolve it with $\mathcal{H}_{\pi/2}$ up to times $t_\mathrm{wait}$ after which we switch on dissipation. The results are displayed in Fig.~\ref{fig:method_waiting_time}. As the waiting time increases, the zero-crossing time grows accordingly. This is expected since the system diffuses polarization away from $r=0$ during $t_\mathrm{wait}$ and becomes less sensitive to dissipation (here acting only at $r=0$).
A less obvious observation is the decrease of the absolute value of the minimally reached total net polarization $\vert\mathrm{min}(\mathcal{I}_x)\vert$ as a function of the waiting time (Fig.~\ref{fig:method_waiting_time} (c)). 
A possible explanation is that with increasing waiting time, diffusion causes domain walls to melt, flattening the polarization profiles; hence, the net amount of polarization left at large distances (i.e., probed at long times) is reduced with increasing waiting time.
This is also supported by simulations performed with a uniformly polarized (within a small region around the NV) initial state where no such decay appears as a function of waiting time. In fact, in this scenario, even a slight increase can be observed. Intuitively, this is expected as for increased $t_\mathrm{wait}$ more polarization is able to diffuse away from the dissipation-active region around the NV. All these results are qualitatively consistent with experimental observations.

The numerical simulations shown in \zfr{fig3}F-G have been performed by means of the LITE algorithm. During the waiting time $J\tw$, the system evolves under the one-dimensional short-range version of $\mHdd$ and is subject to spin diffusion. When the readout protocol is activated, dissipation comes into play (see above). Therefore, we initialize the system with a different number of positively polarized spins and activate dissipation after $J\tw$. To plot \zfr{fig3}F we measure the zero-crossing point $\tzc$, while for \zfr{fig3}G we compute the minimum of the total net polarization.

\tocless\subsection{Details on the Theoretical Results for {Hamiltonian Engineering}}{subsec:Hamiltonian}

\tocless\subsubsection{Energy vs.~Polarization diffusion}{subsubsec:diffusion}

In the {Hamiltonian Engineering} simulations, we find that the variance of the energy distribution grows as ${\sim}\sqrt{t}$ during the late-time dynamics ($Jt{>}10$), as expected for diffusive processes (see SI~\ref{sec:energy_diff}). Since the energy operator has a large overlap with the $\xhat$-polarization at long times, the diffusive scaling becomes visible in the spreading of the polarizing front (dashed tilted line, \zfr{fig5}B); however, we stress that in {Hamiltonian Engineering}, proper diffusion of polarization is absent due to a lack of $\xhat$-polarization conservation.

\tocless\subsubsection{Effect of Dissipation}{subsubsec:dissipation}

Let us emphasize that in contrast to {State Engineering}, the $\tzc$ zero-crossing for {Hamiltonian Engineering} arises even in the absence of dissipation (see SI~\ref{sec:one_d_pi}). As observed in the long-time slices in \zfr{fig5}B, dissipation only serves to extinguish polarization close to the NV (i.e., accelerating the inevitable diffusion-induced polarization inversion).

\tocless\subsubsection{Critical radius $r_c$}{subsubsec:rc2}

In the main text, we argued that the polarization zero-crossing radius is given by the zeroes, $\potential(r_c)=0$, of the effective on-site potential $\potential$. To leading order, this on-site potential is determined solely from the microscopic gradient field induced by the NV together with the details of the kick sequence~(see SI~\ref{subsec:SpinLocking}). Note that, on the experimental timescale ($t{=}180\,$s), total polarization spreading is expected to remain within a radius ${\lesssim}12\,$nm; the individual NV-$\Cs$ systems can be considered isolated.
Thus, assuming that we have full knowledge about the gradient field, we can estimate the crossing radius for the experimental data, see \zfr{fig4}C-D.

In particular, we assume that the gradient field induced by the NV electron follows the dipole-dipole form $B=2 P K_\mathrm{exp}(3\cos^2\theta - 1)/r^3$ with $K_\mathrm{exp}=\mu_0 \hbar \gamma_n\gamma_e/4\pi$ and electron spin polarization of $P({\approx}10\,\%)$. Then, using the experimental parameters from \zfr{fig4}, i.e., $\Omega{\approx}50\,\mathrm{kHz}$, $t_\mathrm{acq}=51\,\mu\mathrm{s}$ and, e.g., $\theta=0.94\,\pi$~($\tkick=95\,\mu\mathrm{s}$), we can estimate the crossing radius $r_c{\approx} 2.7\,\mathrm{nm}\times \sqrt[3]{|3\cos^2\theta - 1|}$. For more details see SI, Sec.~\ref{subsec:crossing_radius}.

Let us emphasize, that for a single NV center and its environment this slice might not be populated by a $\Cs$-nuclear spin. However, the resulting polarization profile still changes sign depending on whether $r{=}\gtrless r_c$.

\nocite{beatrez2022electron}
\nocite{Beatrez21}
\nocite{Awschalom_quenchdec_2008,Jarmola12}
\nocite{breuer2002theory,benatti2005open,rivas2012open}
\nocite{Coish04}
\nocite{breuer2002theory,rivas2012open}
\nocite{rivas2012open}
\nocite{gorini1976n,frigerio1976n,frigerio1977master}
\nocite{Else17,Machado2020,abadal2020floquet}
\nocite{Beatrez21,fleckenstein2021prethermalization,fleckenstein2021thermalization,peng2021floquet}
\nocite{abanin2015heating,abanin2017heating}
\nocite{Beatrez2022,Sahin2022}
\nocite{Magnus54}
\nocite{Achilles2012}
\nocite{Srednicki1995,Deutsch2018,dalessio2016quantum}
\nocite{lazarides2014equilibrium,dalessio2014longtime,luitz2020prethermalization}
\nocite{abanin2015heating,abanin2017heating}
\nocite{kinder2011TDVP,haegeman2011TDVP,haegeman2016TDVP,leviatan2017TDVP}
\nocite{white2018MPO,richter2019ClusterExpansion,Kvorning2022,rakovszky2022DissipationMPS}
\nocite{Kvorning2022,Artiaco2024}
\nocite{petz1986}
\nocite{kinder2011TDVP,haegeman2011TDVP,haegeman2016TDVP,leviatan2017TDVP}
\nocite{Vidal2003,Vidal2004,White2004,PAECKEL2019,JASCHKE201859}
\nocite{LiebRobinson}
\nocite{peng2021floquet}
 \nocite{Kvorning2022,rakovszky2022DissipationMPS}
\nocite{Abanin2021,Hetterich2018}
\nocite{fleckenstein2021thermalization,fleckenstein2021prethermalization}
\nocite{Schulz2019,van2019bloch,Morong2021}
\nocite{Beatrez2022}
\nocite{Abanin2018_LongRange}
\nocite{Pizzi21, ye2021floquet}
\nocite{Loison2004}

\let\oldaddcontentsline\addcontentsline
\renewcommand{\addcontentsline}[3]{}
\bibliography{bib_compressed}
\bibliographystyle{sciencemag}
\let\addcontentsline\oldaddcontentsline

\tocless\section{Acknowledgements}{sec:acknowledgements}

We thank J.~H.~Bardarson, T.~Klein Kvorning, R.~Moessner, C.~Ramanathan, J.~Reimer, D.~Suter for insightful discussions, and J. Mercade (Tabor Electronics) for technical assistance.

\paragraph*{\textbf{Funding:}}

This work was supported by ONR (N00014-20-1-2806), AFOSR YIP (FA9550-23-1-0106), AFOSR DURIP (FA9550-22-1-0156), a Google Faculty Research Award and the CIFAR Azrieli Foundation (GS23-013). 

CF and CA acknowledge funding from the European Research Council (ERC) under the European Union’s Horizon 2020 research and innovation program (Grant Agreement No. 101001902). 

MB and PMS were funded by the European Union (ERC, QuSimCtrl, 101113633). Views and opinions expressed are however those of the authors only and do not necessarily reflect those of the European Union or the European Research Council Executive Agency. Neither the European Union nor the granting authority can be held responsible for them.

The computations with the LITE algorithm were enabled by resources provided by the National Academic Infrastructure for Supercomputing in Sweden (NAISS) and the Swedish National Infrastructure for Computing (SNIC) at Tetralith partially funded by the Swedish Research Council through grant agreements no.\ 2022-06725 and no.\ 2018-05973. 

Classical simulations were performed on the MPI PKS HPC cluster.

\paragraph*{\textbf{Author Contributions:}}

KH, CF, ND, PMS and DM contributed equally to this work. Order in which these authors are listed was decided by dice roll.

\begin{itemize}[itemsep=-0.5em,leftmargin=0.5em, topsep=0em]
    \item Conceptualization: AA, DM, ED, PR, MB, KH, ND, CF, PMS, WB, AN, XL
    \item Methodology: ND, DM, KH, CF, PMS, CA, QRF, WB, AA, MB, ED, AN
    \item Formal Analysis: ND, DM, KH, CF, PMS, CA, QRF, AA, UB, SB,
    \item Validation: KH, CF, ND, PMS, DM, CA, WB, ED, MH, AA, XL
    \item Software: CF, PMS, CA, ND, WB, QRF, SB
    \item Data curation: ND, DM, ED
    \item Investigation: ND, DM, KH, CF, PMS, CA, QRF, WB, AA, ED, UB, SB, XL, AN
    \item Visualization: ND, CF, PMS, KH, DM, UB, AP, MH, QRF, SB, and AA
    \item Funding Acquisition: MM, MB, AA
    \item Resources: MM, ED, DM, MH
    \item Project-Administration: AA, MB, DM, ED 
    \item Supervision: MB, AA, CF, ED
    \item Writing—original draft: AA, MB, PMS, CF, CA, ND, KH, UB, DM, ED, AP
    \item Writing—review \& editing: AA, MM, DM, ND, KH, CF, PMS, MB, CA, ED, XL, UB
\end{itemize}

\paragraph*{\textbf{Competing Interests:}}

The authors declare no competing interests.

\paragraph*{\textbf{Data and materials availability:}}

All data needed to evaluate the conclusions in the paper are present in the paper and/or the Supplementary Materials.
Datasets can be accessed in the Zenodo repository with the identifier \url{https://doi.org/10.5281/zenodo.11658424}. 
The NV diamond sample can be provided by Element Six pending scientific review and a completed material transfer agreement. Requests for the NV diamond should be submitted to: Element Six.

%% file: SI.tex
\clearpage
\onecolumngrid
\begin{center}
\textbf{\large{\textit{Supplementary Information:} \\ \smallskip Nanoscale engineering and dynamical stabilization of mesoscopic spin textures}} \\\smallskip

\hfill \break
\smallskip
Kieren Harkins,$^{1,{\ast}}$  Christoph Fleckenstein,$^{2,{\ast}}$  Noella D'Souza,$^{1,{\ast}}$ Paul M. Schindler,$^{3,{\ast}}$ David Marchiori,$^{1,{\ast}}$ Claudia Artiaco,$^{2}$ \\
Quentin Reynard-Feytis,$^{1}$ Ushoshi Basumallick,$^{1}$ William Beatrez,$^{1}$ Arjun Pillai,$^{1}$ Matthias Hagn,$^{1}$ Aniruddha Nayak,$^{1}$ \\ 
Samantha Breuer,$^{1}$ Xudong Lv,$^{1}$ Maxwell McAllister,$^{1}$ Paul Reshetikhin,$^{1}$ Emanuel Druga,$^{1}$ Marin Bukov,$^{3}$ and Ashok Ajoy,$^{1,4,5}$$^\dagger$\\
\smallskip
\emph{${}^{1}$ {\small Department of Chemistry, University of California, Berkeley, Berkeley, CA 94720, USA.}} \\
\emph{${}^{2}$ {\small Department of Physics, KTH Royal Institute of Technology, SE-106 91 Stockholm, Sweden.}} \\
\emph{${}^{3}$ {\small Max Planck Institute for the Physics of Complex Systems, N\"othnitzer Str.~38, 01187 Dresden, Germany.}}\\
\emph{${}^{4}$ {\small Chemical Sciences Division,  Lawrence Berkeley National Laboratory,  Berkeley, CA 94720, USA.}}\\
\emph{${}^{5}$ {\small CIFAR Azrieli Global Scholars Program, 661 University Ave, Toronto, ON M5G 1M1, Canada.}}

\hfill \break
$^\ast$ These authors contributed equally to this work

$^\dagger$ Corresponding author. Email: ashokaj@berkeley.edu
\end{center}

\twocolumngrid

\section*{The Supplemental Materials include}

\begin{itemize}
    \item Supplementary Text
    \item Figs S1 to S28
    \item Supplemental Movies 1 - 4
\end{itemize}

\beginsupplement
\tableofcontents

\clearpage

\section{Summary}
Given the length of this Supplementary Information, as an aid to the reader, we present a summary of the main results here. Section~\ref{sec:dissipation_sources}, summarizes sources of nuclear relaxation, with a focus on dissipation stemming from the central electron. Section~\ref{sec:transverse longitudinal relaxation} discusses the dephasing time ($T_2^{\ast}$), longitudinal relaxation time ($T_1$), and transverse relaxation under spin-lock driving  ($T_2^{\prime}$) in our system.  Section~\ref{sec:indirect evidence} elucidates indirect experimental signatures of NV-driven transverse relaxation of the $\Cs$ spins. 

Section~\ref{fig:earlytau} shows the expanded data from \zfr{fig2}B for $\tau-$ values up to the onset of the zero-crossing behavior. Decays are featureless until the generated negative polarization is comparable in magnitude to the positive polarization and ZC behavior commences. Normalization sets the maximum value to 1, which does not change the underlying physics.

Section~\ref{sec:wide range of angles} describes spin-lock decays for a wide range of flip angles $\xt\in[0,1.5\pi]$, showing slow featureless decays in all regions away from $\xt{\app}\pi$, and sharp zero-crossings in the latter due to the formation of Hamiltonian engineering driven spin textures.

Section~\ref{sec:z pol} elucidates complementary experimental studies focused on the Hamiltonian engineering method.
In Sec.~\ref{sec:z pol}, we describe experiments that confirm that the loss of signal during the zero-crossing is not due to the spin vector being trivially tilted onto the $\zhat$ axis. This was demonstrated by inserting a $\pi/2$ pulse perpendicular to those in the train of pulses. If the spin vector were in the $\zhat$ axis, this sequence would immediately create observable polarization.

Section~\ref{sec:exp param} examines changes in $\tzc$ to probe the sensitivity of formed spin texture to experimental parameters. Increasing hyperpolarization time in fig.~\ref{fig:SI_pol_time} does not significantly change the location of $\tzc$, showcasing the robustness of Hamiltonian Engineering to different initial states. 

Decreasing the temperature in fig.~\ref{fig:SI_temp} diminishes the effect of electron dissipation to the system, preserving the generated spin texture for longer and moving $\tzc$ to later times in the experimental data. In fig.~\ref{fig:SI_duty}, changing the effective Rabi frequency, by changing the pulse duty cycle while fixing $\xt$, leads to a movement in $\tzc$ as anticipated from theory.

Changing sample orientation creates composite behavior at kick angle $\pi$ qualitatively identical to other Hamiltonian Engineering experiments (compare fig.~\ref{fig:SI_orientation}Aiii and Biii). The multi-dip behavior seen in fig.~\ref{fig:SI_orientation}Bii is consistent with domain boundaries forming from NV centers inequivalently aligned with respect to the applied magnetic field. Lastly, changing the pulse transmit offset frequency (TOF) shifts the Rabi curve of the system by a finite phase, yet zero crossing behavior can appear before or after the bulk $\pi$ value for a given TOF (fig.~\ref{fig:SI_offset}).

Section~\ref{sec:model} details the theory model for dipolar-coupled $\Cs$ nuclear spins and nitrogen-vacancy (NV) center electrons. The system is described as a closed system with a hierarchical energy scale separation, allowing for independent manipulation of the $\Cs$ nuclei and NV electrons. The Hamiltonian captures the couplings between spins and the external magnetic field. Additionally, the open system nature of the experimental setup is discussed in Sec.~\ref{sec:SI-open-system}, where spins and NV centers couple to a phonon bath, resulting in decoherence and dissipation. An effective Lindblad master equation is derived, considering the singular coupling limit, which describes the dissipative dynamics of the spins only. The dissipative terms in the Lindblad equation account for dephasing and dissipation effects operational in experiments.

In Sec.~\ref{sec:effective_Hamiltonian}, we investigate in more detail the dynamics of the $\Cs$ spin system coupled to the NV center. By considering the fast dynamics of the NV spin, we replace the NV operators with their expectation values. This leads to an effective on-site potential for the $\Cs$ spins. We analyze the system's evolution under a kick sequence and derive the effective Hamiltonian using Floquet's theorem. We consider two cases: spin locking with a kick angle different from $\pi$ and near $\pi$. In the former case, the effective Hamiltonian conserves the $\xhat$-polarization, while it is not conserved in the latter case, and the strong NV-induced potential leads to a spatially inhomogeneous effective Hamiltonian. Exact numerical simulations confirm the validity of the derived effective Hamiltonians. In Sec.~\ref{subsec:ETH}, we discuss the dynamics of the $\Cs$ system within the Eigenstate Thermalization Hypothesis (ETH). The system is expected to (ultimately) thermalize to an infinite temperature state due to the absence of energy conservation. However, at high driving frequencies, a pre-thermal plateau emerges before full thermalization. We find that in the $\vartheta{=}\pi$ case, the local polarization follows the local on-site potentials induced by the NV center, which can lead to a sign-inversion of the local polarization, and hence also of the integrated polarization. In Sec.~\ref{subsec:crossing_radius}, we estimate the $r_c$ domain boundary where the local polarization changes sign. Finally, in Sec.~\ref{subsec:simplifications}, we present two simplified, effective Hamiltonians for $\vartheta{\neq}\pi$ and $\vartheta{\approx}\pi$ that aid in a qualitative understanding of the dynamics.

Section~\ref{sec:approximate_dynamics} first introduces the approximate time-evolution algorithm used to perform the quantum dynamics of the simplified one-dimensional toy models used in the rest of this Section. In particular, Sec.~\ref{sec:LITE} summarizes the main features of the local-information time-evolution (LITE) algorithm. LITE is designed to investigate the out-of-equilibrium transport of short-range systems. In contrast to similar algorithms, it preserves local constants of motion. LITE decomposes the system into subsystems and solves the von Neumann equation for each subsystem in parallel. Importantly, it can also simulate open quantum systems described by the Lindblad master equation. In Sec.~\ref{sec:one_d_pi}, we apply LITE to simulate an effective one-dimensional short-range toy model for the case $\vartheta {=} \pi$. Such a toy model is derived from the experimental three-dimensional long-range model within the approximation of sparse density of spins. Thus, the one-dimensional model retains only the dominant nearest-neighbor mutual coupling and the space-dependent on-site potential generated by the NV center. For $\vartheta {\approx} \pi$, the energy diffusion and the behavior of the total $\xhat$-polarization are analyzed, showing a sign inversion of polarization in the presence of a space-dependent potential. The diffusion of energy in the inhomogeneous systems is explored, revealing a slowing down of energy spread that can be explained in terms of a spatially dependent diffusion constant. Dissipation effects are then introduced. It is shown that dissipation favors energy diffusion, inducing the emergence of Gaussian-like energy distributions. The presence of dissipation does not alter the possibility of observing sign inversion in the total $\xhat$-polarization. In Sec.~\ref{sec:one_d_pi_half}, the dynamics of the one-dimensional spin system obtained for a kick angle around $\pi/2$ is investigated. Unlike the case of a kick angle close to $\pi$, the total (net) $\xhat$-polarization is conserved in addition to energy. This leads to the diffusion of polarization throughout the system, making it impossible to detect polarization gradients using global measurements. Only by introducing dissipation can a sign inversion of the integrated polarization be induced. The effects are demonstrated using numerical simulations obtained within the LITE algorithm. In Secs.~\ref{subsec:comparison-energy-polarization} and \ref{subsec:one-dimension-summary}, we highlight the main differences between the two kick angles analyzed and summarize the results.

In Sec.~\ref{sec:dimensionality}, we discuss the role of dimensionality and interaction range in the simplified one-dimensional versions of the long-range three-dimensional quantum system used in the experiments. We argue that, although our simplified models may not capture all the details of the out-of-equilibrium dynamics, we expect them to provide qualitative insights into the three-dimensional system. We discuss the effects of dimensionality on Floquet prethermalization and equilibration dynamics within the prethermal plateau. We also consider the possible effects of many-body localization and the impact of spin-spin couplings and angular dependence.

In Sec.~\ref{sec:classical_simulation}, we perform classical simulations on a three-dimensional diamond lattice to complement the previous analysis. The classical simulation allows us to study larger systems and reach longer time scales, although quantum correlations are neglected. We analyze the spin locking regime near $\vartheta\approx\pi$ and observe the formation of a spatially inhomogeneous local polarization profile. The classical simulation confirms the predictions from the analytical and numerical results obtained for one-dimensional toy models. The simulations show that the local polarization at late times follows the applied local potential up to an overall constant, which is related to the inverse temperature. The classical simulation results provide further support for the formation of a robust spatially inhomogeneous local polarization profile in the experimental system.

\section{\label{sec:dissipation_sources}Dissipation sources in the experiment}

\subsection{\label{sec:transverse longitudinal relaxation}Nuclear relaxation in transverse and longitudinal directions}
\begin{figure}[t]
  \centering
  {\includegraphics[width=0.49\textwidth]{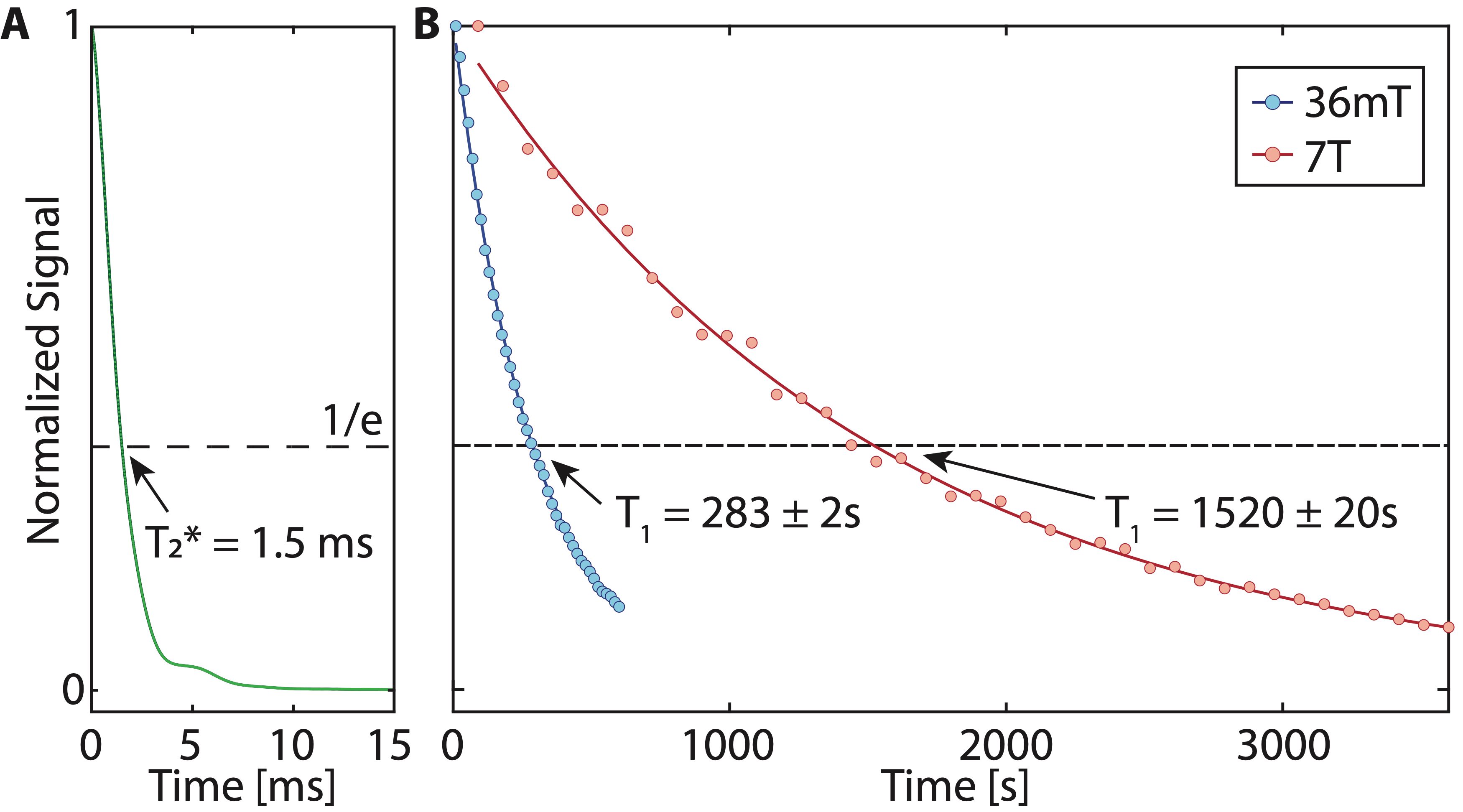}}
  \caption{
  \T{Nuclear dephasing and relaxation}.
    (A) Panel illustrates $T_2^{\ast}$ dephasing of $\Cs$ spins during free induction decay at room temperature. The fast decay observed, $T_2^{\ast}{\app}1.5$ms, is attributed to internuclear interactions. Reproduced from ~\cite{Beatrez21}
    (B) Panel presents $T_1$ decay at room temperature, lasting over twenty-five minutes at 7T. Reproduced from ~\cite{beatrez2022electron}.
    }
    \label{fig:SI_relax}
\end{figure}

\zsl{relax}
In this discussion, we elucidate the mechanisms for $\Cs$ nuclear relaxation for different initial states. Let us begin by examining fig.~\ref{fig:SI_relax}A, which illustrates the $\Cs$ dephasing when prepared in a superposition state, ${\propto}I_x$. In the absence of the applied drive, the measured signal rapidly decays in $T_2^{\ast}{\sim}1.5$ms. This is primarily attributable to interspin interactions, specifically, the formation of many-body spin states invisible under inductive readout. The magnitude of the internuclear dipolar coupling can be estimated from this decay as $\langle J \rangle {=} 660$ Hz.

Moving on to fig.~\ref{fig:SI_relax}B, we consider the spins prepared in the $\xr_z\propto I_z$ state. In this case, the corresponding ($T_1$) relaxation is found to be remarkably long, with $T_1{<}1$hr even at room temperature. Relaxation arises from spin-flipping noise perpendicular to $\zhat$, stemming from the spectral density component that matches the nuclear Larmor frequency ($\omega_L{=}\xg_nB_0$). Phononic contributions likely play a dominant role in these decay channels; their inherently weak nature results in the long $T_1$. Conversely, to assess the contribution of electronic dissipation to this relaxation, it is worth noting that the corresponding spectral density is centered at $(T_{1e})^{-1}$, which is orders of magnitude separated from $\xo_L$. Hence, this contribution can be considered to be extremely weak. This is supported by experiments in \zfr{fig3} of the main paper. 

Lastly, we direct our attention to $T_2^{\prime}$ relaxation, as shown in fig.~\ref{fig:SI_relax_time}. Here spins are prepared in a superposition state $\xr_I{\propto} I_x$ and subjected to periodic driving, which preserves them along $\xhat$. However, this effective Hamiltonian is only accurate to leading order in a Magnus expansion, and higher-order terms can induce heating to infinite temperature. Relaxation additionally arises from $\zhat$-oriented noise spectral matched to the energy gap in the rotating frame, $\xO_{\R{eff}}{=} \Omega (t_p/\qt)$, where the latter factor is the pulsing duty cycle. In our experiments, $(T_{1e})^{-1}{\sim}100$kHz and exhibits a significant contribution at $\xO_{\R{eff}}$. Consequently, the electron can play a role as a \I{``dissipator"} for the spins.

Note that while the lattice contains both NV and P1 center electrons, the NV center is present at the same location $r{=}0$ for every $\Cs$ ensemble considered while the P1s are randomly positioned. Thus, upon ensemble averaging, the P1s contribute to only a background rate of $T_2^{\prime}$ relaxation, while the NV-driven relaxation is discernible in a spatially dependent manner.

\subsection{\label{sec:indirect evidence}Indirect evidence for electron dissipation on the \texorpdfstring{$\Cs$}{TEXT} nuclear spins}

\begin{figure}[t]
  \centering
{\includegraphics[width=0.49\textwidth]{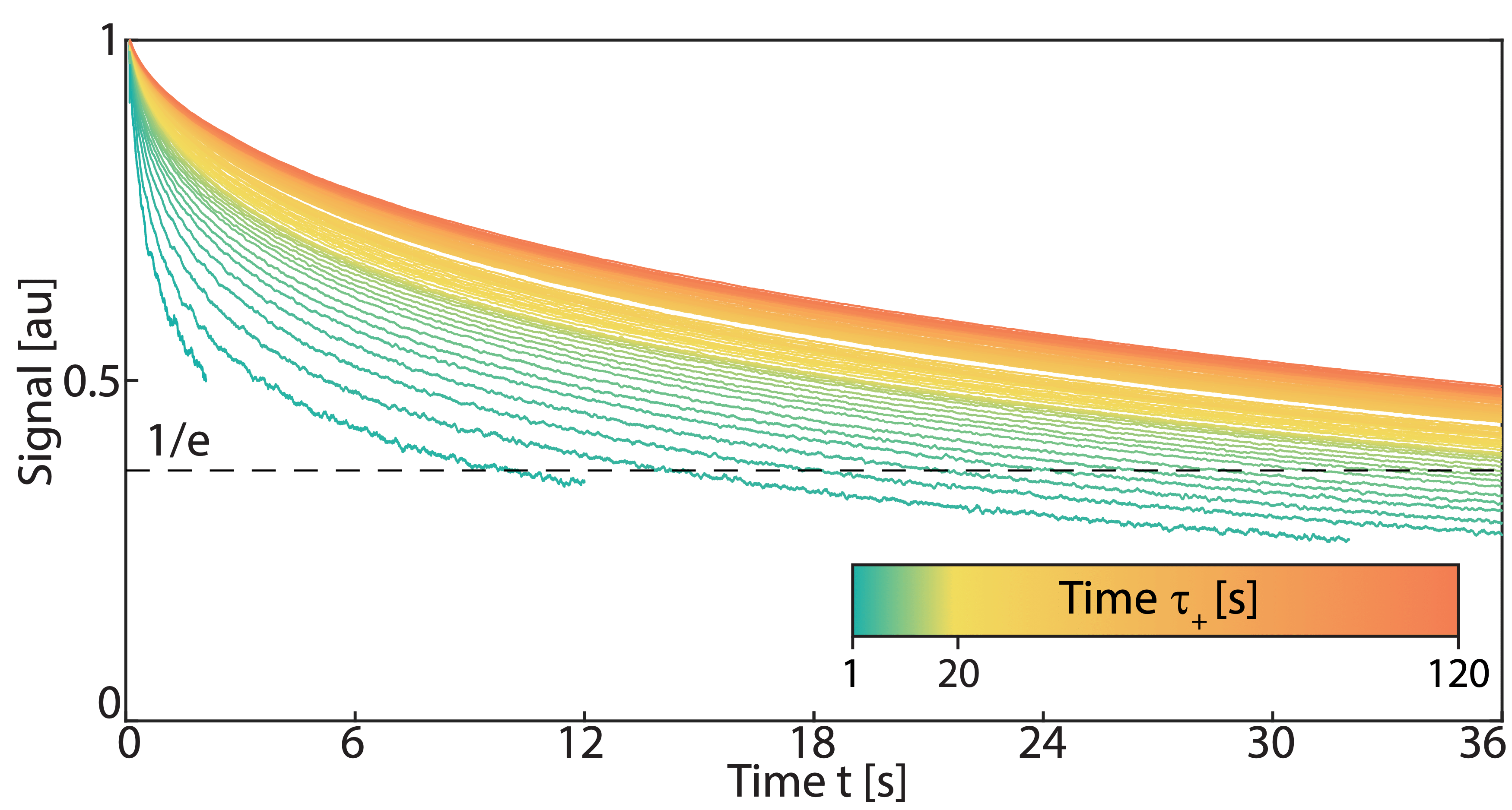}}
  \caption{ \T{Indirect evidence for electron dissipation}. Normalized spin lock curves plotted against increasing polarization time $\qt_+$, reproduced from ~\cite{beatrez2022electron}. Data demonstrates that $T_2’$ values increase with $\qt_+$, indicating that $\Cs$ nuclei located further away are less affected by relaxation from the NV, while proximal spins experience faster NV relaxation. This indirect evidence is further supported by the formation of shells, as depicted in \zfr{fig2}.}
  \label{fig:SI_relax_time}
\end{figure}

In previous work, we presented preliminary evidence supporting the role of the electron spin as a nanoscale dissipator for neighboring nuclear spins~\cite{beatrez2022electron}. Here, we provide a concise summary of these findings. Specifically, we observed that the relaxation profiles of $\Cs$ spins in the $T_2^{\prime}$ regime appear to lengthen when a greater degree of hyperpolarization (achieved by employing longer $\tau_+$ durations) is injected into the spins. This is shown in fig.~\ref{fig:SI_relax_time}, where the colorbar represents $\qt_+$. We interpreted this behavior by considering that, with longer $\tau_+$ durations, the polarization has more time to diffuse within the lattice, thus reaching nuclei less affected by relaxation originating from the NV center. Instead, the observation of sharp zero-crossings in \zfr{fig2} of the paper provides more \I{direct} evidence for this, and allows it to be exploited for the readout of spin texture.

\begin{figure}[t]
  \centering
  {\includegraphics[width=0.49\textwidth]{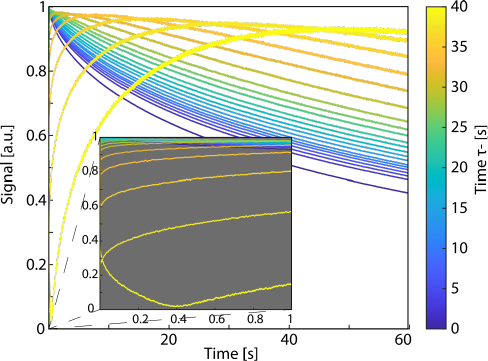}}
  \caption{
  \T{State engineering spin-lock decays for $0<\tau_{-}<40\,$s} correspond to the region up to the onset of zero-crossing (ZC) behavior. The maximum value is normalized to 1. Curves are relatively featureless until the ZC onset, as generated negative polarization is quickly relaxed by the NV- center. Zoom into the first $1\,$s of the dataset shows ZC onset first begins at early experiment readout times ($<1\,$s).
  }
  \label{fig:SI_earlytau}
\end{figure}

\begin{figure*}[t]
  \centering
 {\includegraphics[width=0.99\textwidth]{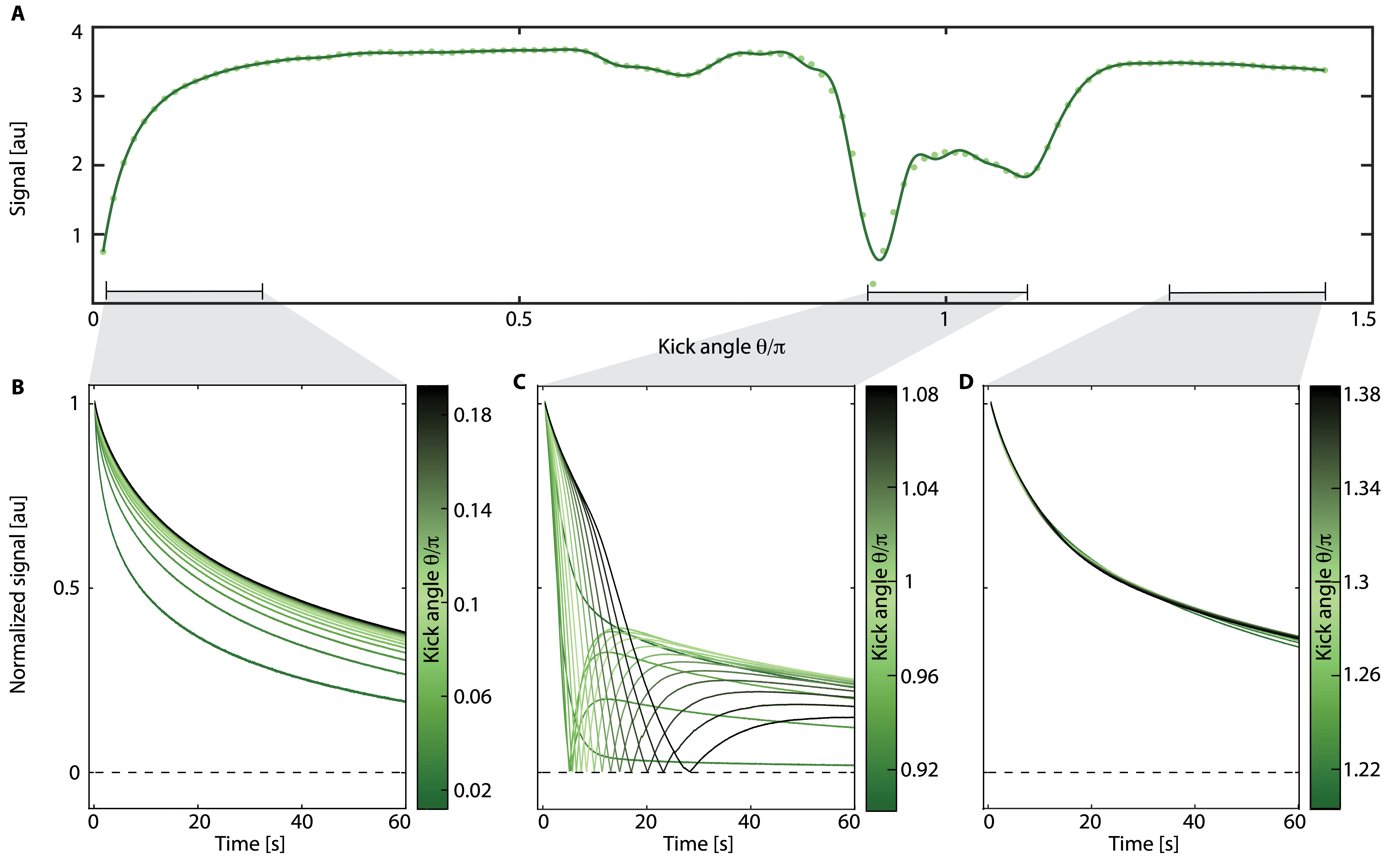}}
   \caption{\T{Data for a wide range of spin-locking angles}, $\xt{\in}[0,1.5\pi]$. (A) Integrated signal $\Sint$ over the first $60\,$s of experiment time 
   Shaded regions show representative regions centered at $\xt{=}0.1\pi$, $\pi$, and $1.3\pi$, respectively. (B-D) Corresponding decay profiles for the shaded regions in (A) are shown, with $\xt$ values indicated by the colorbar. Prethermal decays are featureless and follow stretched exponential decays on either side away from $\xt{=}\pi$. Near $\pi$, however, there is a significant change in the signal profile, associated with the formation of shell-like spin textures. Full movie corresponding to this dataset is available in Movie S4.
   }
   \label{fig:SI_angle}
\end{figure*}

\section{State-engineered spin-lock decays before onset of zero-crossing}
\zsl{earlytau}

Here we expand upon the discussion of state-engineering data from main paper \zfr{fig2}.
Examining individual traces within the range of $\tau_{-} = 0$ to $30$ seconds in fig.~\ref{fig:SI_earlytau} reveals featureless, monotonic decays.
The key physics we investigate emerges as the integrated signal in \zfr{fig2}A nears zero ($\tau_{-}>30\,$s), where the traces exhibit a cusp-like shape and undergo a zero crossing. 
To aid in visualizing the spin-lock signals’ time-domain profiles, we normalize each trace by setting its maximum value to one.
This normalization allows all traces to be plotted together on a single graph, and enhances the clarity of the varying time-domain signal profile with changing $\tau_-$. Importantly, this normalization holds little physical meaning and serves purely as a visual aid to highlight changes in the signal profile over time.
The initial signal begins at less than $1$ for $\tau-$ values $>30\,$s.
This is because the negative and positive polarization created in the initial state are nearly equivalent in magnitude, so the zero-crossing condition is achieved very soon in the experiment time (close to $t=0$).
In a wider x axis range ($[0,\,60]$) the signal appears to start at or near 0, but zooming in to an x axis range of $[0,1]$ shows that these datasets are just an early onset of the zero crossing.

\section{\label{sec:wide range of angles}Prethermal signal decays at a wide range of flip angles}
\zsl{wide_angle}
We present here an analysis of spin-lock decay behavior for a wide range of flip angles, $[0, 1.5\pi]$, as a comparison to the results in the main text which focused on the proximity of $\xt{=}\pi$ (\zfr{fig4}) and $\xt{=}\pi/2$ (\zfr{fig2}). Experiments here (fig.~\ref{fig:SI_angle}) were conducted at 115K, similar to conditions in \zfr{fig4}. fig.~\ref{fig:SI_angle}A shows the integrated signal $S_\R{int}$ for the range of angles, and fig.~\ref{fig:SI_angle}B-D shows representative slices of signal $S$ at specific values of $\xt$ in three ranges (shaded). A full movie of the dataset is available in Movie S4.

Data in fig.~\ref{fig:SI_angle}B,D illustrates that far from $\pi$, the spin-lock profiles `exhibit slow decays at $T_2’$ that follow stretched exponential behavior. In the regime away from $\pi$, there is no inversion of the spin signal with changing $\xt$. However, as observed in fig.~\ref{fig:SI_angle}C (also \zfr{fig4}), in the proximity of $\pi$, there are sharp zero crossings corresponding to the formation of spin-shell texture. Overall, these results demonstrate unexpected sign inversions for the case of CPMG pulse trains, which to our knowledge, has not been reported previously.

\begin{figure}[t]
  \centering
  {\includegraphics[width=0.49\textwidth]{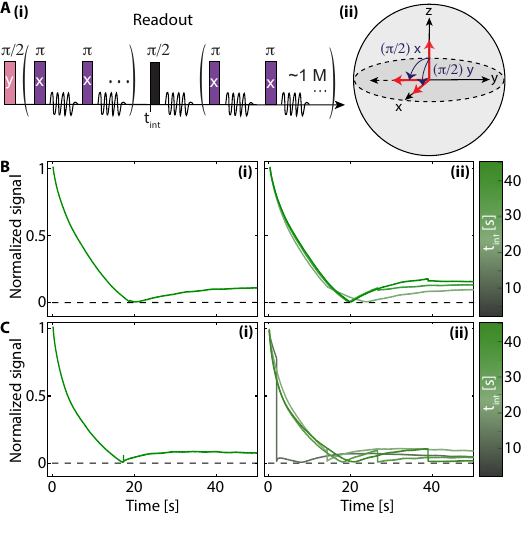}}
  \caption{\T{``Interrupted” periodic driving experiments} investigating if zero-crossing observed in \zfr{fig4} is caused by polarization tipping to the $\zhat$ axis.  
    (A) i) \I{Schematic} of the experiment shown in the inset, consisting of a spin-locking train interrupted by an orthogonal $\pi/2$ pulse of phase x or y at $t=t_{int}$ that reorganizes population from the $\zhat$ axis to the $\xy$ plane. 
    (B-C) i) Signal at $t_{int}{=}t_{zc}$ for case of  $\left.\pi/2\right|_x$ in (B) or $\left.\pi/2\right|_y$ in (C) interrupt pulses. The measured $\zhat$ amplitude for $\left.\pi/2\right|_x$ pulses does not significantly change, indicating that zero-crossing in \zfr{fig4} does not result from spins tipping out of the $\xy$ plane. Residual $\zhat$-component can be attributed to slight off-resonance in the pulses.  
    (B-C) ii) Signal decays for the interrupted driving experiments where $t_{int} \neq t_{zc}$. In (B), the $\left.\pi/2\right|_x$ pulse caused minimal disruption to the signal decays. However, in (C) the $\left.\pi/2\right|_y$ pulse reorganized the population of spins from the $\xhat$ axis to the unobservable $\zhat$ axis, causing an abrupt loss of signal.
}
\label{fig:SI_interrupt}
\zfl{SI_interrupt}
\end{figure}

\section{\label{sec:z pol}Experiments probing \texorpdfstring{$\zhat$}{TEXT} polarization}
\zsl{interrupt}

In the experiments describing shell formation through Hamiltonian Engineering (\zfr{fig4} of the main paper), we observed an apparent change in the polarization direction from $\xhat$ to $-\xhat$, which was attributed to the thermalization of spins under the electron gradient induced potential. In this section, we aim to provide additional evidence to support this. Specifically, we conducted measurements probing the $\zhat$ component of the spin vector at various time points $t$ along experimental traces in \zfr{fig4}B. This allows us to examine whether the observed zero-crossings in \zfr{fig4} are trivially a result of the spin vector tilting towards the $\zhat$ axis, where the spins become unobservable.

To further investigate this, we performed additional experiments as depicted in fig.~\ref{fig:SI_interrupt}A. The spins are subject to the drive as before, but it is interrupted at time $t{=}t_c$ with $\pi/2$ pulses applied in either the $\xhat$ direction (fig.~\ref{fig:SI_interrupt}A) or the $\yhat$ direction (fig.~\ref{fig:SI_interrupt}B). These pulses reorganize the spin populations from the $\yz$ or $\xhat-\zhat$ planes into the $\xy$ plane at $t_c$. Setting $t_c{=}\tzc$, we can precisely probe the $\zhat$ component at the zero-crossing point. Moreover, by varying $t_c$, we can track the $\zhat$ component throughout the observed dynamics in fig.~\ref{fig:SI_interrupt}A-B.

Specifically, fig.~\ref{fig:SI_interrupt}A(i) and fig.~\ref{fig:SI_interrupt}B(i) display data obtained at $t_c{=}\tzc$ slices for both cases. Notably, no significant $\zhat$ signal was observed, with the residual signal primarily attributed to the frequency offset of the pulses employed in these experiments. To further support this observation, the inset fig.~\ref{fig:SI_interrupt}B(ii) illustrates the variation of the $\zhat$ amplitude as a function of $t_c$. It reveals that the $\zhat$ component undergoes negligible change with $t$. Consequently, fig.~\ref{fig:SI_interrupt} provides compelling evidence that the zero-crossing observed in \zfr{fig4} arises from the thermalization dynamics of the spins, rather than from the polarization vector deviating away from the $\xy$ plane.

\begin{figure}[t]
  \centering
  {\includegraphics[width=0.49\textwidth]{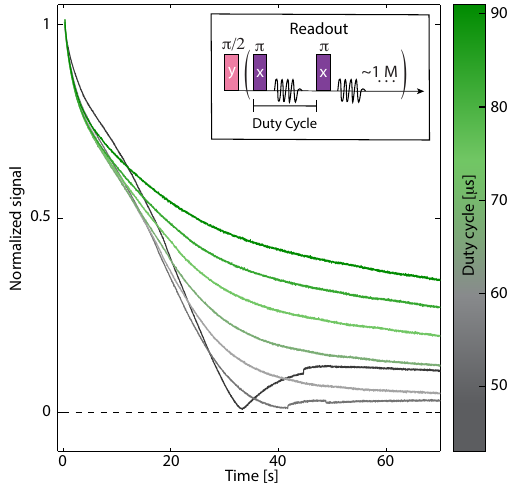}}
  \caption{\T{Variation of spin-lock profiles with pulse duty cycle}.  Spin-lock profiles for different effective Rabi frequencies (see colorbar) applied to the nuclear spins and $\xt{\sim}\pi$, similar to \zfr{fig4} of the main text. $\xO_{\R{eff}}$ is controlled here by varying the pulse duty cycle, while maintaining a fixed flip angle $\xt$. Inset shows pulse sequence schematic, with definition of the duty cycle being the total time between the start of one pulse and the next. 
  }
\label{fig:SI_duty}
\zfl{SI_duty}
\end{figure}

\begin{figure}[t]
  \centering
  {\includegraphics[width=0.49\textwidth]{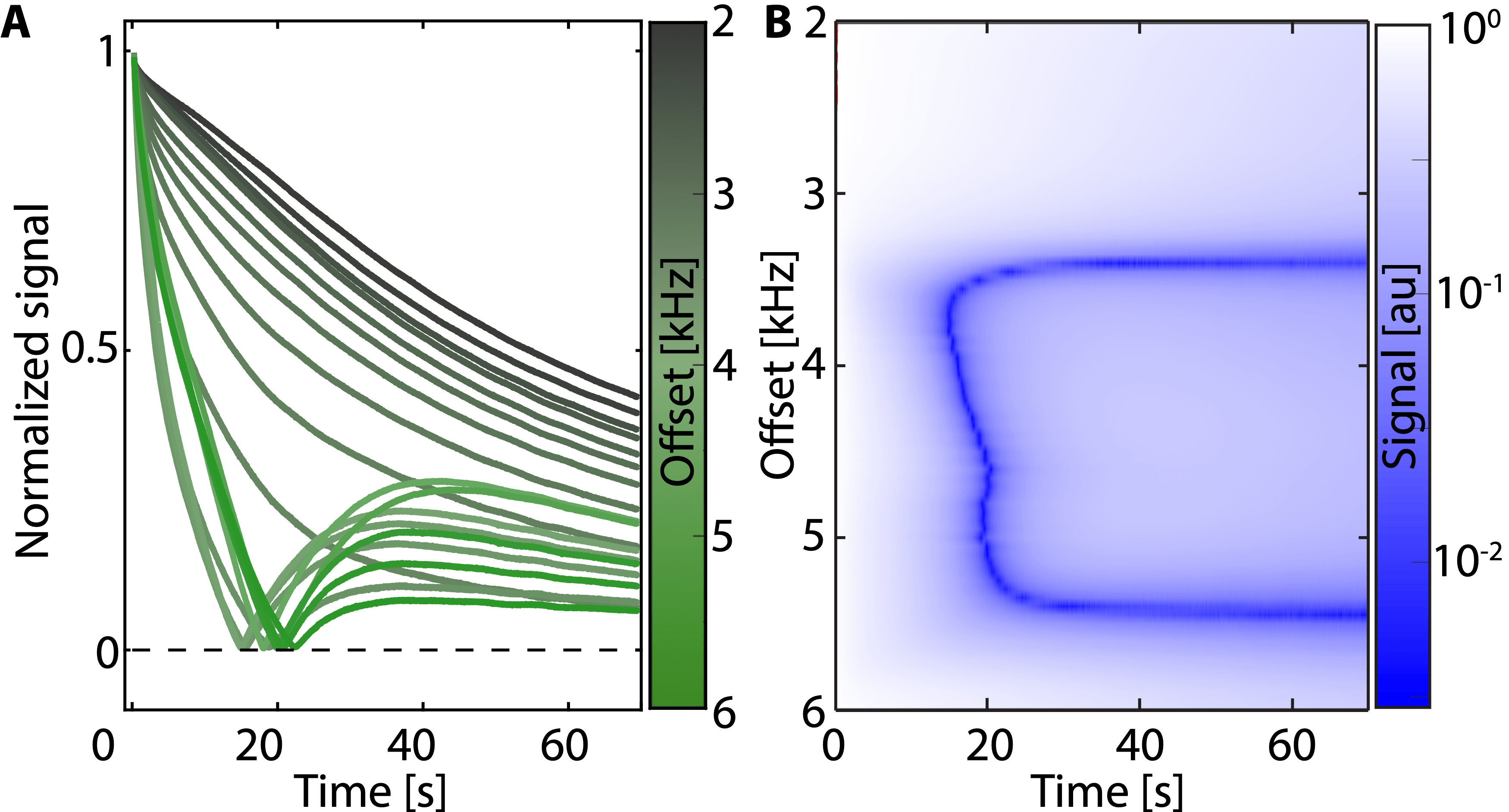}}
  \caption{\T{Variation of spin-lock profiles with frequency offset}.  (A) Spin-lock profiles at different frequency offset values for the drive (see colorbar). (B) Movement of zero-crossing with offset. The 2D color plot shows offset data from (A) plotted on a logarithmic scale. Zero crossing appears as an abrupt decrease in signal (colored blue). There is a wide range of offsets ($\sim$2 kHz) for which a zero crossing is observed.
  }
  \label{fig:SI_offset}
\zfl{SI_offset}
\end{figure}

\section{\label{sec:exp param}Hamiltonian engineering: variation of \texorpdfstring{$\tzc$}{TEXT} with experimental parameters}
\zsl{offset}

Accurately estimating the domain boundary of the generated spin texture based solely on the zero-crossing position $\tzc$ poses a significant challenge. $\tzc$ relies on the complex interplay between electron dissipation and spin diffusion, making it inadequate for providing an accurate measurement of the critical radius $r_c$ associated with this boundary. However, by considering the variations in experimental parameters, it is possible to anticipate the trends in the movement of $r_c$ (and, consequently, $\tzc$), which can then be validated through experimental investigations.

Consider first the influence of the nuclear Rabi frequency $\xO_{\R{eff}}$. Since the effective flip angle $\xt$ is determined as an interplay between the electronic gradient field and the strength of the $\xhat$ Rabi field, changing  $\xO_{\R{eff}}$ would directly affect $r_c$. We anticipate from  theory that decreasing $\xO_{\R{eff}}$ should result in a leftward shift in $\tzc$. In experiments (see fig.~\ref{fig:SI_duty}), we manipulate $\xO_{\R{eff}}$ by modifying the duty cycle of the drive, as schematically denoted in the inset of fig.~\ref{fig:SI_duty}. Experimental data in fig.~\ref{fig:SI_duty} then confirms this theoretical prediction. 

Similarly, changing pulse frequency offset at fixed $\xO$ should result in a shift of the $\Cs$ spin layer that has a flip angle matching $\pi$, thereby changing $r_c$ and $\tzc$. In the rotating frame of the periodic drive, changing offset refers to applying a global $\zhat$-oriented field to the $\Cs$ nuclei. Since this field can either enhance or counteract the local field from the NV center gradient, the precise magnitude of the shift to $r_c$ is challenging to predict. However, qualitative trends in these shifts are easily observable. For instance, fig.~\ref{fig:SI_offset}A demonstrates the effect of changing the offset at a fixed length $t_p$ of the applied pulses. A change in $\tzc$ is observed, and for large offsets, the zero-crossing no longer appears in the 70s experiment time. This is consistent with effectively shifting the position of the $\Cs$ slice corresponding to $r_c$. We find that the $\xt {=}\pi$ slice occurs \I{after} the parabola-like $\tzc$ curve observed in \zfr{fig4}C for off-resonant pulses.  This again points to shifting $r_\mathrm{c}$ via the resonance offset field. Nevertheless, the physics underlying the formation and thermalization of the spinning shell remains unchanged in this case.

\begin{figure}[t]
  \centering
  {\includegraphics[width=0.49\textwidth]{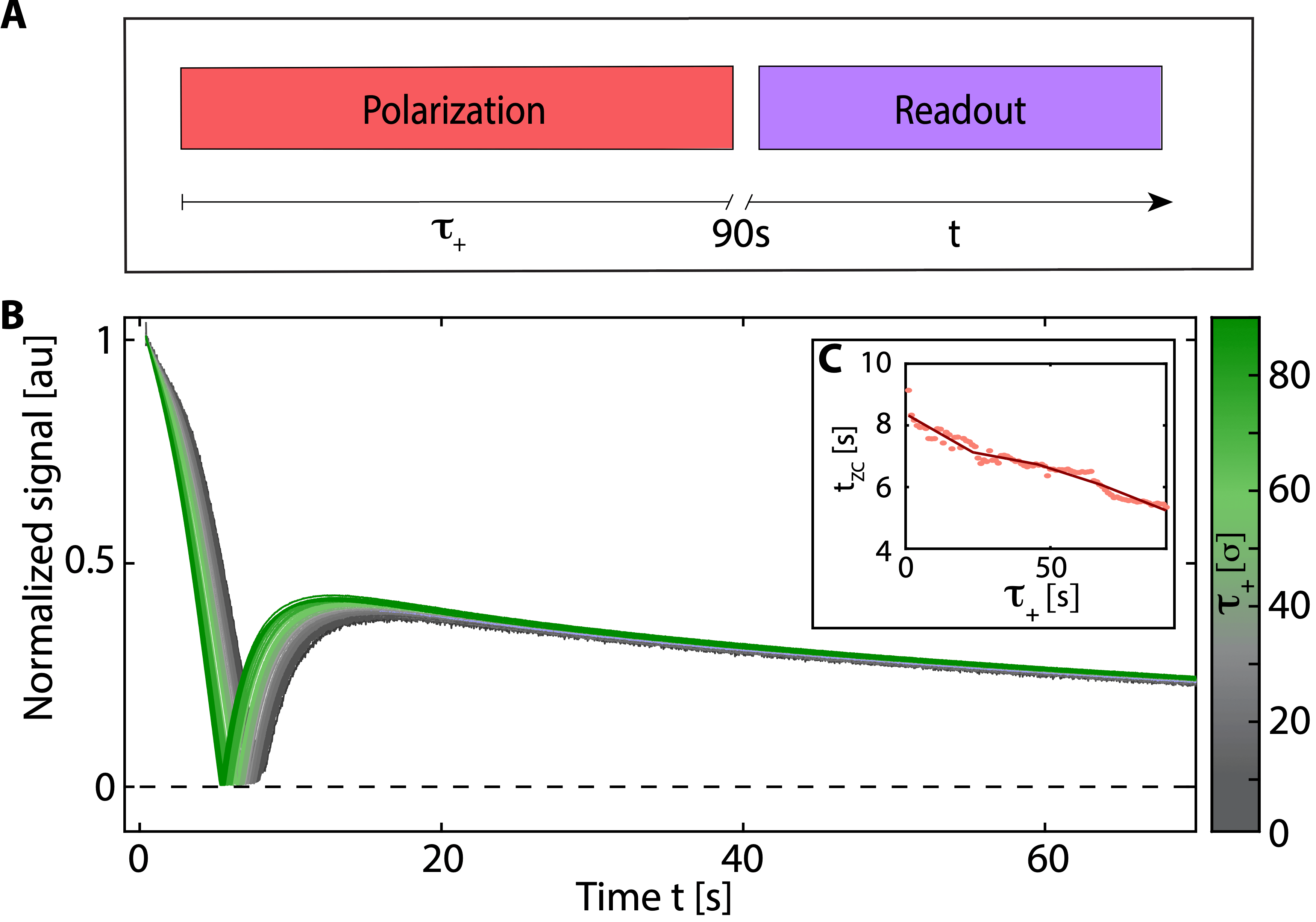}}
    \caption{\T{Effect of hyperpolarization injection $\qt_+$ on spin-shell texture.} (A) Schematic of the experiment sequence, where injection time $\qt_+$ is varied. After shuttling time $t_s{=}90$s the sample is subjected to a spin-locking train with $\xt{\sim}\pi$ for readout, identical to main text \zfr{fig4}A. (B) Normalized signal decays for different values of $\qt_+$, as indicated by the colorbar. (C) Analysis of the zero-crossing times $\tzc$ with $\qt_+$. Solid line is a spline guide to the eye. }
    \label{fig:SI_pol_time}
\zfl{SI_pol_time}
\end{figure}

\begin{figure*}[t]
  \centering
  {\includegraphics[width=0.99\textwidth]{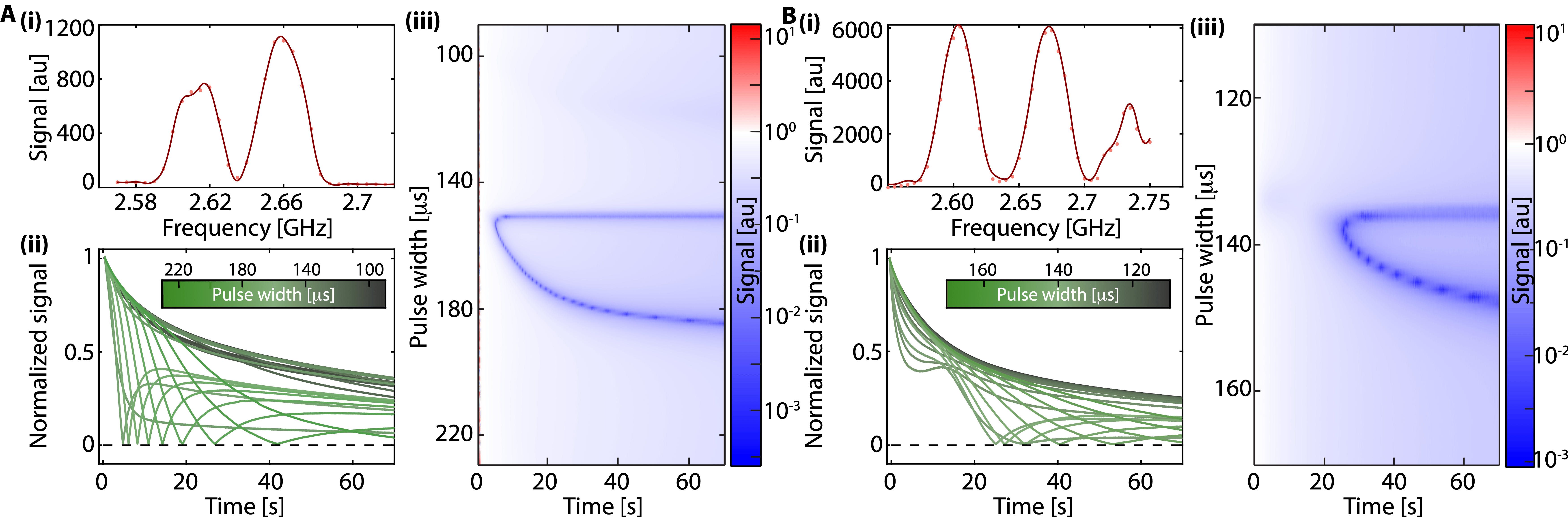}}
  \caption{\T{Spin-texture for two different sample orientations.} (A-B)i \I{EPR spectra.} Indirectly obtained EPR spectrum via $\Cs$ DNP, produced by sweeping MWs in narrow 25MHz windows. The integrated signal of resulting NMR decay curves is plotted as a function of the center frequency of the swept MWs. Solid line is a spline guide to the eye. (A-B)ii Representative signal decays in the proximity of $\xt{=}\pi$ showing zero crossings. (A-B)iii Color plots showing full data similar to \zfr{fig4}E of the main paper, highlighting the movement of the zero crossings with $\xt$ for the two orientations.
  Panels (A-B)iii indicate physics underlying {Hamiltonian Engineering} (\zfr{fig4}) is qualitatively independent of sample orientation.}
  \label{fig:SI_orientation}
\zfl{SI_orientation}
\end{figure*}

\begin{figure}[t]
  \centering
  {\includegraphics[width=0.49\textwidth]{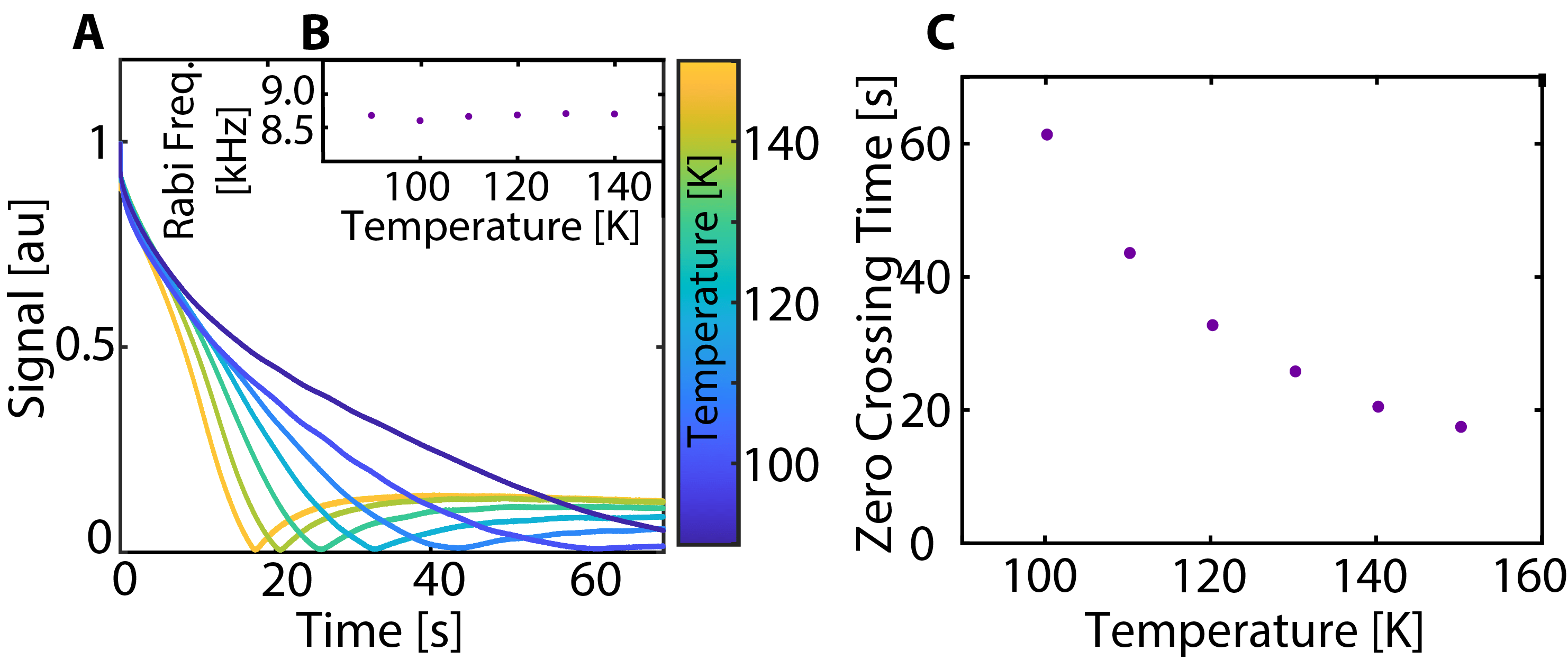}}
    \caption{\T{Variation of spin-shell profiles with temperature.}  (A) Representative spin-lock decays with the temperature (see colorbar). Experiments are performed at a fixed flip angle ($\xt{\sim}\pi$), and (B) identical Rabi frequency. (C) Variation of zero-crossing time $\tzc$ with temperature.
    }
\zfl{SI_temp}
\end{figure}

\subsection{Experiments with changing hyperpolarization time}
\zsl{hyp_time}

We examine the relationship between results displayed in \zfr{fig4} of the main paper and the amount of hyperpolarization injected into the $\Cs$ nuclear spins. Specifically fig.~\ref{fig:SI_pol_time} describes experiments varying the amount of time $\qt_+$ for which hyperpolarization is injected. Temperature is maintained consistent at 115K, flip angle here is 160$^\circ$, and shuttling time $t_s{=}90\,$s. 

The amplitude signal obtained in these experiments (fig.~\ref{fig:SI_pol_time}B) shows the expected zero crossing and subsequent phase inversion. The zero crossing time $\tzc$ does not significantly change with increased hyperpolarization time $\qt_+$. This stands in contrast to the movement of the zero crossing with variations in the flip angle (\zfr{fig4}) or temperature (fig.~\ref{fig:SI_temp}). These results support the fact that spin texture engineered via \T{Hamiltonian Engineering} is qualitatively independent of the initial state of nuclear spins.

\subsection{Experiments with different orientations}
\zsl{orient}
Here we examine the effect of changing the diamond sample orientation to results described in \zfr{fig4} of the main paper. We compare two different sample orientations, where the NV center families have different directions relative to $B_0$. Figure~\ref{fig:SI_orientation}A-B shows the DNP-measured EPR spectra of the NV centers in both cases. For experiments similar to \zfr{fig4}B, we find qualitatively identical development of spin-shell texture and zero-crossings in both cases, as described in the colorplots in fig.~\ref{fig:SI_orientation}Aiii,Biii. The physics underlying \zfr{fig4} can therefore be considered independent of the exact sample orientation. 

\subsection{Temperature dependence of spin texture}
\zsl{temp}
We now consider the effect of changing temperature on the observed spin texture signal via \T{Hamiltonian Engineering} (\zfr{fig4} of the main paper). A key advance enabled by the instrumentation (fig.~\ref{fig:method_instrument}) introduced in this paper is the ability to study the role of temperature (in 77K-RT range), which previously has been shown to have a strong effect on rates of electronic relaxation $T_{1e}$ (of both NV and P1 centers).

Figure~\ref{fig:SI_temp} presents experiments conducted at six representative temperatures. To ensure consistency, we maintain the same nuclear Rabi frequency for all experiments. Results indicate that the time at which the zero-crossing occurs ($\tzc$) increases as the temperature decreases. This can be seen as a direct reflection of the increase in the NV center $T_{1e}$, which in turn decreases the strength of the dissipation acting on the $\Cs$ nuclear spins (see also \zsr{SI-open-system}). We note that Ref. \cite{Awschalom_quenchdec_2008,Jarmola12} found that for samples with comparable NV center densities as the one employed here, $T_{1e}$ increases sharply close to 100K and can exceed $T_{1e}{>}1$s under these conditions.



\section{\label{sec:model}Theory model}

\subsection{Closed system}
\label{sec:closed_system}

As outlined in the main text, the system of interest is given by dipolar coupled $\Cs$ nuclear spins with the (rotating-frame) Hamiltonian
\begin{equation}
\label{eq:H_dd}
    \mHdd = \sum_{k<\ell}b_{k\ell}\left(3 I_{k}^zI_{\ell}^z-\mathbf{I}_k\cdot\mathbf{I}_\ell\right),
\end{equation}
where $\mathbf{I}_k=\left(I_k^x,I_k^y,I_k^z\right)^T$, the spin-$1/2$ operators $I_k^\eta= \sigma_k^\eta/2$ describes the nuclear spins, and $\sigma^\eta$ ($\eta\in \{x,y,z\}$) are the Pauli matrices. The couplings between different $\Cs$ spins strongly depends on their distance and the angle with the $\zhat$-axis via $b_{k\ell} = J_\mathrm{exp} (3\cos^2(\theta_{k\ell})-1)/{r}_{k\ell}^3$ where $\cos(\theta_{k\ell}) {=} \mathbf{B}\cdot\mathbf{r}_{k\ell}/(\vert \mathbf{B}\vert \vert \mathbf{r}_{k\ell}\vert )$; $\mathbf{r}_{k\ell}$ is the inter-spin vector and $\mathbf{B}=(0,0,B)^T$ is the externally applied magnetic field and $J_\mathrm{exp}{=}\mu_0 \hbar \gamma_n^2/4\pi$, with the gyromagnetic ratio of the neutron $\gamma_n$ and the vacuum permeability $\mu_0$. The $\Cs$-spins relevant for the experiment are randomly placed on a diamond lattice in the proximity of NV-centers. The NV-centers are approximately described by an electronic two-level system, which is itself coupled to all $\Cs$ nuclear spins via dipole-dipole interactions:
\begin{eqnarray}
\label{eq:H_NV}
 \mathcal{H}_{\mathrm{NV}}&=& \epsilon_{\mathrm{NV}} S^z \, , \\
 \label{eq:H_coupling}
 \mathcal{H}_{\text{NV-}^\mathrm{13}C} &=& \sum_{j} K_{j} \left(3S^z I_j^z - \mathbf{S} \cdot \mathbf{I}_j\right).
\end{eqnarray}
Here $S^\eta=\sigma^\eta/2$ with $\alpha \in \{x,y,z\}$ are (pseudo-)spin-$1/2$ operators describing the two-level system of the NV-center. $K_j= K_\mathrm{exp}(3\cos^2(\theta_j)-1)/{r}_{j}^3$ with the vector $\mathbf{r}_j$ between the NV center and the $j$-th $\Cs$ nuclear spin and the corresponding angle $\cos(\theta_{j})= \mathbf{B}\cdot\mathbf{r}_{j}/(\vert \mathbf{B}\vert \vert \mathbf{r}_{j}\vert ) $ between $\mathbf{r}_j$ and the external magnetic field $\mathbf{B}$, and $K_\mathrm{exp} {=}\mu_0 \hbar \gamma_n \gamma_e/4\pi$, where $\gamma_e$ is the gyromagnetic ratio of the electron.

Note that the system obeys the hierarchy of energy scales ${\epsilon_{\mathrm{NV}}\gg K \gg J}$, where $J$ and $K$ are characteristic energy scales associated with the Hamiltonians of Eq.~\eqref{eq:H_dd} and \eqref{eq:H_coupling}. The separation of energy scales is important as it allows to independently kick (i.e., apply strong short pulses to) the $\Cs$ spins while leaving the NV two-level system unaffected. Moreover, it allows us to consider the singular-coupling limit in order to derive an effective Lindblad master equation for the $\Cs$ spins coupled to the NV center (see below). The kick Hamiltonian of interest, created from an external microwave field at resonant Larmor-frequency of the external magnetic field, is given in the rotating frame by
\begin{equation}
    \mathcal{H}_x(t) = \begin{cases}
        \Omega \sum_j I_j^x &  n\period< t < n\period + \tkick \,, \hspace{.2cm} n \in \mathbb{N} \\
        0                   &  \mathrm{otherwise}
    \end{cases}  \, , \label{eq:Hx}
\end{equation}
with the driving period $\period$ and pulse width duration $\tkick$.

For the combined system of NV and $\Cs$ nuclear spins, it is useful to perform an interaction-picture transformation with $\Hdd+\mathcal{H}_{\text{NV-}^\mathrm{13}\text{C}}$ as the interaction Hamiltonian. Using the rotating-wave approximation, we readily obtain 
\begin{eqnarray}
\label{eq:RWA}
    \widetilde{\mathcal{H}}_\mathrm{tot}  = \Hdd +  \sum_{j} 2K_{j} S^z I_j^z \, ,\label{eq:H_total_RWA}
\end{eqnarray}
in the rotating frame with respect to $\exp(-i t \mathcal{H}_\mathrm{NV})$.

\subsection{Open system}
\zsl{SI-open-system}

In the experimental platform, both the NV center and the $\Cs$ spins are not completely isolated. They are coupled to a bath of phonons generated by the diamond lattice; such a bath causes decoherence and dissipation. 
Thus, a comprehensive theory of the system should take these effects into consideration.
Following the rotating-wave approximation, Eq.~\eqref{eq:RWA}, the microscopic Hamiltonian of the total system composed of the NV center and the $\Cs$ spins reads
\begin{align}
\label{eq:H_tot}
    \mathcal{H}_{\mathrm{tot}} &= \Hdd + \mathcal{H}_\mathrm{NV} +  \sum_{j} 2K_{j} S^z I_j^z \\
    &= \sum_{j<k} I_{jk} \left(3I_j^z J_k^z - \mathbf{I}_j \cdot\mathbf{I}_k\right) + \epsilon_{\mathrm{NV}}S^z + \sum_{j} 2K_{j} S^z I_j^z ,
\end{align}
in the lab frame.
The first term represents the Hamiltonian of the $\Cs$ spins, the second is the Hamiltonian of the NV center, and the last is the effective interaction between the two.
As mentioned, the total system obeys the following hierarchy of energy scales: 
\begin{equation}
\label{eq:hierarchy-energy}
    \epsilon_{\mathrm{NV}} \gg K \gg J,
\end{equation}
which implies that the timescale intrinsic to the NV center is much shorter than any other timescale of the problem.
This allows us to assume that the NV center is strongly coupled to the thermal phonon bath while the $\Cs$ spins are only weakly coupled to it. 
Therefore, we also assume that the NV center is always in thermal equilibrium with the phonons. 
For simplicity, we consider the NV at infinite temperature: $\rho_\mathrm{NV}(t) = \frac{1}{2} \identity_\mathrm{NV},$ for all times $t$; however, our results can be easily generalized to any temperature.

The thermal-equilibrium assumption for the NV allows us to derive an effective quantum master equation for the reduced system of the $\Cs$ spins by tracing out the NV center (and, indirectly, the phonons to which it is coupled) in the form of a Lindblad master equation. 
The latter is the most general Markovian master equation. 
It describes the equation of motion of a reduced system in contact with a thermal bath which is memoryless, i.e., whose timescale is much shorter than any other timescale in the problem, as the NV center in our system.
Moreover, the hierarchy of energy scales~\eqref{eq:hierarchy-energy} indicates the validity of the so-called \textit{singular coupling limit}, that enormously simplifies the microscopic derivation of the Lindblad equation~\cite{breuer2002theory,benatti2005open,rivas2012open}. 
Notice that previous works~\cite{Coish04} have investigated the opposite scenario in which one is interested in integrating out the nuclear spins to derive a master equation for the NV electron only. 
From Eq.~\eqref{eq:hierarchy-energy}, it is clear that this opposite scenario generates a highly non-Markovian master equation for the NV electron.

The derivation of the Lindblad equation for the $\Cs$ spins proceeds as follows. 
Let us rewrite the total system Hamiltonian~\eqref{eq:H_tot} in the more general form as
\begin{equation}
\label{eq:H-singular-coupling}
    \mathcal{H}_{\mathrm{tot}} = \mathcal{H}_\mathrm{^{13}C} \otimes \identity_\mathrm{NV} + \alpha^{-2} \identity_\mathrm{^{13}C} \otimes \mathcal{H}'_\mathrm{NV} + \alpha^{-1} V',
\end{equation}
where we have renormalized the bath and interaction Hamiltonian as $\mathcal{H}'_\mathrm{NV} = \alpha^2 \mathcal{H}_\mathrm{NV}$ and $V' = \sum_{j} 2 K'_{j} S^z I_j^z$ with $K'_j = \alpha K_j$.
Note that, by considering a generic system Hamiltonian $H_\mathrm{^{13}C}$ we emphasize that the Lindblad master equation derived in the singular coupling limit does not depend on it, as will be clear in the following.
%
The parameter $\alpha$ is the inverse of the coupling strength; in the singular coupling limit we eventually take $\alpha \to 0$.
Equation~\eqref{eq:H-singular-coupling} highlights the hierarchy of energy scales of our model~\eqref{eq:H_tot}. 

We start by considering the Nakajima-Zwanzig equation~\cite{breuer2002theory,rivas2012open}
\begin{eqnarray}
    \nonumber
    \frac{d}{dt}\mathcal{P}\tilde{\rho}(t) = \alpha^{-1} \mathcal{P} \mathcal{V}(t) \mathcal{P}\tilde{\rho}(t) + \alpha^{-1} \mathcal{P} \mathcal{V}(t) \mathcal{G}(t,0)\mathcal{Q}\tilde{\rho}(0)\\
    + \alpha^{-2} \int_0^t du \mathcal{P}\mathcal{V}(t) \mathcal{G}(t,u) \mathcal{Q} \mathcal{V}(u) \mathcal{P} \tilde{\rho}(u).
    \label{eq:Nakajima-Zwanzig}
\end{eqnarray}
where $\tilde{\cdot}$ denotes the time-evolved operator in the interaction picture; $\mathcal{P}$ and $\mathcal{Q}$ are two orthogonal projector operators (i.e., $\mathcal{P}^2 = \mathcal{P}$, $\mathcal{Q}^2 = \mathcal{Q}$, and $\mathcal{PQ} = \mathcal{QP} = 0$) given by $\mathcal{P} \rho = \mathrm{Tr}_\mathrm{NV}(\rho) \otimes \rho_\mathrm{NV}$ and $\mathcal{Q}\rho = (\identity - \mathcal{P}) \rho$; $\mathcal{V}(t) \,(\cdot) = -i [\tilde{V}'(t),\cdot]$; $\mathcal{G}(t,u)=\mathcal{T} e^{\int_u^tdt'\mathcal{Q}\mathcal{V}(t')}$ is the propagator with $\mathcal{T}$ the time-ordering operator.

The integro-differential equation~\eqref{eq:Nakajima-Zwanzig} is exact and describes the equation of motion of the relevant subspace $\mathcal{P}\tilde{\rho}(t)$ in the interaction picture. 
Unfortunately, it is usually as difficult to solve as the von Neumann equation describing the dynamics of the total system. 
Hence, we need to make a few more assumptions to proceed further:
First, we assume that 
\begin{equation}
    \label{eq:first-assumption-singular-coupling}
    \mathcal{P}\mathcal{V}(t)\mathcal{P}=0,
\end{equation}
which corresponds to $\mathrm{Tr}_\mathrm{NV}\bigl( \tilde{V}'(t) \rho_\mathrm{NV} \bigr) =0$.
The Hamiltonian~\eqref{eq:H_tot} fulfils this condition since $\mathrm{Tr}_\mathrm{NV}\bigl( S^z \rho_\mathrm{NV} \bigr) = \mathrm{Tr}_\mathrm{NV}\bigl( S^z/2 \bigr) = 0$.
Nonetheless, if the condition is not fulfilled one can shift the system Hamiltonian $\mathcal{H}_\mathrm{^{13}C}$ such that it is satisfied~\cite{rivas2012open}.
Second, we assume that 
\begin{equation}
    \label{eq:second-assumption-singular-coupling}
    \rho(0) = \rho_\mathrm{^{13}C}(0) \otimes \rho_\mathrm{NV} (0),
\end{equation}
which implies that we have control over the system of the $\Cs$ spins (for example, we can prepare them in a pure state at $t=0$).
This assumption is known as the Born approximation and it is necessary to obtain a universal dynamical map~\cite{rivas2012open}.

Assumptions~\eqref{eq:first-assumption-singular-coupling} and~\eqref{eq:second-assumption-singular-coupling} make the first and second term on the right-hand side of Eq.~\eqref{eq:Nakajima-Zwanzig} vanish; the latter now becomes
\begin{equation}
\label{eq:Nakajima-Zwanzig_2}
    \frac{d}{dt}\mathcal{P} \tilde{\rho}(t) = \alpha^{-2} \int_0^t du \mathcal{K}(t,u)\mathcal{P}\tilde{\rho}(u),
\end{equation}
where $\mathcal{K}(t,u)=\mathcal{P} \mathcal{V}(t) \mathcal{G}(t,u)\mathcal{Q}\mathcal{V}(u)$ is the memory kernel.
Since the NV is always in thermal equilibrium, the memory kernel becomes homogeneous: $\mathcal{K}(t,u)=\mathcal{K}(t-u)$.

Equation~\eqref{eq:Nakajima-Zwanzig_2} is in general non-Markovian as the state of the system at time $t$ depends on all the states at former times from 0 to $t$. 
By definition, a Markovian master equation for the system is obtained if the kernel $\mathcal{K}(t-u)$ behaves as a delta function with respect to $\tilde{\rho}(u)$.
This is verified in the singular coupling limit, where the hierarchy of energy scales~\eqref{eq:hierarchy-energy} implies that the typical timescale at which the kernel vanish, $\tau_\mathrm{NV}$, is much smaller than the typical variation timescale of the system, $\tau_\mathrm{^{13}C}$; thus, $\tau_\mathrm{NV}/\tau_\mathrm{^{13}C} \to 0$ and $\mathcal{K}(t-u) \propto \delta(t-u)$. This can be seen by integrating Eq.~\eqref{eq:Nakajima-Zwanzig_2} and taking the zeroth-order expansion in $1/\alpha$ of $\mathcal{K}(t-u)$---in fact, it can be proven that higher order terms vanish (see Refs.~\cite{gorini1976n,frigerio1976n,frigerio1977master})---we obtain
\begin{align}
    \mathcal{P}\tilde{\rho}(t) =& \mathcal{P} \tilde{\rho}(0)  \\
    &+ \alpha^{-2} \int_0^t ds \int_0^s du \mathcal{P} \mathcal{V}(s)\mathcal{V}(u) \mathcal{P} \tilde{\rho}(u) + \mathcal{O}(\alpha^{-3}), \nonumber
\end{align}
which is equivalent to
\begin{multline}
    \tilde{\rho}_\mathrm{^{13}C}(t) = \tilde{\rho}_\mathrm{^{13}C}(0) \\+ \alpha^{-2} \sum_{nm} \int_0^t ds \int_0^s du \frac{4 K'_n K'_m}{r^3_n r^3_m}
    \biggl( C_{ij}(s-u) \bigl[ \tilde{I}_j^z(u) \tilde{\rho}_\mathrm{^{13}C}(u), \tilde{I}^z_i(s) \bigr] \\
    + C^*_{nm}(s-u)  \bigl[ \tilde{I}^z_n(s), \tilde{\rho}_\mathrm{^{13}C}(u) \tilde{I}_m^z(u)  \bigr]  \biggr) + \mathcal{O}(\alpha^{-3}),
    \label{eq:Nakajima-Zwanzig_3}
\end{multline}
where the correlation functions are defined as
\begin{equation}
    C_{nm}(s-u) = \mathrm{Tr} \bigl( \tilde{S}^z(s-u) S^z \rho_\mathrm{NV} \bigr).
\end{equation}
By Fourier-decomposing the NV operator $\tilde{S}^z(s-u)$ and taking into account the factor $\alpha^{-2}$ in the free NV evolution, we get
\begin{equation}
\label{eq:bath-correlation}
    C_{nm}(s-u) = \int_{-a}^a d\omega \, e^{-\frac{i \omega (s-u)}{\alpha^2}} \mathrm{Tr}\bigl( S^z(\omega) S^z \rho_\mathrm{NV} \bigr),
\end{equation}
where the integration limits $a$ denotes the spectral support. 
We see that, when $\alpha \to 0$, the integral in Eq.~\eqref{eq:bath-correlation} tends to a delta function: $C_{nm}(s-u) \propto \delta(s-u)$.
Thus, by making the change of variable $w = \alpha^{-2} u$, and taking the limit $\alpha \to 0$, Eq.~\eqref{eq:Nakajima-Zwanzig_3} becomes a Markovian master equation for the $\Cs$-spins; in the Schr\"odinger picture, it reads
\begin{multline}
    \frac{d}{dt} \rho_\mathrm{^{13}C}(t) = -i \bigl[ \mathcal{H}_\mathrm{^{13}C} + \mathcal{H}_\mathrm{LS}, \rho_\mathrm{^{13}C}(t) \bigr] \\
    + \sum_{nm} \frac{4 K'_n K'_m}{ r^3_n r^3_m} \gamma_{nm} \biggl( I^z_m \rho_\mathrm{^{13}C}(t) I^z_n - \frac{1}{2} \{ I^z_n I^z_m, \rho_\mathrm{^{13}C}(t) \} \biggr),
\end{multline}
where the Lamb shift Hamiltonian is
\begin{equation}
    \mathcal{H}_\mathrm{LS} = \sum_{nm} L_{nm} I^z_n I^z_m.
\end{equation}
Here $\gamma_{nm}$ and $L_{nm}$ represents the Hermitian and anti-Hermitian components of the integral over time of the bath correlation functions~\eqref{eq:bath-correlation}:
\begin{equation}
    \int_0^\infty ds \, C_{nm}(s) = \frac{\gamma_{nm}}{2} + i L_{nm}
\end{equation}
Hence, by inserting Eq.~\eqref{eq:bath-correlation}, we have $\gamma_{nm} \propto 1 $ and $L_{nm} = 0$.
By absorbing constants of order unity into $K'_i K'_j$, we arrive at the final expression for the $\Cs$ master equation:
\begin{multline}
    \label{eq:lindblad-13C}
    \frac{d}{dt} \rho_\mathrm{^{13}C}(t) = -i \bigl[ \mathcal{H}_\mathrm{^{13}C}, \rho_\mathrm{^{13}C}(t) \bigr] \\
    + \sum_{nm} \frac{K'_n K'_m}{ r^3_n r^3_m} \biggl( I^z_m \rho_\mathrm{^{13}C}(t) I^z_n - \frac{1}{2} \{ I^z_n I^z_m, \rho_\mathrm{^{13}C}(t) \} \biggr).
\end{multline}

A few comments are in order. The hierarchy of energy scales~\eqref{eq:hierarchy-energy} and, thus, the validity of the singular coupling limit approximation for this model greatly simplify the derivation of the Lindblad master equation. In particular, the Lindbladian term on the second line of Eq.~\eqref{eq:lindblad-13C} is determined only by the bath Hamiltonian $\mathcal{H}_\mathrm{NV}$ and the interaction Hamiltonian $\alpha V'$ of Eq.~\eqref{eq:H-singular-coupling}, while it is completely agnostic of the system Hamiltonian $\mathcal{H}_{^{13}C}$. This is particularly useful if the system can be governed by different unitary dynamics $\mathcal{H}_{^{13}C}$, e.g., in the prethermal regimes determined by different kick angles (see Secs.~\ref{sec:one_d_pi} and~\ref{sec:one_d_pi_half}). 
Moreover, we note that the jump operators in Eq.~\eqref{eq:lindblad-13C} act as dephasing operators if the spins are polarized in the $\zhat$ direction, while they generate both dephasing and dissipation for a system that is polarized in the $\xhat$ direction. This result is in agreement with the experimental observations. In particular, it is at the origin of the non-decaying behavior of the normalized signal $S$ as a function of $\tw$ for $\tau_-{=}0$ observed in \zfr{fig3}E of the main text. Indeed, when $\tau_-{=}0$ the whole system is positively polarized along the $\zhat$ axis. Therefore, during $\tw$ no diffusion nor dissipation affect the system, giving a uniform behavior for any $\tw$.
Finally, we note that the strength of the dissipative terms $K'_n K'_m/(r^3_n r^3_m)$ decays faster with the distance from the NV center with respect to the dipole interaction. This allows us to consider the simplified dissipative one-dimensional toy model studied in Sec.~\ref{sec:approximate_dynamics}.

Let us emphasize that, the system also has a finite population of P1 centers~($20\,$ppm) that exceeds the NV-center population. However, the P1 centers do not contribute to the local magnetic field. As they are randomly distributed around each NV, their average effect can be modelled by a global spin depolarization, i.e., by rescaling the strength of the dissipative term. Since we are not interested in precise quantitative results the effect of P1 centers can be safely neglected.

\section{\label{sec:effective_Hamiltonian} Effective Hamiltonian analysis}

\begin{figure}
    \centering
   {\includegraphics[width=0.49\textwidth]{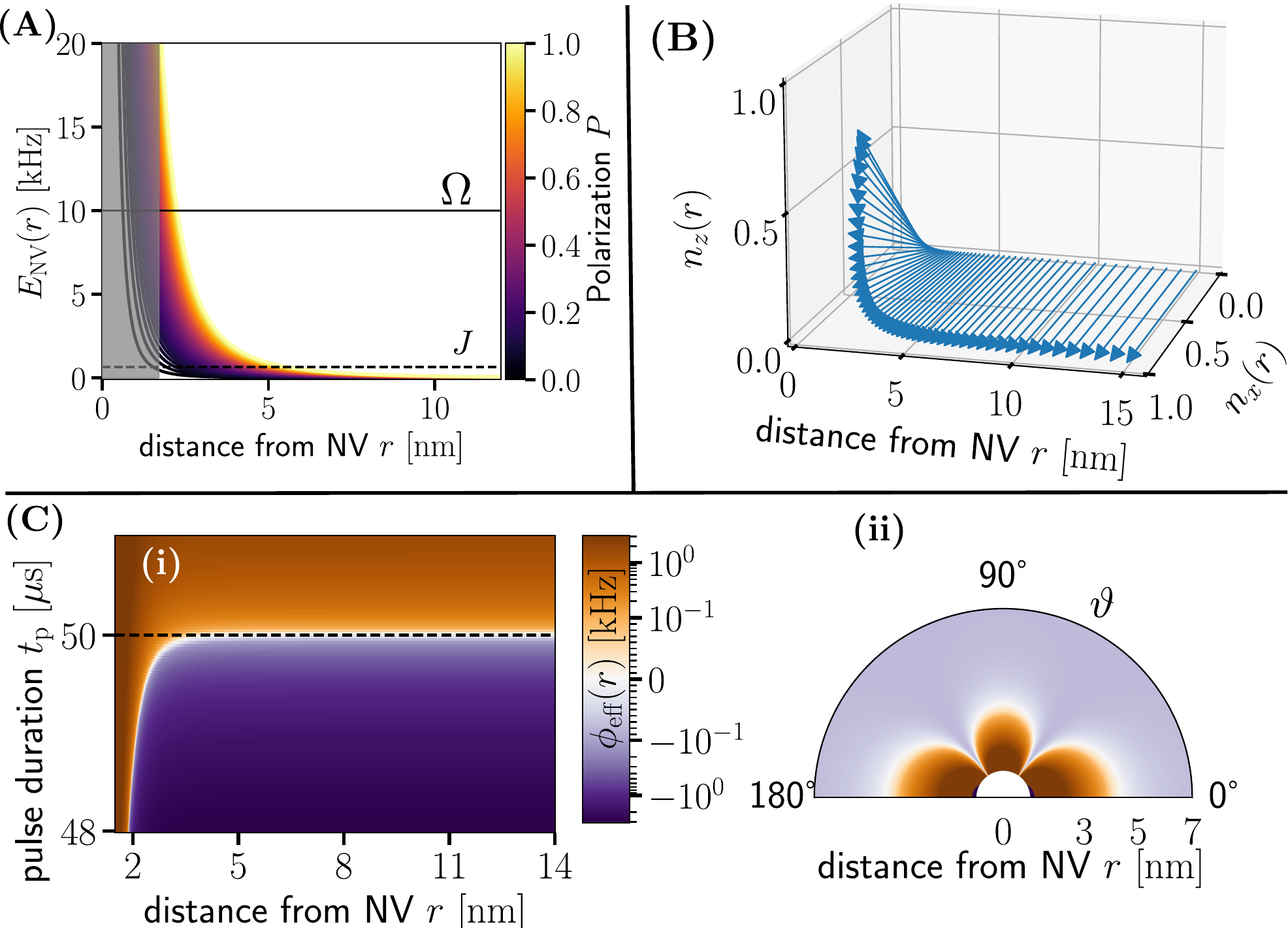}}
    \caption{
    \textbf{Effective Hamiltonian Summary.} 
    ${(\mathrm{A})}$ Energy scale $E_\mathrm{NV}(r)$ of magnetic field induced by the NV spin as a function of the distance from the NV, different colors correspond to different polarization of the NV spin $P=0-1$. Gray regime corresponds to the experimentally inaccessible frozen core.
    ${(\mathrm{B})}$ Spin locking at $\vartheta {\napprox} \pi$: Effective Hamiltonian at small tilt of the components of the conserved axis $\widetilde{I}$, Eq.~\eqref{eq:conserved_axis},  for spin locking at $\vartheta{\neq}\pi$ as a function of distance from NV.
    ${(\mathrm{C})}$ spin locking at $\vartheta {\approx} \pi$: ${(\mathrm{i})}$ Effective on-site potential $\potential_\mathrm{eff}(r)$, Eq.~\eqref{eq:effective_kickangle}, as a function of the distance $r$ from NV and pulse width $\tkick$, with 
    ${(\mathrm{ii})}$ Angle and radial distribution of effective on-site potential $\potential(r,\, \theta)$ for $\tkick=49.5\,\mu\mathrm{s}$. Zero crossing radius $r_c$ is given by $r_c(\theta) \approx 3.4\,\mathrm{nm}\times \sqrt[3]{| 3\cos(\theta)^2 - 1|}$.
    \textit{Summary.--} In the case of the spin locking sequence at $\vartheta{\neq}\pi$, the spatially inhomogenous on-site potential induced by the NV spin only leads to a tilt in the spin locking axis but no qualitative change in the physics compared to the case of absent on-site potential.
    However, for spin locking at $\vartheta{\approx}\pi$ the strong NV-induced on-site potential also leads to spatially inhomogenous on-site potential in the effective Hamiltonian with a possible sign change in space.
    \textit{Parameters.--} We choose Rabi frequency $\Omega=10\,\mathrm{kHz}$, interpulse time $\tdd=40\,\mu\mathrm{s}$ and for (B), (C) NV electron polarization $P=0.1$.
    }
    \label{fig:Heff_summary}
\end{figure}

Since the NV system experiences fast dynamics compared to the $\Cs$-spins we neglect the NV-spin degree of freedom and replace it by its mean value.
Hence, we replace the operator $S^z$ in Eq.~\eqref{eq:H_total_RWA} by its thermal expectation value $P=P(T)$, which depends on the temperature $T$, thus, leading to an effective on-site potential $\nvfield_{j} = 2 K_j P $ for the $\Cs$ nuclear spins. 
Thereby, we also neglect the effect of dissipation.

The kick sequence (see main text \zfr{fig1}B) at stroboscopic times $n\period {\approx} n (\tdd + \tkick)$, $n \in \mathds{N}$, can then be written as 
\begin{equation}
\label{eq:U_full}
    U = e^{-i \Hfull \tdd} e^{-i \Hkick \tkick}  \, ,
\end{equation}
with 
\begin{equation}
\label{eq:kick_evolution}
    \begin{aligned}
        \Hfull   &= \Hdd + \Hfield \, ,\\
        \Hkick   &= \mathcal{H}_x + \Hfield\, \,,\\
        \Hfield  &= \sum_j \nvfield_{j} I_j^z \, .
    \end{aligned}
\end{equation}
In Eq.~\eqref{eq:U_full} we additionally neglect the dipole-dipole coupling $\Hdd$ during the kicks, since its magnitude is much smaller compared to $\mathcal{H}_x$, i.e., $\norm{\Hdd} {\ll} \norm{\mathcal{H}_x}$. However, we consider the effective on-site potential during the kicks, since it can be of comparable strength at short distances from the NV, see also fig.~\ref{fig:Heff_summary}~(A) for a schematic of the on-site potential and other energy scales. Spins within a radius of $\leq 1.7\,\mathrm{nm}$, indicated as a shaded area in fig.~\ref{fig:Heff_summary} (A), are far detuned such that they are not contributing to the bulk dynamics and also cannot be measured by NMR experiments.

The kicked evolution in Eq.~\eqref{eq:U_full} corresponds to a periodic evolution with period $\period$. Therefore, we can exploit Floquet's theorem, which states that stroboscopic dynamics ($t=n\tau$, $n\in\mathbb{N}$) are described by an effective Hamiltonian 
\begin{equation}
    U(n\tau) = U(\tau)^n = U_F^n
\end{equation}
with
\begin{equation}
\label{eq:floquet_theorem}
    U_F = e^{-i \Hfull \tdd} e^{-i \Hkick \tkick} \equiv e^{-i \Hfloquet \period} \, .
\end{equation}
Although the exact effective Hamiltonian $\Hfloquet$ is, in general, a highly non-local object, analytical progress can be made by considering a high-frequency (small $\period$) expansion~\cite{Bukov15} up to some order $O(\period^n)$,
\begin{equation}
\label{eq:HFE}
    \Hfloquet = \sum_{n=0}^\infty \period^n \mathcal{H}_F^{[n]} = \Hfloquet^{(n)} + O(\period^{n+1})\, ,
\end{equation}
valid up to some finite number of drive cycles. For sufficiently small period $\tau$ we expect that this Hamiltonian is a good description of the prethermal plateau~\cite{Bukov15,Else17,Machado2020,abadal2020floquet,Beatrez21,fleckenstein2021prethermalization,fleckenstein2021thermalization,peng2021floquet} observed in the experiment. Any deviations from the effective Hamiltonian are expected to result in Floquet heating to a featureless infinite temperature state at long times~\cite{abanin2015heating,abanin2017heating}.

Before proceeding further, we outline the rest of this section and summarize the main results.
We start with a description in Sec.~\ref{subsec:ed_simulation} of the numerical algorithm used to support the analytical findings obtained in the rest of this section.
Then, in Sec.~\ref{subsec:SpinLocking} we will derive the effective Hamiltonian $\Heff$ to lowest order in the driving period $\period$. The derived effective Hamiltonians serve as the starting point of all other types of numerical and theoretical analysis.

Using a toggling frame expansion, we derive two effective Hamiltonians capturing the time evolution at stroboscopic times for the two regimes of a kick-angle $\vartheta\neq\pi$ and near $\vartheta\approx\pi$ in the presence of the hyperfine splitting field $\eta_j$.
In agreement with earlier findings~\cite{Beatrez2022,Sahin2022}, for $\vartheta\neq\pi$ we find that the effective Hamiltonian conserves the $\xhat$-polarization, while the hyperfine splitting $\eta_j$ only leads to a quantitative change in axis $\xhat\to\Tilde{\xhat}$.
In the case, $\vartheta\approx\pi$, this conservation law is broken and the hyperfine splitting induces an effective spatially inhomogenous potential dominating the relaxation dynamics of the polarization.

In Sec.~\ref{subsec:ETH} we analyze the long-time dynamics generated by the derived effective Hamiltonians. We show that, for finite systems, the long-time expectation values are well described by ETH-like arguments. In particular, we find that, in the $\vartheta=\pi$ case, the local polarization follows the local on-site potential induced by the electron which can lead to a sign-inversion of the local polarization and hence -- also of the integrated polarization.
We close this section by introducing simplified toy-model effective Hamiltonians; they capture the main physics but, due to their reduced complexity, make the qualitative analysis easier to understand.

\subsection{\label{subsec:ed_simulation} Exact diagonalization simulation}

Throughout this section, we support our findings by numerically exact \change{simulations of the full time-dependent problem}. 
However, these simulations are limited to small system sizes of $L{=}16$ spins. 
As we anticipate to find that the distance from the NV plays a crucial role, we will consider in the following a one-dimensional geometry to maximize the radial extension. Concretely, in this section we consider spins, labeled by $j=1,\,\dots,\,L$, to have positions on a one-dimensional chain described by $x_j = x_0 + a j + \delta x_j$, with lattice spacing $a=1$, offset distance $x_0=5$, some independent identically distributed random displacements $\delta x_j$, drawn from a normal distribution $\mathcal{N}(\mu{=}0,\, \sigma{=}0.01)$ with zero mean and standard deviation $\sigma=0.01\,a$. 
However, if not explicitly stated otherwise, we will consider the full time evolution described by Eq.~\eqref{eq:kick_evolution}, i.e. the spins interact via long-range interactions which decay as $1/r^3$ in the spin-spin distance $r$.
In addition we choose ${\Rabi/J=10}$, ${\detuning/J=0}$, ${J\tdd=0.1}$ and $\tkick$ is chosen such that $\Rabi \tkick=\xt$ reproduces the desired angle $\xt$, where $J$ is the median coupling described below.

The median coupling $J$ serves as an experimentally observable measure for the energy scale of the system, determined as the inverse $1/e$ decay time of a $\xhat$-polarized initial state in the absence of spin locking. Therefore, we can match numerics qualitatively with experimental results by re-expressing our quantities with this time scale.

\subsection{\label{subsec:SpinLocking} Derivation of Effective Hamiltonians}

\begin{figure}
    \centering
   {\includegraphics[width=0.49\textwidth]{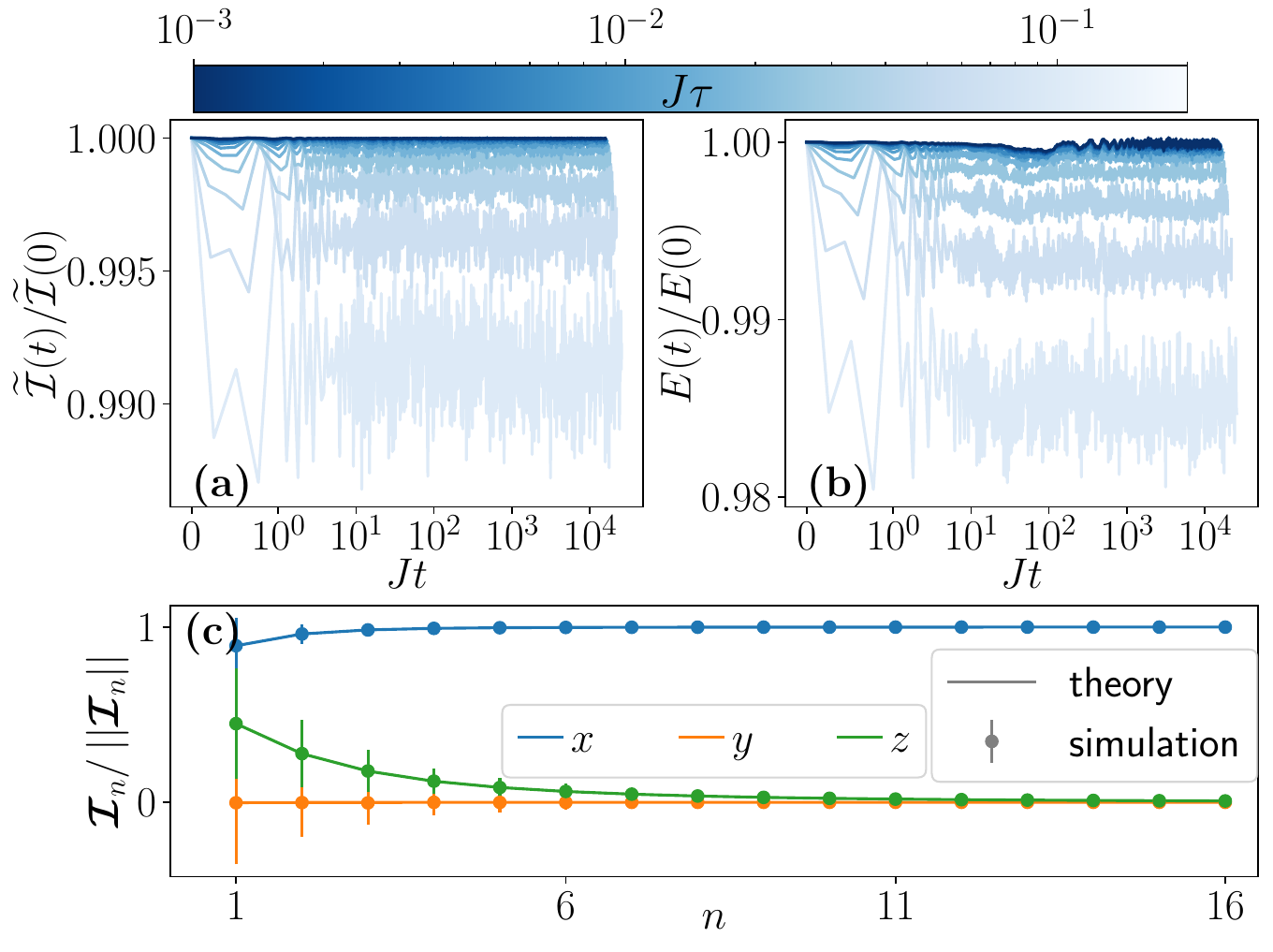}}
    \caption{
    \textbf{Spin Locking $\vartheta\neq\pi$}.
    Conservation of new spin direction (a) and the effective Hamiltonian $\mathcal{H}_\mathrm{eff}$~\eqref{eq:Heff_nopi} (b) with respect to time evolved state $\widetilde{\mathcal{I}}=\left< \widetilde{I}\right>$~\eqref{eq:conserved_axis} for different periods (colorbar).
    (c) Site resolved polarizations $\boldsymbol{\mathcal{I}}_n=\langle \mathbf{I}_n\rangle$ for smallest period $J\period{=}10^{-3}$ at late times; circles correspond to simulation results and lines are the results from ETH arguments, Eq.~\eqref{eq:thermal_state_nopi}.
    Details of simulation are described in Sec.~\ref{subsec:ed_simulation}.
    \textit{Summary.--}
    The dynamics for intermediate times are well described by the effective Hamiltonian Eq.~\eqref{eq:Heff_nopi} and the late time expectation value is in agreement with the ETH analysis, Eq.~\eqref{eq:local_polarization}. Deviations from the effective Hamiltonian description are parametrically suppressed in the driving period.
    }
    \label{fig:SL_nopi}
\end{figure}

A common technique to derive the lowest-order Floquet Hamiltonian is the so-called \textit{Floquet Magnus expansion}~\cite{Magnus54} which here is equivalent to using the Baker-Campbell-Hausdorff (BCH) formula~\cite{Achilles2012}: 
 \begin{equation}
    \label{eq:BCH}
    \exp(A) \exp(B) = \exp(A+B) + O(\norm{A}\cdot \norm{B}) \, ,
\end{equation}
which is a good approximation if $\norm{A}, \norm{B} \ll 1$.
However, with $\norm{\Hkick\tkick}>1$, as $\Rabi \tkick {>} 1$, this assumption is not met; moreover all powers $\Hkick^n$, $n=0,\,1,\,\dots$, of $\Hkick$ are of similar magnitudes $\norm{\Hkick}^n{\approx} \norm{\Hkick}^1$. Therefore, we cannot directly apply the BCH Formula~\eqref{eq:BCH} but would formally need to resum an infinite number of contributions stemming from the kicks. To this end, we will perform a so-called toggling frame transformation.

Notice also that, in the vicinity of the NV centre, the magnetic on-site potential contribution from the NV is strong compared to the dipole-dipole couplings between the nuclear spins. Since these terms commute, $[\Hdd,\, \Hfield  ]{=}0$, we can separate them into two exponentials $e^{-iHt} {=} e^{-i\Hdd t} e^{-i \Hfield t}$; hence, we can combine all strong contributions into a single one-particle unitary
\begin{equation}
\label{eq:U_singleparticle}
    U_\mathrm{SP} {\equiv} e^{-i \Hsp \tdd} {=}e^{-i \Hfield  \tdd} e^{-i \tkick \Hkick} \, ,
\end{equation}
where we defined $\Hsp \tdd {=}\sum_{j} \vartheta_j \nhat_j \cdot \mathbf{I}_j$ with
 \begin{equation}
    \label{eq:effective_kickangle}
     \begin{aligned}
         \vartheta_j =& 2 \arccos\left[ \cos(\alpha_j/2) \cos(\nvfield_j \tdd/2) \right. \\
                &\hspace{.1cm} \left. -\, \frac{\nvfield_j \tkick}{\alpha_j}\sin(\alpha_j/2) \sin(\nvfield_j \tdd/2) \right]\, ,
     \end{aligned}
 \end{equation}
and $\alpha_j {=} \tkick \sqrt{\Omega^2 + \nvfield_j^2}$. 
Note that, the time $\tdd$ appearing in the definition of $\Hsp$~\eqref{eq:U_singleparticle} is chosen such that the following equations take a simpler form. As only the product $\Hsp \tdd$ is physical we can choose a different time by rescaling $\Hsp$.

The components of the normalized direction vector $\nhat_j=n^x_j \xhat + n^y_j \yhat + n^z_j \zhat$ are given by 
\begin{equation}
\label{eq:direction_vec}
    \begin{aligned}
        n^x_j &= \frac{\Omega \tkick}{\alpha_j } \frac{\sin(\alpha_j/2) \cos(\nvfield_j \tdd/2)}{\sin(\vartheta_j/2)},   \\
        n^y_j &= \frac{-\Omega \tkick}{\alpha_j} \frac{\sin(\alpha_j/2) \sin(\nvfield_j \tdd/2)}{\sin(\vartheta_j/2)},  \\
        n^z_j &= \frac{\nvfield_j  \tkick}{\alpha_j} \frac{\sin(\alpha_j/2) \cos(\nvfield_j \tdd/2)}{\sin(\vartheta_j/2)} + \frac{\cos(\alpha_j/2) \sin(\nvfield_j \tdd/2)}{\sin(\vartheta_j/2)}.
    \end{aligned}
\end{equation}

\subsubsection{\label{subsec:SL_notpi}Spin locking for state engineering~\texorpdfstring{($\xt \napprox \pi$)}{TEXT}}

We can account for the strong kicks by transforming to the toggling frame considering the unitary evolution over $N$ cycles
\begin{equation}
\label{eq:toggling_frame}
    \begin{aligned}
        U^N &= \prod_{\ell=1}^N e^{-i \tdd \Hdd } U_\mathrm{SP} \\
            &= U_\mathrm{SP}^N \prod_{\ell=1}^N U_\mathrm{SP}^{-\ell} e^{-i \tdd \Hdd } U_\mathrm{SP}^\ell \\
            &= U_\mathrm{SP}^N \prod_{\ell=1}^N e^{-i \tdd \tilde{\mathcal{H}}_{\mathrm{dd}, \ell} } \, ,
    \end{aligned}
\end{equation}
where in the second line we introduced $N$ identities $U_\mathrm{SP}^{-k}U_\mathrm{SP}^k{=}\identity$ and in the last line we defined the rotated Hamiltonians $\tilde{H}_{\mathrm{dd}, \ell} {=} U_\mathrm{SP}^{-\ell}  \Hdd U_\mathrm{SP}^\ell$.

Notice that $\norm{\tilde{H}_{\mathrm{dd}, \ell}}{=}\norm{\Hdd}$ since they are related by a unitary transformation. Therefore, all contributions in the final expression in Eq.~\eqref{eq:toggling_frame} are small, so we can apply the BCH formula~\eqref{eq:BCH} to obtain
\begin{equation}
\label{eq:Heff_formally}
    \Heffnopi = \sum_{j=1}^{L} \frac{(\vartheta_j N)\mathrm{mod}(2\pi)}{N} \nhat_j \mathbf{I}_j + \Heffdd \, ,
\end{equation}
where $\Heffdd{=}\sum_{\ell=1}^{N} \tilde{\mathcal{H}}_{\mathrm{dd}, \ell}$.
In the following, we will evaluate the second term in Eq.~\eqref{eq:Heff_formally} explicitly in the limit of a large number of cycles $N{\to}\infty$. To this end, let us first write the dipole-dipole term as
\begin{equation}
    \Hdd = \sum_{k<\ell} b_{k\ell} \mathbf{I}_k^T \mathbf{D}  \mathbf{I}_\ell
\end{equation}
where $^T$ denotes transpose, and we introduced the diagonal matrix $\mathbf{D} = (-\xhat\xhat^T -  \yhat\yhat^T + 2\zhat\zhat^T) $. Then, the action of $U_\mathrm{SP}$ amounts to the matrix-vector product:
\begin{equation}
    \label{eq:pauli_rotation}
    U_\mathrm{SP}^{-\ell} \mathbf{I}_j U_\mathrm{SP}^{\ell} = \mathbf{r}(\ell \vartheta_j, \nhat_j) \mathbf{I}_j \, ,
\end{equation}
with the $3\times 3$ rotation matrix $\mathbf{r}(\vartheta, \nhat)$ rotating about the axis $\nhat$ by the angle $\vartheta$.
Hence, we have $\sum_{\ell=1}^{N} \tilde{H}_{\mathrm{dd}, \ell} {=} \sum_{k<l} b_{kl} \mathbf{I}_k \mathbf{M}_{kl} \mathbf{I}_l$, with the matrix $\mathbf{M}_{kl}=N^{-1}\sum_{\ell=1}^N  \mathbf{r}^T(\ell \vartheta_k, \nhat_k) \mathbf{D} \mathbf{r}(\ell \vartheta_l, \nhat_l)$.

In order to evaluate $\mathbf{M}_{kl}$, we make use of the Rodrigues representation of rotation matrices:
\begin{equation}
    \label{eq:Rodrigues_rotation}
    \mathbf{r}(\vartheta, \nhat) = 
     \nhat \nhat^T 
    + \cos(\vartheta) (1 - \nhat \nhat^T)
    + \sin(\vartheta) \epsilon(\nhat)  \, ,
\end{equation}
where $\epsilon_{ij}(\nhat){=}\sum_k \epsilon_{ikj} \nhat_k$ and $\epsilon_{ijk}$ is the fully antisymmetric Levi-Civita symbol.
Thus, $\mathbf{M}_{kl}$ reads as
\begin{equation}
\label{eq:side_Mkl} 
    \begin{aligned}
        \mathbf{M}_{kl} = \frac{1}{N} &\sum_{\ell=1}^{N} \left( \left[ \nhat_k \nhat_k^T 
    + \cos(\ell\vartheta_k) (1 - \nhat_k \nhat_k^T)
    + \sin(\ell \vartheta_k) \epsilon(\nhat_k)\right] \right. \\
        & \left. \mathbf{D} \left[ \nhat_l \nhat_l^T 
    + \cos(\ell\vartheta_l) (1 - \nhat_k \nhat_l^T)
    + \sin(\ell \vartheta_l) \epsilon(\nhat_l)\right] \right).
    \end{aligned}
\end{equation}
The sums of the $\sin$ and $\cos$ contributions in $\mathbf{M}_{kl}$ can be evaluated as
\begin{equation}
    \begin{aligned}
        \mathcal{G}_s(N, \vartheta) &= \sum_{j=1}^N \sin(j \vartheta) = \frac{\sin(N\vartheta/2)}{N\sin(\vartheta/2)} \sin((N-1)\vartheta/2), \\
        \mathcal{G}_c(N, \vartheta) &= \sum_{j=1}^N \cos(j \vartheta) = \frac{\sin(N\vartheta/2)}{N\sin(\vartheta/2)} \cos((N-1)\vartheta/2) \, .
    \end{aligned}
\end{equation}

Using the relation $\lim_{N\to \infty} \mathcal{G}_{s,c}(N, \vartheta) = 0$ for $\vartheta\neq 2\pi k$, $k\in \mathbb{N}$,
all contributions appearing in Eq.~\eqref{eq:side_Mkl} which are not proportional to unity, $\cos(\vartheta_j)\cos(\vartheta_k)$, or $\sin(\vartheta_j)\sin(\vartheta_k)$ vanish in the large $N$ limit.

In summary, in the large $N$ limit, we find the following expression for the leading-order Floquet Hamiltonian: 
\begin{equation}
\label{eq:Heff_nopi}
    \Heffnopi = \sum_k \frac{(\vartheta_k N)\mathrm{mod}(2\pi)}{N} \nhat_k \cdot \mathbf{I}_k +  \sum_{k<l} b_{kl} \mathbf{I}_k \mathbf{M}^{(0)}_{kl} \mathbf{I}_l \, ,
\end{equation}
with
\begin{align*}
    \mathbf{M}^{(0)}_{kl} =& \nhat_k \nhat_k^T \mathbf{D} \nhat_l \nhat_l^T \\
                                &+ \mathcal{G}_c(\Delta_{kl} \vartheta) (1 - \nhat_k \nhat_k^T) \mathbf{D} (1 - \nhat_l \nhat_l^T) \\
                                &+ \mathcal{G}_c(\Delta_{kl} \vartheta) \epsilon(\nhat_k)^T \mathbf{D}\epsilon(\nhat_l)\, .
\end{align*}

Notice that Eq.~\eqref{eq:Heff_nopi} is invariant under the basis transformation $U(\vartheta){=}\exp(-i\vartheta \sum_k \nhat_k \mathbf{I}_k)$ for any $\vartheta$. This can be immediately checked by applying the transformation rules
\begin{align*}
\left(1-\nhat_k\nhat_k^T\right)\mathbf{I}_k 
    &\overset{U(\vartheta)}{\longrightarrow} \cos(\vartheta) \left(1-\nhat_k\nhat_k^T\right) \mathbf{I}_k + \sin(\vartheta) \epsilon(\nhat_k)  \mathbf{I}_k \, ,
\\
\epsilon(\nhat_k) \mathbf{I}_k   
    &\overset{U(\vartheta)}{\longrightarrow}
    -\sin(\vartheta) \left(1-\nhat_k\nhat_k^T\right) \mathbf{I}_k + \cos(\vartheta) \epsilon(\nhat_k) \mathbf{I}_k \, ,
\\
\nhat_k^T \mathbf{I}_k &\overset{U(\vartheta)}{\longrightarrow} \nhat_k^T \mathbf{I}_k \, ,
\end{align*}
to Eq.~\eqref{eq:Heff_nopi}, leading to $\Heff \overset{U(\vartheta)}{\longrightarrow}\Heff$.
Therefore, the total rotated polarization 
\begin{equation}
\label{eq:conserved_axis}
    \widetilde{I} = \sum_{k} \nhat_k \cdot \mathbf{I}_k \, , 
\end{equation}
is a quasi-conserved quantity, i.e. $[\widetilde{I} ,\, \Heffnopi] = 0$.

In the absence of the local on-site potential from the NV ($\nvfield_j {=}0$), which enters the expression for $\nhat_k$, the conserved quantity simplifies to the total $\xhat$-polarization, $\widetilde{I}{=}I_x$. This is consistent with previous expressions \cite{Beatrez2022,Sahin2022} where the spatially-dependent on-site potential of the NV was neglected. 
By contrast, in the presence of the NV-induced on-site potential ($\nvfield_j {\neq}0$), the conserved quantity is locally tilted away from the $\xhat$-axis close to the NV, see fig.~\ref{fig:Heff_summary}(b) and fig.~\ref{fig:SL_nopi}(c).

In fig.~\ref{fig:SL_nopi}(a,b) we compare the dynamics generated by the lowest order Floquet Hamiltonian $\Heff$ derived above, against the exact kicked quantum simulation performed for $L{=}16$ spin; in particular, we measure the conservation of the effective axis and the effective Hamiltonian. 
We find that both quantities are conserved up to errors which parametrically scale down in $\tdd$, as expected from the regime of validity of the lowest order approximation.

In summary, the presence of the NV-induced on-site potential $\nvfield_j $ only leads to minor quantitative corrections to the spin-locking sequence far away from $\vartheta {=} \pi$.
In the next section, we show that this is not true for spin locking in the vicinity of $\vartheta {=} \pi$, where the situation is much more intriguing.

\subsubsection{\label{subsec:SL_pi}Spin locking for Hamiltonian engineering \texorpdfstring{($\xt \approx \pi$)}{TEXT}}

\begin{figure}
    \centering
   {\includegraphics[width=0.49\textwidth]{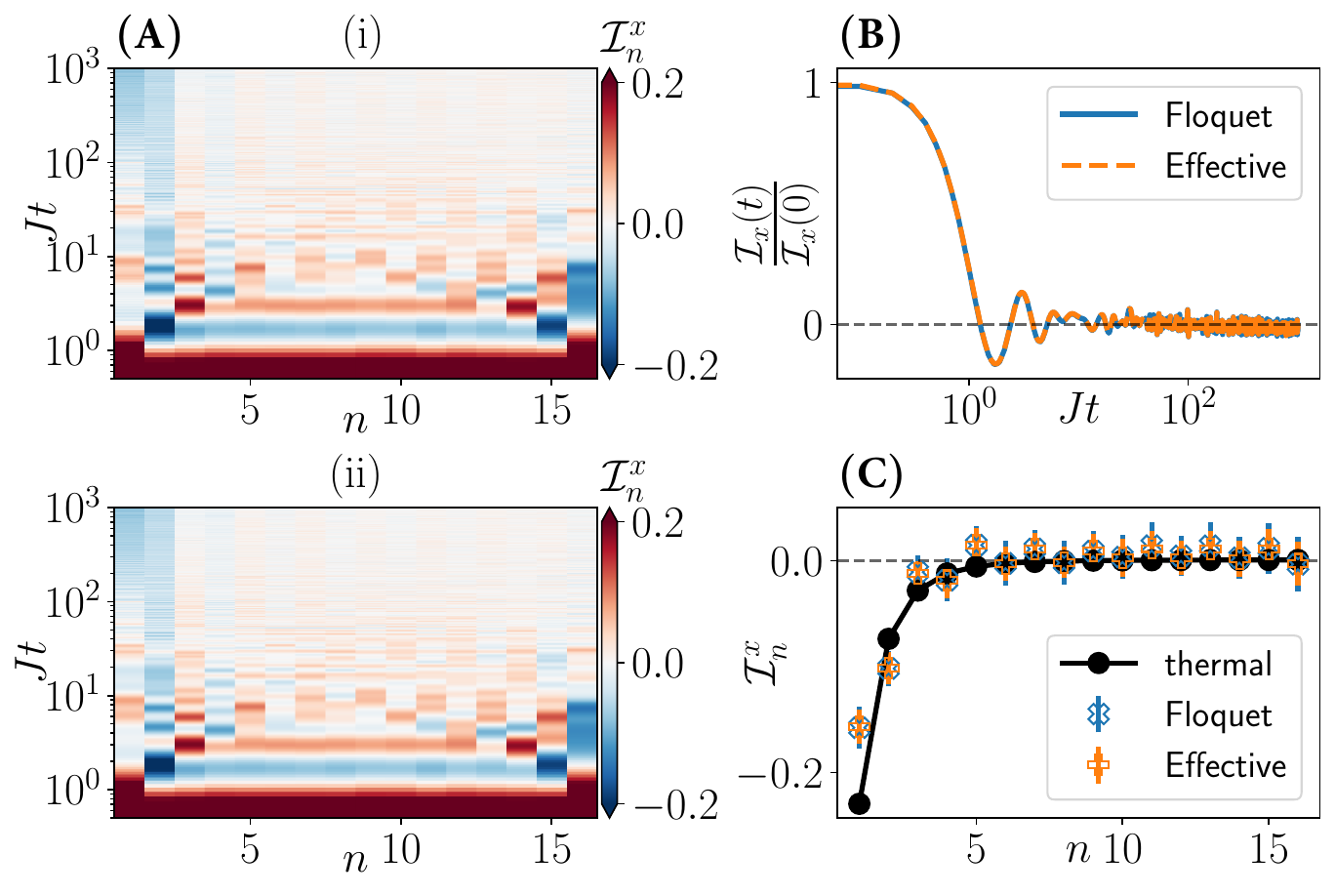}}
    \caption{\textbf{Spin Locking at $\vartheta\approx \pi$}--
    (a) (i) Exact Floquet evolution for $L=16$ spins $\xhat$-polarization $\left<\sigma^x_n\right>(t)$ of each spin, indexed by $n$, as a function of time.
    (ii) Same as (i) but for effective Hamiltonian~\eqref{eq:Heff_pi}.
    (b) Total $\xhat$-polarization $\mathcal{I}_x{=}\sum_{n=1}^N \exptval{\sigma_n^x}$ as a function of time for full time-dependent~(blue) and effective Hamiltonian~(orange) evolution.
    (c) Spatially resolved late time expectation value of the $\xhat$-polarization for full time-dependent~(blue crosses) and effective Hamiltonian~(orange pluses) evolution compared to the ETH thermal expectation values~(black circles) with finite inverse temperature $\beta\approx -0.45\,J$. Error bars correspond to standard deviation around mean value over the last $1000$ cycles.
    \textit{Summary}-- 
    The full evolution is well described by the effective Hamiltonian approach for all considered times. Moreover, pre-thermalization follows the predictions from ETH arguments for the effective Hamiltonian as discussed around Eq.~\eqref{eq:local_polarization_pi}.
    \textit{Parameters}--See Sec.~\ref{subsec:ed_simulation} and ${J\tkick\approx 0.341}$ is chosen such that the crossing appears at the ${n=7}$ spin.
    }
    \label{fig:SL_pi}
\end{figure}

The analysis performed in Sec.~\ref{subsec:SL_notpi} breaks down as the spatially-dependent kick-angle $\vartheta_j$, see Eq.~\eqref{eq:effective_kickangle}, approaches $\pi$, as we can no longer assume $\lim_{N\to \infty} \mathcal{G}_{s,c}(N, \vartheta) {=} 0$.
In this section, we will analyze this case in detail.

Let us first focus on the fine-tuned case $\vartheta{=}\pi$. Notice that, for $\vartheta{=}\pi$ the kick unitary squares to the identity, $U_\mathrm{kick}^2=\identity$.
Therefore, the toggling frame unitary from Eq.~\eqref{eq:toggling_frame},
\begin{equation*}
    U^N = U_\mathrm{SP}^N \prod_{n=1}^N U_\mathrm{SP}^{-n} e^{-i \Hdd \tdd} U_\mathrm{SP}^n \, ,
\end{equation*}
reduces to
\begin{equation*}
    U^N = \prod_{n=1}^{N/2} e^{-i \Hdd \tdd} U_\mathrm{SP}^\dagger e^{-i \Hdd \tdd} U_\mathrm{SP} \, .
\end{equation*}
Thus, at exactly $\vartheta{=}\pi$ the lowest order Floquet Hamiltonian is simply given by
\begin{equation}
    \Bar{H}  \equiv \frac{1}{2} \left(\Hdd + U_\mathrm{SP}^\dagger \Hdd U_\mathrm{SP}\right). 
\end{equation}

If $\vartheta{\approx} \pi$ but not exactly $\pi$, we may split the unitary $U_\mathrm{SP}{=}\exp(-i \delta \vartheta \nhat\cdot \mathbf{I})\exp(-i \pi \nhat\cdot \mathbf{I})$ into a $\vartheta{=}\pi$ and a $\delta \vartheta{=}(\vartheta - \pi)$ contribution.
Notice that the $\delta \vartheta$-contribution is small by assumption. Hence, to lowest order $O(\delta \vartheta, J\period)$ we may include this contribution into the dipole-dipole Hamiltonian $\exp(-i \Hdd \tdd) {\to} \exp(-i \left[ \Hdd \tdd + \delta \vartheta \nhat\cdot \mathbf{I}\right]) + O(\delta \vartheta, J\period)$. 
Thus, the lowest-order toggling frame expansion reads as
\begin{equation}
\label{eq:Heff_pi}
    \Heffpi = \bar{H} +  \sum_{k=1}^L \boldsymbol{\potential}_{\mathrm{eff},k} \cdot \mathbf{I}_k \, ,
\end{equation}
where $\boldsymbol{\potential}_{\mathrm{eff},k} = \delta \vartheta_k \nhat_k/\period$.

Let us emphasize that around $\vartheta{=}\pi$ the effective Hamiltonian does not conserve the (tilted) total net polarization $\sum_{k=1}^L \boldsymbol{\potential}_{\mathrm{eff},k} {\cdot} \mathbf{I}_k$.
This is in stark contrast to the results for spin locking far away from $\vartheta{=}\pi$, see discussion after Eq.~\eqref{eq:conserved_axis}.

In fig.~\ref{fig:SL_pi} we compare the exact evolution at $\vartheta{\approx}\pi$ with the effective Hamiltonian evolution~\eqref{eq:Heff_pi} for a system of $L{=}16$ spins. 
The effective Hamiltonian dynamics capture the exact evolution for all observed times. At even longer times we expect Floquet heating to become dominant, leading to a heat death of the driven system; notice that this does not happen in the dynamics generated by the effective Hamiltonian shown in fig.~\ref{fig:SL_pi}.

\subsection{\label{subsec:ETH} Analysis based on the Eigenstate Thermalization Hypothesis}

Any generic many-body system is expected to thermalize according to the eigenstate-thermalization hypothesis~\cite{Srednicki1995,Deutsch2018,dalessio2016quantum} (ETH). As a periodically driven system does not conserve energy it is expected to thermalize to a featureless, infinite temperature state ${\propto \identity}$ for long times~\cite{lazarides2014equilibrium,dalessio2014longtime,luitz2020prethermalization}.
However, heating is suppressed at high driving frequency, leading to a separation of time scales in the high-frequency regime and to the build-up of a so-called prethermal plateau~\cite{abanin2015heating,abanin2017heating}: the system first (pre-)thermalizes with respect to the low order effective Hamiltonian, before eventually fully thermalizing at long times towards the infinite temperature state.

In this section, we analyze the prethermal plateau characterized by the effective Hamiltonians derived in Sec.~\ref{subsec:SpinLocking}.
ETH suggests that if a system evolves under some generic unitary evolution $U(t)$, and this unitary preserves some local operators $\{ O_{j} \}_{j=1}^{n_O}$, then the system is expected to thermalize to a state which is locally equivalent to
\begin{equation}
\label{eq:thermal_state}
    \rho(\lambda_1,\,\dots,\, \lambda_{n_O}) = e^{\sum_{j=1}^{n_O} \lambda_{j} O_j }/Z,
\end{equation}
where $Z=\mathrm{Tr}(e^{\sum_{j=1}^{n_O} \lambda_{j} O_j })$. The parameters $\lambda_1,\,\dots,\, \lambda_{n_O}$ are Lagrange multipliers fixed by the initial conditions using the self-consistency relation
\begin{equation*}
    \left< O_j \right>_{\psi_0} \overset{!}{=} \mathrm{Tr}\left( \rho(\lambda_1,\,\dots,\, \lambda_{n_O}) O_j\right) \, ,
\end{equation*}
where $\ket{\psi_0}$ is the initial state.

\paragraph*{Thermalization at $\vartheta \neq \pi$. --}
In the regime $\vartheta\neq \pi$, both the effective Hamiltonian~$\Heff$ and the rotated polarization $\widetilde{I}_x$ are (prethermally, i.e., quasi-) conserved quantities, as $\left[\widetilde{I}_x, \Heff\right]=0$~(see discussion around Eq.~\eqref{eq:conserved_axis}). Therefore, using Eq.~\eqref{eq:thermal_state} expectation values of local observables in the (pre-)thermal plateau are obtained from 
\begin{equation}
\label{eq:thermal_state_nopi}
    \rho_{\vartheta\neq\pi}(\beta, \mu) = e^{-\beta \Heffdd + \mu \widetilde{I}_x }/Z \, ,
\end{equation}
where we collected all contributions ${{\propto} \widetilde{I}_x}$ and with ${Z{=}\mathrm{Tr}(e^{-\beta \Heffdd + \mu \widetilde{I}_x })}$. The parameters $\beta$ and $\mu$ are fixed by the energy and polarization of the initial state $\ket{\psi_0}$:
\begin{equation}
\label{eq:LagrangeMultipliers_nopi}
    \begin{aligned}
        \left< \Heff \right>_{\psi_0}      &= \Tr{ \Heff \rho_{\vartheta\neq\pi}} \\ 
        \left< \widetilde{I}_x\right>_{\psi_0} &= \Tr{ \widetilde{I}_x \rho_{\vartheta\neq\pi}} .
    \end{aligned}
\end{equation}

In particular, the local polarizations $\left< \sigma_n^\alpha\right>$, $n=1,\,\dots\,L$ and $\alpha=x,\,y,\,z$, at long times are expected to be described by the thermal expectation value
\begin{equation}
\label{eq:local_polarization}
    \begin{aligned}
        \mathcal{I}_n^\alpha = \langle {I}_n^\alpha \rangle
        &= \mathrm{Tr}\left(I_n^\alpha \rho_{\vartheta\neq\pi}(\beta, \mu)\right) \\
        &\approx \mathrm{Tr}\left(I_n^\alpha e^{\mu \widetilde{I}_x}/Z \right) \\
        &= \nhat^\alpha_n \tanh({\mu}/2) \, ,
    \end{aligned}
\end{equation}
where we used in the second line, that $\beta$ is expected to be small, $\beta{\ll}1$. Further we exploit the tracelessness of all non-trivial pauli strings 
\begin{equation}
\label{eq:pauli_trace}
    \mathrm{Tr}\left( \sigma^\mu \otimes \sigma^\nu \otimes \dots \sigma^\rho \right) \neq 0 \Longleftrightarrow \sigma^\mu=\sigma^\nu= \dots =\sigma^\rho=\identity \, .
\end{equation}
Therefore, the spins also locally align with the rotated axis $\nhat_n$, i.e.
\begin{equation}
\label{eq:local_direction}
    \frac{\boldsymbol{\mathcal{I}}_n }{\norm{\boldsymbol{\mathcal{I}}_n}} = \nhat_n \, .
\end{equation}

In fig.~\ref{fig:SL_nopi}(c) we compare the numerically exact, dynamics at long times against the results expected from the ETH analysis~(Eq.~\eqref{eq:local_direction}). We find excellent agreement between the numerically exact evolution results and the analytical predictions in Eq.~\eqref{eq:local_direction}.

\paragraph*{Thermalization around $\vartheta=\pi$. --} In the regime $\vartheta \approx \pi$ only the Hamiltonian $\Heffpi$ is conserved in the prethermal plateau. 
Thus, the (pre-)thermal steady state for a finite system at long times is expected to be locally equivalent to the thermal density matrix
\begin{equation}
\label{eq:thermal_state_pi}
    \rho_{\pi}\left(\beta\right) = e^{-\beta \Heffpi }/Z \, ,
\end{equation}
where $Z=\mathrm{Tr}(e^{-\beta \Heffpi })$. The inverse temperature $\beta$ is determined due to (prethermal) energy conservation by the energy expectation in the initial state $|\psi_0\rangle$:
\begin{equation}
\label{eq:LagrangeMultipliers_pi}
    \langle\psi_0| \Heffpi |\psi_0\rangle \overset{!}{=} \mathrm{Tr}\left( \Heffpi \rho_{\pi}\left(\beta\right)\right) .
\end{equation}
Notice that this so-called temperature is an intrinsic property of the evolution and the initial state, and is not related to the actual temperature the experiment is operated at.

Using a high-temperature, $\beta\norm{\Heffpi} \ll 1$, expansion of the density matrix $\rho_\pi\approx (\identity - \beta\Heffpi )/Z$, where now $Z=\mathrm{Tr}(\identity)=2^L$, in Eq.~\eqref{eq:thermal_state_pi} we can compute the local polarization
\begin{equation}
\label{eq:local_polarization_pi}
    \begin{aligned}
        \mathcal{I}_n^\alpha
        &= \mathrm{Tr}\left( \rho_\pi I_n^\alpha \right)\\
        &\approx -\beta \mathrm{Tr} \left( \Heffpi I_n^\alpha\right)/Z\\
        &= -\beta \potential^{\alpha}_{\mathrm{eff},n}/2 \\
        &\propto \potential^{\alpha}_{\mathrm{eff},n} \, ,
    \end{aligned}
\end{equation}
where we used in the last equality the definition of $\Heffpi$, Eq.~\eqref{eq:Heff_pi}, and tracelessness of all non-trivial pauli-strings, Eq.~\eqref{eq:pauli_trace}.
In addition, the inverse temperature can be obtained similarly from Eq.~\eqref{eq:LagrangeMultipliers_pi}
\begin{equation}
\label{eq:inverse_temperature}    
    \begin{aligned}
        \beta  = -\frac{\langle\psi_0| \Heffpi |\psi_0\rangle}{\mathrm{Tr}\left(\Heffpi  \Heffpi \right)/Z}\, .
    \end{aligned}
\end{equation}

In fig.~\ref{fig:SL_pi}(c) we compare the exact long time steady states for numerically exact and effective Hamiltonian evolution with the thermal expectation value Eq.~\eqref{eq:local_polarization_pi}. We find excellent agreement between all three methods.
In particular, let us emphasize that due to $\potential^{\alpha}_{\mathrm{eff},n}\propto \delta \vartheta_n = \vartheta_n - \pi$ the effective magnetic on-site potential acting on the spins experiences a sign change around ${\vartheta_n=\pi}$ as a function of the distance from the NV. Therefore, since $\left< I_n^\alpha\right> \propto \potential^{\alpha}_{\mathrm{eff},n}$, see Eq.~\eqref{eq:local_polarization_pi}, the final polarization profile is inhomogenous in space and can have positive and negative distribution although the initial state is fully positively polarized.
This is a key result of this work as it allows for engineering desired non-homogeneous states via Hamiltonian-engineering.

\subsection{\label{subsec:crossing_radius} Estimating the crossing radius \texorpdfstring{$r_c$}{TEXT}}

\begin{figure}
    \centering
   {\includegraphics[width=0.5\textwidth]{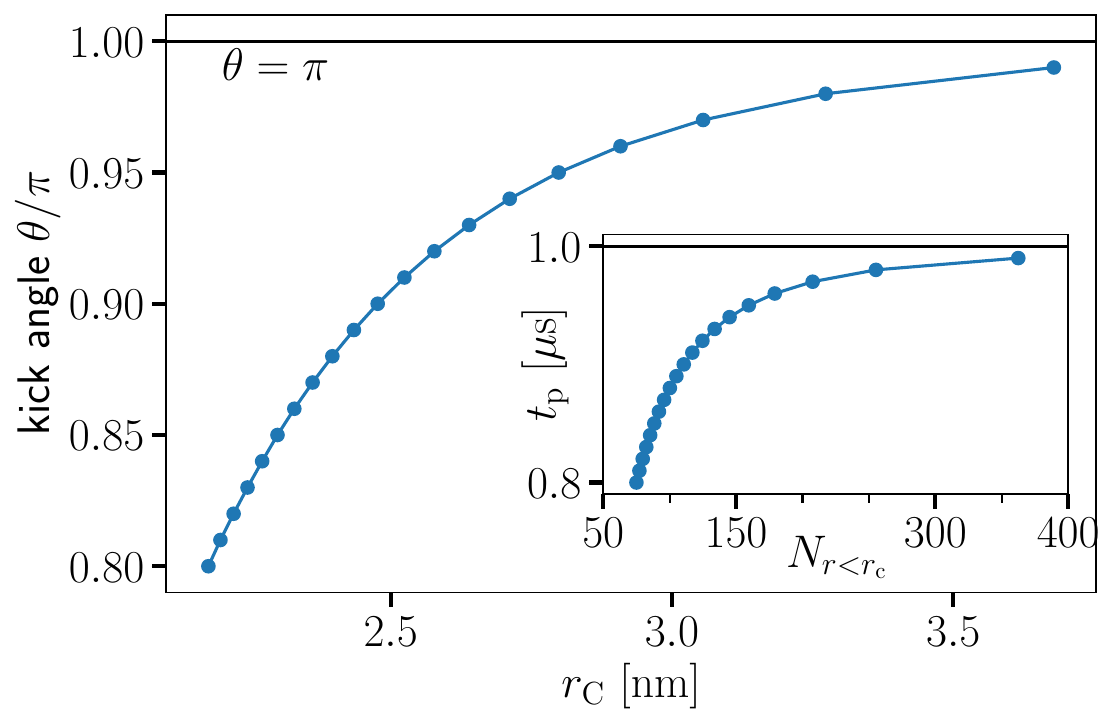}}
    \caption{
    \textbf{Estimation of Crossing Radius:} Estimated crossing radii $r_\mathrm{c}$ as functions of the kick angle $\xt$ for experimental data in~\zfr{fig4}, obtained as described in Sec.~\ref{subsec:crossing_radius}. \textit{Inset:} Estimated number of spins within crossing radius for given kick angle $\xt$.
    \textit{Summary.--} We can use the kick angle $\xt$ to tune the microscopic structure of the polarization profile.
    }
    \label{fig:crossingradius}
\end{figure}

In the previous section we have shown that combining Floquet engineering with ETH-like arguments leads to a (pre-)thermal steady states which has a spatially inhomogenous polarization profile with positive and negative contributions. In particular the polarization profile follows the effective on-site potential $\mathcal{I}_n^\alpha \propto \potential_{\mathrm{eff}, n}^\alpha$, which in turn is determined by the microscopic on-site potential $B_n$ induced by the NV electron and the kick angle $\xt$ via Eq.~\eqref{eq:effective_kickangle}. 
Therefore, by tuning the kick-angle $\xt$ close to $\xt=\pi$ we can engineer states with robust domain-wall like spatial polarization profiles. 
The boundary of the domain wall is given by the crossing radius $r_\mathrm{c}$ which defines the distance from the NV at which the spatial profile has zero polarization
\begin{equation}
\label{eq:rc}
    \potential(r=r_\mathrm{c}) \equiv 0 \,.
\end{equation}
It is directly obtained from setting $\norm{\boldsymbol{\potential}_{\mathrm{eff}}(r)}=0$, i.e. $\delta \theta(r=r_\mathrm{c})=0$. Using Eq.~\eqref{eq:effective_kickangle} we find the implicit equation for the crossing radius
\begin{equation}
\label{eq:rc_imnplicit}
   \tan\left(\frac{\tkick \sqrt{\Omega^2 + \nvfield(r)^2}}{2}\right) \tan\left(\frac{\nvfield(r)\tdd}{2}\right) 
   =  \frac{\sqrt{\Omega^2 + \nvfield(r)^2}}{\nvfield(r)} \, .
\end{equation}
Note that in $3D$ the on-site potential $\nvfield=\nvfield(r,\theta)$ has a radial and angular dependence. However, if the detuning $\delta$ vanishes, which we generally assume to be the case, the angular dependence for the crossing radius simply reduces to a multiplicative factor $r_\mathrm{c}(\theta)=r_{\mathrm{c},0}\times \sqrt[3]{|3\cos^2(\theta) -1 |}$, where $r_\mathrm{c}(\theta)$ is determined from setting $\sqrt[3]{|3\cos^2(\theta) -1 |}=1$.

Using the parameters in the experiment, \zfr{fig4}, i.e. the Rabi frequency $\Omega\approx 50\,\mathrm{kHz}$ and $\tdd\approx 51\,\mu\mathrm{s}$ we can estimate the crossing radius for the different pulse widths $\tkick$, i.e. different kick angles $\xt=\Omega \tkick$, which lead to a spin-polarization inversion. The results are shown in fig.~\ref{fig:crossingradius}. There we also estimated the average number of spins $N_{r<r_\mathrm{c}}$ which are within the crossing radius, using a $\Cs$-spin density of ${{\approx} 1/\mathrm{nm}^3}$. 
For example using $\xt=0.94\,\pi$ we find a crossing radius of $r_\mathrm{c}(\theta)\approx 2.8 \times \sqrt[3]{|3\cos^2(\theta) -1 |}$ encompassing on average around ${\app}150$ spins.
Let us emphasize, that the results above are not limited to finite size systems and are also expected to hold in the thermodynamic limit.

\subsection{\label{subsec:simplifications} Simplifications}

In this section, we repeatedly found that the non-equilibrium dynamics for intermediate times are well-captured by effective Hamiltonians, in particular Eqs.~\eqref{eq:Heff_nopi} and \eqref{eq:Heff_pi}, derived using a high-frequency expansion. Therefore, in the following, we consider only time-evolution generated by these time-independent effective Hamiltonians. The corresponding results are expected to describe the intermediate time dynamics well but ignore any Floquet heating effects. However, Floquet heating will only lead to an onset of slow homogeneous decay of all local observables at times beyond the prethermal plateau.

Moreover, in the following, we will consider simplified versions of the effective Hamiltonians derived above. While the details of Eqs.~\eqref{eq:Heff_nopi} and \eqref{eq:Heff_pi} are needed to derive quantitative results, like the crossing radius, Eq.\eqref{eq:rc}, their key qualitative properties can be described by simplified models. In particular, the key property of Eq.~\eqref{eq:Heff_nopi} is the conservation of a (tilted) global polarization, see discussion around Eq.~\eqref{eq:conserved_axis}. However, the tilt leads to only minor quantitative changes in the exact dynamics so we can neglect the tilt for a qualitative understanding of the dynamics. Analogously, the main properties of Eq.~\eqref{eq:Heff_pi} consist in the absence of a conserved polarization axis and in the presence of a spatially inhomogeneous on-site potential which can take positive and negative values and is not conserved by the Hamiltonian. 

Therefore, in what follows, we will study the simplified effective Hamiltonians
\begin{equation}
\label{eq:Heff_nopi_simple}
    \Heffnopi \approx \sum_{j<k} J_{jk} \left(\frac{3}{2} \left[I_j^z I_k^z + I_j^y I_k^y  \right]- \mathbf{I}_j \cdot\mathbf{I}_k\right),
\end{equation}
for the regime far away from $\vartheta\approx \pi$, and
\begin{equation}
\label{eq:Heff_pi_simple}
   \Heffpi \approx \sum_{j<k} J_{jk} \left(3 I_j^z I_k^z - \mathbf{I}_j \cdot\mathbf{I}_k\right) + \sum \potential_k I_k^x
\end{equation}
for $\vartheta\approx \pi$, where we also dropped the $_\mathrm{eff}$ subscript on the on-site potential and removed all terms not poiting along the $\xhat$-direction. Moreover, in the following we will not use the exact expression for $\potential_k$ given below Eq.~\eqref{eq:Heff_pi}, but rather use different approximations including the various properties of $\potential_k$ without using the complicated exact expression.

In addition, let us emphasize that while the theoretical analysis focuses on a single \textit{cluster} consisting of a single NV center and its surrounding \C\ spins, the experiment is performed on a macroscopic sample that contains many such clusters. However, on the time scales relevant for the experiment these clusters can be considered well separated. Therefore, the experimentally observed data corresponds to an ensemble average of many of such clusters. Note that the microscopic details of each of these clusters might be different, since (i) the random network of \C-atoms around each NV will be different, (ii) the effectiveness of the hyperpolarization scheme and hence the initial state depends on the orientation of the NV center, and (iii) due to the spatial inhomogenities of the external magnetic field the kick angle $\theta$ might vary by a few percent, possibly affecting the location of the zero crossing. 
As we are only interested in quantitatively understanding the experimental observation, we will neglect these ensemble effects throughout.

\section{\label{sec:approximate_dynamics}One-dimensional approximate quantum dynamics}
While the ETH analysis outlined in Sec.~\ref{subsec:ETH} is able to explain some aspects of the experimentally observed signatures, it fails, for instance, to explain the slow transient dynamics.
To fully understand the qualitative physics behind the experimentally observed signatures, we need a more comprehensive theory that takes into account the effects of both diffusion and dissipation. The description of diffusive effects requires large systems which typically exceed those accessible in an exact diagonalization (ED) approach by at least one order of magnitude. To circumvent the system size constraint imposed by ED, several algorithms were introduced in the context of approximate quantum simulations and dynamics \cite{kinder2011TDVP,haegeman2011TDVP,haegeman2016TDVP,leviatan2017TDVP,white2018MPO,richter2019ClusterExpansion,Kvorning2022,rakovszky2022DissipationMPS}. 

In this work, we employ a new algorithm -- the local-information time-evolution (LITE) algorithm -- that was recently developed by some of the authors to deal with the out-of-equilibrium transport of one-dimensional systems~\cite{Kvorning2022,Artiaco2024}. The crucial difference with respect to other algorithms is that LITE preserves all local constants of motion with a support below a specified truncation scale. In the following, we apply LITE to perform a series of numerical simulations using a one-dimensional short-range toy model akin to the true long-range three-dimensional experimental Hamiltonians of the different regimes (i.e., the different effective Hamiltonians found in Sec.~\ref{sec:effective_Hamiltonian}).

In Subsec~\ref{sec:LITE} we give a brief introduction to the LITE algorithm. We then study the two approaches of {Hamiltonian Engineering} and {State Engineering} in great detail in Subsecs.~\ref{sec:one_d_pi} and \ref{sec:one_d_pi_half}, respectively. We close this section by a short comparison between the two cases, Subsec~\ref{subsec:comparison-energy-polarization}, and a brief summary, Subsec~\ref{subsec:one-dimension-summary}.

\subsection{Introduction to the LITE algorithm}
\label{sec:LITE}

Let us briefly outline the basic concepts behind the LITE algorithm. For a detailed introduction to the topic we refer the reader to Refs.~\cite{Kvorning2022,Artiaco2024}. Our goal is to solve the von Neumann equation
\begin{eqnarray}
\label{eq:vonNeumann}
\partial_t \rho = -i \left[\mathcal{H},\rho\right] ,
\end{eqnarray}
for the density matrix $\rho$ under the generic local Hamiltonian $H$. The algorithm uses a decomposition of the full system into smaller subsystems. Each subsystem can be characterised by two indices, $\ell$ and $n$. The integer number $\ell$ denotes the scale or range of the subsystem, i.e., one plus the number of neighboring sites within the subsystem under consideration. The integer or half-integer number $n$ defines the center of the subsystem. For example, the subsystem $\mathcal{C}_n^\ell$ is defined by the $\ell +1$ spins centered around the physical lattice site $n$. By virtue of partial trace operations, one can write the von Neumann equation for the subsystem $\mathcal{C}_n^\ell$ as
\begin{eqnarray}
\label{eq:subsys_vonNeumann}
\partial_t \rho^{\ell}_n & = & -i \left[H^{\ell}_n, \rho_n^{\ell}\right] \\
&& - i \mathrm{Tr}_L\left(  \left[\mathcal{H}_{n-1/2}^{\ell+1}- \mathcal{H}_n^{\ell},\rho_{n-1/2}^{\ell+1} \right] \right) \nonumber \\
&& - i \mathrm{Tr}_R\left(  \left[\mathcal{H}_{n+1/2}^{\ell+1}- \mathcal{H}_n^{\ell},\rho_{n+1/2}^{\ell+1} \right] \right) \nonumber.
\end{eqnarray}
Here, $\rho_n^{\ell}$ ($\mathcal{H}_n^\ell$) denotes the subsystem density matrix (subsystem Hamiltonian) associated with $\mathcal{C}_n^\ell$. Note that $\mathrm{Tr}_L$ and $\mathrm{Tr}_R$ denote partial trace operations over the leftmost ($L$) and the rightmost ($R$) spin, respectively. In Eq.~\eqref{eq:subsys_vonNeumann} we have assumed a nearest neighbor Hamiltonian. To solve Eq.~\eqref{eq:subsys_vonNeumann}, we require knowledge of the density matrices of the subsystems $\mathcal{C}_n^\ell$, $\mathcal{C}_{n+1/2}^{\ell+1}$, and $\mathcal{C}_{n-1/2}^{\ell+1}$. Thus, Eq.~\eqref{eq:subsys_vonNeumann} is not closed, and to solve the equations of motion for the subsystem $\mathcal{C}_{n}^{\ell} $ we need to solve them for higher level subsystems as well. Therefore, the problem is as complex as the one in Eq.~\eqref{eq:vonNeumann}.

The LITE algorithm solves Eq.~\eqref{eq:subsys_vonNeumann} at a subsystem scale $\ell^*$ smaller than the original system size. This is done in a two-step approach: First, if quantum entanglement is only present at small scales $< \ell^*$, thanks to Petz recovery maps \cite{petz1986}, we can exactly recover the density matrices at scales $> \ell^*$ from the density matrices at scale $\ell^*$ and numerically solve Eq.~\eqref{eq:subsys_vonNeumann} for a small time increment $\delta t$. Over time, entanglement spreads and quantum mutual information builds up on increasing scales. Thus, to continue the time-evolution using the above recipe, the scale $\ell^*$, which is the length at which we solve the equations of the kind of \eqref{eq:subsys_vonNeumann}, has to be increased accordingly.

The important second step of the algorithm is activated when $ \ell^* $ has reached a maximum length scale $\ell_\mathrm{max}$ (which is the largest value we are able to handle efficiently, given a fixed amount of computation power and time) and a finite portion $q_\mathrm{max}$ of the total information has accumulated at $\ell_\mathrm{max}$. Then, the algorithm removes mutual information at a truncation length $\ell_{\mathrm{min}}<\ell_\mathrm{max}$ so that time evolution can be continued without further increasing $\ell^*$, and the entanglement blockade can be (approximately) bridged. Importantly, this removal of quantum information has to be done so that the density matrices in the smaller subsystem, as well as the information currents, remain unaffected. In contrast to many other established algorithms (such as those based on the time-dependent variational principle \cite{kinder2011TDVP,haegeman2011TDVP,haegeman2016TDVP,leviatan2017TDVP} or time-evolving block decimation \cite{Vidal2003,Vidal2004,White2004,PAECKEL2019,JASCHKE201859}), all steps involved in the present algorithm preserve all local constants of motion up to scale $\ell_\mathrm{min}$. This makes LITE particularly well suited to investigate hydrodynamics effects. The truncation length $\ell_{\mathrm{min}}$ is the most important parameter of the algorithm, and should be chosen as large as possible (while keeping $\ell_\mathrm{min} < \ell_\mathrm{max}$); $q_\mathrm{max}$, instead, has to be empirically chosen. For the model investigated in the present study, we find $q_\mathrm{max}\sim1\%$ as the optimal value. For a detailed introduction to the algorithm, we refer the reader to Ref.~\cite{Kvorning2022,Artiaco2024}.
 
The LITE algorithm can be straightforwardly generalized to open quantum systems described by the Lindblad master equation,
\begin{eqnarray}
    \label{eq:lindblad}
    \partial_t \rho = -i \left[\mathcal{H},\rho\right] + \sum_j \gamma_j \left(L_j \rho L_j^{\dagger} - \frac{1}{2}\lbrace L_j^{\dagger}L_j, \rho \rbrace \right),
\end{eqnarray}
where $L_j$ ($L_j^{\dagger}$) denote Lindblad jump operators describing the system-environment interaction, and $\gamma_j$ are the respective coupling constants. Assuming on-site jump operators, the corresponding subsystem equation reads as

\begin{eqnarray}
\label{eq:subsys_Lindblad}
\partial_t \rho^{\ell}_n & = & -i \left[H^{\ell}_n, \rho_n^{\ell}\right] +  \sum_{j \in \mathcal{C}_n^\ell} \gamma_j\left(L_j \rho_n^\ell L_j^{\dagger} -\lbrace L_j^{\dagger}L_j, \rho_n^\ell \rbrace \right) \nonumber \\
&& -  i \mathrm{Tr}_L\left(  \left[\mathcal{H}_{n-1/2}^{\ell+1}- \mathcal{H}_n^{\ell},\rho_{n-1/2}^{\ell+1} \right] \right) \nonumber \\
&& -  i \mathrm{Tr}_R\left(  \left[\mathcal{H}_{n+1/2}^{\ell+1}- \mathcal{H}_n^{\ell},\rho_{n+1/2}^{\ell+1} \right] \right) .
\end{eqnarray}

The LITE algorithm removes information that is exclusively found at scales $>\ell_\mathrm{min}$, while keeping all (sub-)subsystem density matrices on smaller scales and corresponding information currents fixed. Thus, at the time of the (first) removal, the state of the system on scales $<\ell_\mathrm{min}$ coincides with the exact state of the system. However, removing information changes the information flow or, in other words, the time dependence of the information currents that flow between different scales. As the information flow might be accelerated or slowed down, this changes the microscopic dynamics at later times. The parameter that controls how much this distorted distribution of information affects expectation values of local observables is $\ell_\mathrm{min}$. 
In the limit $\ell_\mathrm{min}\rightarrow \infty$, the algorithm becomes exact. Therefore, the corresponding scaling in $\ell_\mathrm{min}$ allows to extract asymptotic values of local observables. At (intermediate) times $t\sim \ell_\mathrm{min}/v_{LR}$ (with the Lieb-Robinson speed $v_{LR}$ \cite{LiebRobinson}), the distortion of the information distribution is expected to be the largest and so are the deviations from the exact dynamics. 

At late times, the state of the system on small scales can be well approximated by a local Gibbs state. This is the case also for an infinite system, where the corresponding thermalization time diverges, so a true steady state cannot be reached: The ongoing thermalization process translates into information currents flowing from small to large scales and applying the local Gibbs approximation~\cite{Kvorning2022} for different snapshots in time yields different results. In this regime, a removal of information at scales larger than those present in the Hamiltonian has only minor effects on the dynamics. When the system is finite and equilibrium is reached, no information currents flow between small and large scales and the local Gibbs approximation becomes exact. At this stage also the LITE algorithm becomes exact even at finite $\ell_\mathrm{min}$ (up to errors imprinted into the dynamics at earlier times).

\subsection{Energy diffusion around \texorpdfstring{$\xt \approx \pi$}{TEXT}: Hamiltonian engineering}
\label{sec:one_d_pi}

The three-dimensional long-range Hamiltonian of Eq.~\eqref{eq:Heff_pi_simple} is not suitable to be analyzed with the LITE algorithm since correlations are expected to appear on all length scales at very short times, and there are no Lieb-Robinson bounds restricting the spread of information and entanglement in the system. Furthermore, the three-dimensional nature of the experimental (effective) Hamiltonian poses a challenge. However, on a qualitative level, the relevant physical processes are not exclusively related to three-dimensional long-range systems but can also be observed in much simpler models. To obtain a qualitative picture within the constrains of the LITE algorithm, we condense the problem into an effective numerically tractable one-dimensional short-range Hamiltonian that we subsequently use as a toy model.

\subsubsection{Toy model Hamiltonian}
\label{sec:toy_model_pi}
Let us assume a very sparse density of $\Cs$ spins so that for any $\Cs$ spin there is exactly one other spin with a dominant mutual coupling. In that case, we can reduce the three-dimensional long-range model to an effective one-dimensional short-range nearest-neighbour model keeping only the dominant couplings
\begin{eqnarray}
\label{eq:short_range_model}
    \mathcal{H}_\pi = \sum_k J_k \left(3I^z_k I_{k+1}^z -\mathbf{I}_k \cdot \mathbf{I}_{k+1}\right)+ \potential_k I_k^x.
\end{eqnarray}
While this might appear as a very crude approximation at first sight, it is indeed a valid approximation in related systems (see for example Ref.~\cite{peng2021floquet}). To mimic the random spin position of $\Cs$ atoms in the original model, we use $J_k = J_0 + W_k$ where $W_k \in [-W,W]$ is a uniformly distributed random number.

We emphasize that the goal of this section is not to find a comprehensive quantitative agreement with the experimental results; rather, we aim to understand the fundamental physical processes that can potentially have a qualitative influence on the experiment: While quantitative details of the many-body dynamics might differ drastically between long and short-range systems in low and high dimensions, the qualitative behavior we discuss below applies in all cases provided basic principles, such as ergodicity, hold. 

For the short-range nearest-neighbor model of Eq.~\eqref{eq:short_range_model} we now approximately solve the von Neumann equation using the LITE algorithm introduced in Refs.~\cite{Kvorning2022,Artiaco2024}.

\subsubsection{Initial states}
\label{sec:initial_states_pi}
\begin{figure}
    \centering
   {\includegraphics[width=0.45\textwidth]{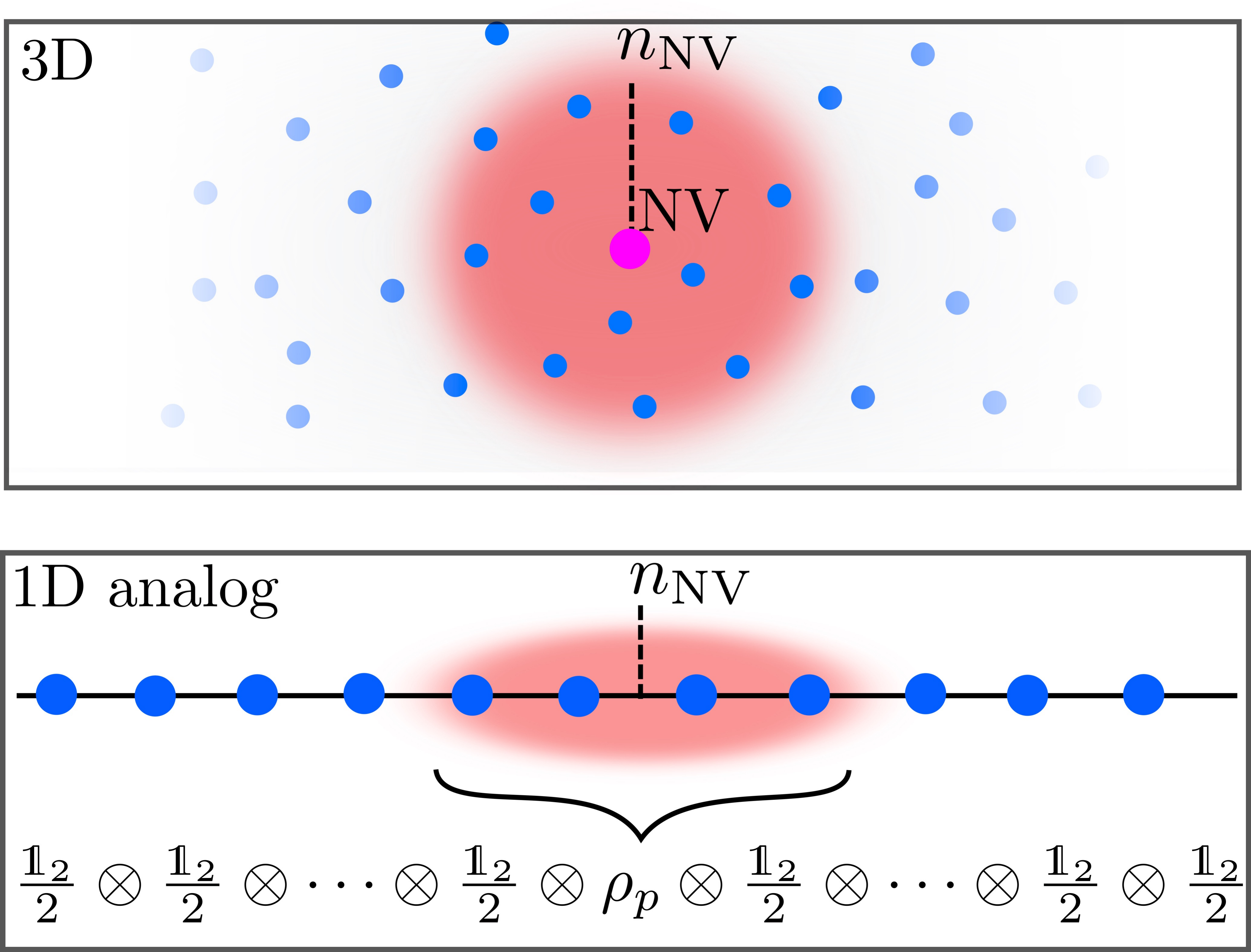}}
    \caption{\textbf{System} -- Schematic of the region of polarization in the experimental (three-dimensional) system and the theoretical (one-dimensional) model: the green ball indicates the NV center and the blue region is polarized after hyperpolarization. The orange dots depict $\Cs$nuclear spins. In the theoretical model with $n_{\mathrm{NV}}$ we denote the center of the polarized region of the initial state. The NV-center is not explicitly part of the model but its action on the $\Cs$ spins is modeled by on-site potentials and dissipation.
    }
    \label{fig:NV_initial_state}
\end{figure}

By virtue of hyperpolarization, the experimentally prepared initial state resembles a mixed product state with a finite polarization of $\Cs$  nuclei in the proximity of a NV center (see fig.~\ref{fig:NV_initial_state}). The absence of long-range correlations in combination with their asymptotic nature make such states well-suited as initial states in the LITE algorithm. In analogy to the experiment we thus choose an initial state of the form 
\begin{eqnarray}
\label{eq:initial_state}
    \rho_{\mathrm{init}} = \frac{\identity_2}{2} \otimes \frac{\identity_2}{2} \otimes \dots\otimes  \frac{\identity_2}{2} \otimes \rho_{\mathrm{p}} \otimes \frac{\identity_2}{2} \otimes\dots \otimes \frac{\identity_2}{2} \otimes \frac{\identity_2}{2},
\end{eqnarray}
where
\begin{eqnarray}
\label{eq:rho_p}
    \rho_\mathrm{p} = \bigotimes_{\substack{ j \in \{\text{region} \\
    \text{of polarization}\}}} \frac{1 }{2} \left(\identity_{2} + p \sigma_j^x \right)
\end{eqnarray}
describes the initially polarized subsystem with a $\xhat$- polarization $p$ for each spin in the region of polarization. $\identity_2$ is an identity matrix of dimension 2. In the experiment, hyperpolarization induces polarization of nuclear $\Cs$ spins in the vicinity of NV-centers while spins far away from NV-centers can be assumed to be in an infinite-temperature mixed state. The initial state in Eq.~\eqref{eq:initial_state} is thus similar to the initial state of the experiment: here the (imaginary) NV-center is located at the center of the polarized region defined by $\rho_\text{p}$ (Eq.~\eqref{eq:rho_p}, see also fig.~\ref{fig:NV_initial_state}).

Note that infinite-temperature density matrices ($\propto \identity$) are invariant under time evolution with any Hamiltonian. Thus, on short timescales all the dynamics generated from time evolving the state of Eq.~\eqref{eq:initial_state} is expected to happen close to the region defined by $\rho_\text{p}$; subsystems far away from $\rho_\text{p}$ remain in an infinite-temperature (time-invariant) state. Thus, initial states of the form of Eq.~\eqref{eq:initial_state} in principle allow us to investigate an (effectively) infinitely extended system. In this case, the (effective) system size is adaptive and updated during time evolution (using the routines in LITE), capturing only those parts which are sufficiently different from the time-invariant infinite-temperature state.

\subsubsection{Energy diffusion}
\label{sec:energy_diff}
Assuming no dissipative processes and disregarding the slow heating due to the periodic drive (which is a good approximation within the prethermal plateau, see Sec.~\ref{sec:effective_Hamiltonian}), the total energy of the (closed) quantum system described by Eq.~\eqref{eq:short_range_model} is (quasi-) conserved, and the system undergoes equilibration. Evidently, an initial state of the form of Eq.~\eqref{eq:initial_state} possesses a spatially dependent distribution of energy density: if the total energy of $\rho_{\mathrm{init}}$ is non-zero with respect to $H$ (which is the case when $\potential_k$ is non-zero in the region of polarization), then the energy density is initially localized in the region of polarization. 

To define a local energy, let us rewrite the Hamiltonian $\mathcal{H}_\pi$ as a sum of local terms
\begin{eqnarray}
\label{eq:H_partition}
    \mathcal{H}_\pi = \sum_k h_{k+1/2} \equiv \sum_n h_n,
\end{eqnarray}
where 
\begin{eqnarray}
    h_{k+1/2} &=&  J_{k} \left(3 I_{k}^zI_{k+1}^z-\mathbf{I}_{k} \cdot \mathbf{I}_{k+1}\right)+  \frac{1}{2}\potential_{k}I_{k}^x + \frac{1}{2}\potential_{k+1}I_{k+1}^x.
\end{eqnarray}
with $k$ being a physical site index. The expectation value of $h_{k+1/2}$ can be interpreted as the energy located at the two sites $k$ and $k+1$, i.e., the energy of the subsystem with center $n=k+1/2$. Thus, in the system under consideration~\eqref{eq:short_range_model}, $h_n \equiv \mathcal{H}_n^1$.

Equation~\eqref{eq:H_partition} allows us to define the variance of the energy distribution as
\begin{eqnarray}
    \sigma_E^2 =  \sum_n (n-\overline{n})^2 \frac{\langle h_{n} \rangle}{\langle H \rangle},
\end{eqnarray}
where $\overline{n}=  \sum_n n \langle h_n \rangle/\langle H \rangle$ can be seen as the spatial expectation value of energy. $\sigma_E^2$ tracks the spread of the local energy distribution and, thus, contains essential information about ongoing equilibration processes. For example, in diffusive systems, the diffusion equation predicts a linear growth of $\sigma_E^2$ with time, $\sigma_E^2 \propto t$ (or correspondingly for the standard deviation $\sigma_E \propto \sqrt{t}$). By contrast, in ballistic systems, $\sigma_E^2$ is expected to grow as $\sigma_E^2 \sim t^2$. Thus, $\sigma_E^2$ might be used to distinguish distinct energy transport regimes in the out-of-equilibrium dynamics of many-body systems \cite{Kvorning2022,rakovszky2022DissipationMPS}.

\begin{figure}
    \centering
   {\includegraphics[width=0.5\textwidth]{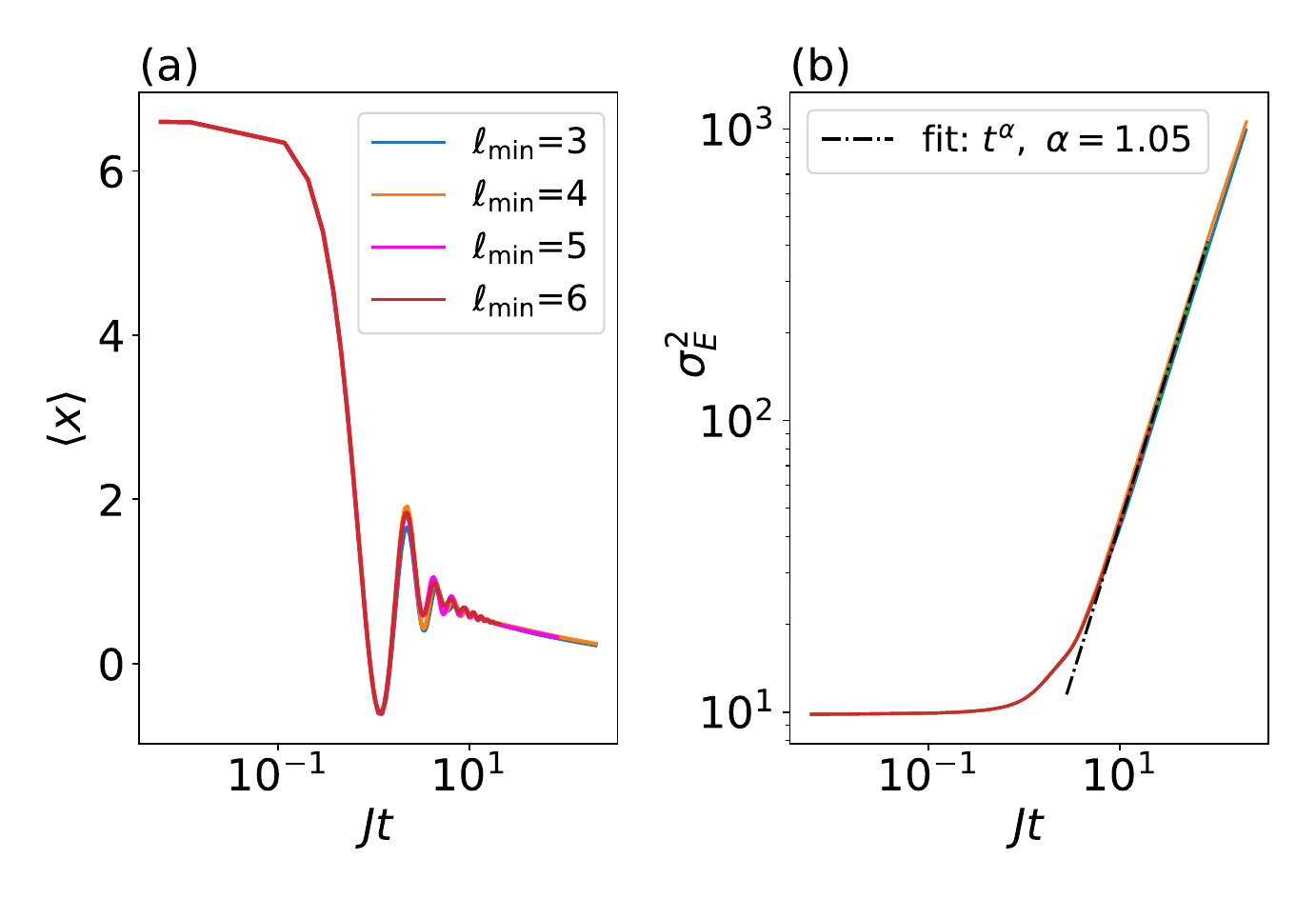}}
    \caption{\textbf{Diffusive energy spread} -- Approximate time-evolution of $\rho_\mathrm{init}$ under $\mathcal{H}_\pi$ where the region of polarization initially extends over $N_p=11$ neighbouring spins with $p=0.6$. (a) Time-evolution curve of the total $\xhat$-polarization $\mathcal{I}_x = \sum_k \sigma_k^x$ for different truncation lengthscales $\ell_\mathrm{min}$. The dashed-dotted line corresponds to the expected steady value at vanishing disorder (i.e., $J_n=J_0$, see Eq.~\eqref{eq:x_local_gibbs}). (b) Time-evolution of $\sigma_E^2$. The dashed-dotted line is a fit to the curve with truncation value $\ell_\mathrm{min}=5$. The system size is effectively infinite (i.e. adaptively increased according to the spread of energy throughout the evolution). Further parameters are $J_0=-0.025$, $W=0.3J_0$ and $\potential_k=0.05\pi$; $\ell_\mathrm{max}=\ell_\mathrm{min}+2$ and $q_\mathrm{max}=0.01$. The characteristic single particle energy is $J=4J_0=-0.1$.
    \textit{Summary.--} Time evolution with respect to $\mathcal{H}_\pi$ does not conserve $\xhat$-polarization but is characterized by energy diffusion.
    } 
    \label{fig:free_diffusion}
\end{figure}

\subsubsection{\label{sec:const_onsite} Constant on-site potential \texorpdfstring{$\potential$}{TEXT}}

In fig.~\ref{fig:free_diffusion} we show the time evolution of the total (or net) $\xhat$-polarization  $\mathcal{I}_x = \sum_k I_k^x $ as well as the corresponding value of $\sigma_E^2$ for the weakly-disordered Hamiltonian of Eq.~\eqref{eq:short_range_model} with constant on-site potential $\potential_k = \potential \neq 0$. The initial state is of the form of Eq.~\eqref{eq:initial_state}. The polarized region at $t=0$ extends over $N_p=11$ spins with a polarization $p=0.6$ per spin. We keep the disorder in $J_k$ small to avoid many-body localization effects~\cite{Abanin2021}, which reflects the regime of the experiment. There are two notable points in fig.~\ref{fig:free_diffusion}: (i) apart from deviations at intermediate times $J\tau \sim 1-10$ the truncation length-scale $\ell_\mathrm{min}$ has little effect on the late time dynamics of short-range observables. In this regime, the total $\xhat$-polarization essentially takes a steady value. (ii) The energy, on the other hand, continues to spread at a rate  $\sigma_E^2  \propto t $, as expected for a diffusive system. The observed scaling also indicates that many-body localization effects are not present in the current parameter regime (related models indeed develop many-body localization in the strongly disordered regime \cite{Abanin2021,Hetterich2018}). Note that whether the energy spread happens ballistically, superdiffusively, or diffusively is not of significant importance for our main result. The primary observation is that the energy density spreads in the system in the first place; the particular scaling at which this happens might influence the corresponding timescales, though. However, particularly in light of the enormous differences to the actual experimental model, we do not expect to quantitatively capture the values measured in the experiment. 

Yet, we can get an idea of the relevant physical mechanisms. In that respect, analyzing the late-time state ($Jt> 10$ in fig.~\ref{fig:free_diffusion}) is very instructive: The initial state in Eq.~\eqref{eq:initial_state} has a finite $\xhat$-polarization and a finite energy $E$ -- since $\potential$ is non-zero in the region of initial polarization. At late times, the variance of the energy distribution grows as $\sim \sqrt{t}$ (see fig.~\ref{fig:free_diffusion}) as expected for diffusive processes. We can thus expect the energy distribution to resemble a Gaussian in this regime:
\begin{eqnarray}
\label{eq:h_n_expect}
     \mathrm{Tr}\left[ h_n \rho(t)\right]\approx  \mathrm{Tr}\left[ \mathcal{H}_\pi \;\rho_\mathrm{init}\right]p_E(n,t),
\end{eqnarray}
where
\begin{eqnarray}
\label{eq:gaussian}
    p_E(n,t)=\frac{1}{\sigma_E(t)\sqrt{2\pi}}\exp\left(  -\frac{(n-n_{\mathrm{NV}})^2}{2\sigma_E^2(t)}\right).
\end{eqnarray}
We explicitly added the time dependence in $\rho(t)$ and $\sigma_E(t)$ to emphasize that the system is \textit{not} in a steady state. In particular, this implies that there is some non-zero information (and information currents) on all lengthscales.

Nevertheless, on small scales (compared to the scales present in $\mathcal{H}_\pi$) we can approximate the state $\rho(t)$ as a $\ell$-local Gibbs state
\begin{eqnarray}
\label{eq:gibbs}
    \rho_n^\ell(t) &=& \mathrm{Tr}_{\overline{\mathcal{C}_n^\ell}}\left[\rho(t)\right] \approx \frac{1}{Z_n^\ell}\exp\left(- \beta_n(t) \mathcal{H}_n^\ell \right) \\
    &=& \frac{\identity_{\mathcal{D}_n^\ell}}{\mathcal{D}_n^\ell} - \frac{\beta_n(t) \mathcal{H}_n^\ell}{\mathcal{D}_n^\ell} + \mathcal{O}(\beta_n^2)\nonumber,~~~
\end{eqnarray}
with the subsystem partition function $Z_n^\ell= \mathrm{Tr}\left(\exp\left(- \beta_n(t) \mathcal{H}_n^\ell \right)\right)$; $\rho_n^\ell(t)$ is the density matrix at time $t$ in the subsystem $\mathcal{C}_n^\ell$, defined by $\ell +1 $ spins centered around $n$. Likewise, $\mathcal{H}_n^\ell$ is the subsystem Hamiltonian (of the full Hamiltonian $\mathcal{H}_\pi$) associated with subsystem $\mathcal{C}_n^\ell$. $\mathrm{Tr}_{\overline{\mathcal{C}_n^\ell}}$ is a partial trace over the complementary subsystem $\overline{\mathcal{C}_n^\ell}$. $\mathcal{D}_n^\ell=2^{\ell+1}$ is the dimension of the subsystem $\mathcal{C}_n^\ell$. 

In the second line of Eq.~\eqref{eq:gibbs}, we assumed $\beta_n(t) \ll 1$ (and $Z_n^\ell \approx \mathcal{D}_n^\ell$). Using Eq.~\eqref{eq:gibbs} in Eq.~\eqref{eq:h_n_expect}, we obtain an expression for the local inverse temperature
\begin{eqnarray}
\label{eq:beta_of_t}
    \beta_n(t) \approx
    \frac{-4}{\frac{3}{2}J_{n}^2+\frac{1}{2}(\potential_{n}^2+\potential_{n+1}^2)} 
    \mathrm{Tr}\left[ \mathcal{H}_\pi \rho_\mathrm{init}\right]\;
    p_E(n,t),~~~
\end{eqnarray}
where, in view of subsequent discussions, we keep the $n$-index despite $\potential$ being assumed to be constant here.

Equation~\eqref{eq:gibbs} produces the (almost) exact expectation values for local observables on scales less than, or at most equal to, those present in the Hamiltonian itself. However, the subsystem density matrix $\rho_n^\ell(t)$ from Eq.~\eqref{eq:gibbs} does not reproduce the same dynamics observed at time $t$: if we stop the dynamics at time $t$, replace the actual state of the system by the snapshot approximation of Eq.~\eqref{eq:gibbs} at time $t$, and restart the evolution with the exchanged state, then the subsequent dynamics differs from the one created by the actual state \cite{fleckenstein2021thermalization,fleckenstein2021prethermalization}. The differences arise as the local Gibbs state of Eq.~\eqref{eq:gibbs} is a maximum-entropy state with just enough information to reproduce the correct expectation values of local constants of the motion and information decays rapidly on larger scales. In the actual time-evolved state at time $t$ this is not the case (i.e., the state is not a maximum-entropy state at time $t$). Consequently, the local Gibbs state is incapable of capturing correct expectation values on scales larger than those present in the Hamiltonian.

Note that this analysis is subtly different from the ETH discussion employed in Sec.~\ref{subsec:ETH}, which is based on equilibrium arguments. In contrast, here the system is inherently in a non-steady state as energy continues to diffuse. Nonetheless, expectation values of local observables can yield steady values, as we now show. Applying the local Gibbs approximation, the total $\xhat$-polarization is given by
\begin{eqnarray}
\label{eq:x_of_t_gibbs}
    \mathcal{I}_x(t) &=& \mathrm{Tr}\left[\left(\sum_n I_n^x\right)\rho(t)\right] \approx \sum_n -\beta_n(t)\frac{\mathrm{Tr}\left[I_n^x \mathcal{H}_n^\ell\right] }{\mathcal{D}_n^\ell} \\
    &=& \mathrm{Tr}\left[ \mathcal{H}_\pi \rho_\mathrm{init}\right]\sum_n \frac{\potential_n}{\frac{3}{2}J_n^2+\frac{1}{2}(\potential_{n}^2+\potential_{n+1}^2)} p_E(n,t).\nonumber
\end{eqnarray}
For vanishing disorder $J_n \rightarrow J_0$ and a constant on-site potential $\potential_n{\rightarrow}\potential $, $\mathcal{I}_x$ becomes steady:
\begin{eqnarray}
\label{eq:x_local_gibbs}
        \mathcal{I}_x 
    = \frac{\potential}{\frac{3}{2}J_0^2+\potential^2} \left(N_p p\potential -\frac{J}{2}p^2(N_p-1)\right)
\end{eqnarray}
where we used $ \mathrm{Tr}\left[  \mathcal{H}_\pi\rho_{\mathrm{init}}\right] = \left(N_p p\potential -\frac{J}{2}p^2(N_p-1)\right)/2 $ with $N_p$ the number of initially polarized spins. The steady value of $\mathcal{I}_x $ is indicated by the dashed-dotted line in fig.~\ref{fig:free_diffusion} (a) for $J_n=J_0$. Note that the data shown in fig.~\ref{fig:free_diffusion} are obtained using disordered $J_n$, hence the small deviations. 

In the (three-dimensional, long-range) experimental system spin-spin interaction terms average to zero and yield no contribution to the initial state energy (see Sec.~\ref{sec:closed_system}). In contrast, in our one-dimensional short-range toy model, we obtain contributions from the spin-spin interaction terms to the initial energy $\propto Jp^2$. Thus, in the following, we consider the limit $p\rightarrow 0$ and keep only terms up to linear order in $p$.

\subsubsection{\label{sec:space_dep_phi}Space-dependent on-site potential \texorpdfstring{$\potential_n$}{TEXT}}

\begin{figure}
    \centering
   {\includegraphics[width=0.5\textwidth]{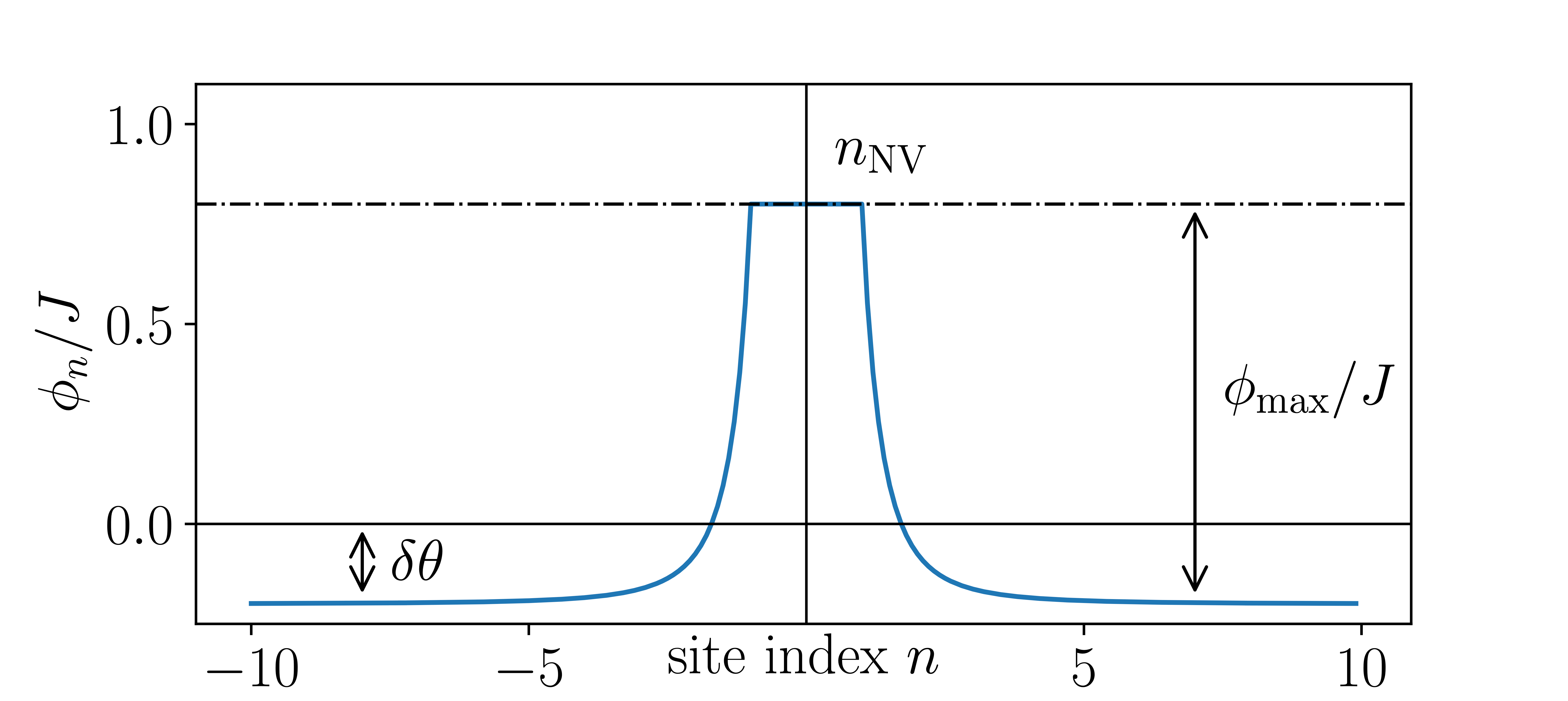}}
    \caption{
    \textbf{Schematic of $\potential_n$} -- The pure $1/(\vert n-n_{\mathrm{NV}}\vert)^3$ dependence is shifted by $\delta\vartheta$ and truncated at $\potential_\mathrm{max}-\delta\vartheta$ as $n \rightarrow n_{\mathrm{NV}}$. The distances $n-n_{\mathrm{NV}}$ are measured in units of a lattice constant $a$, which (unless stated explicitly different) is set to $1/\sqrt[3]{\pi}$.
    }
    \label{fig:phi_of_r}
\end{figure}

\begin{figure}
    \centering
 {\includegraphics[width=0.5\textwidth]{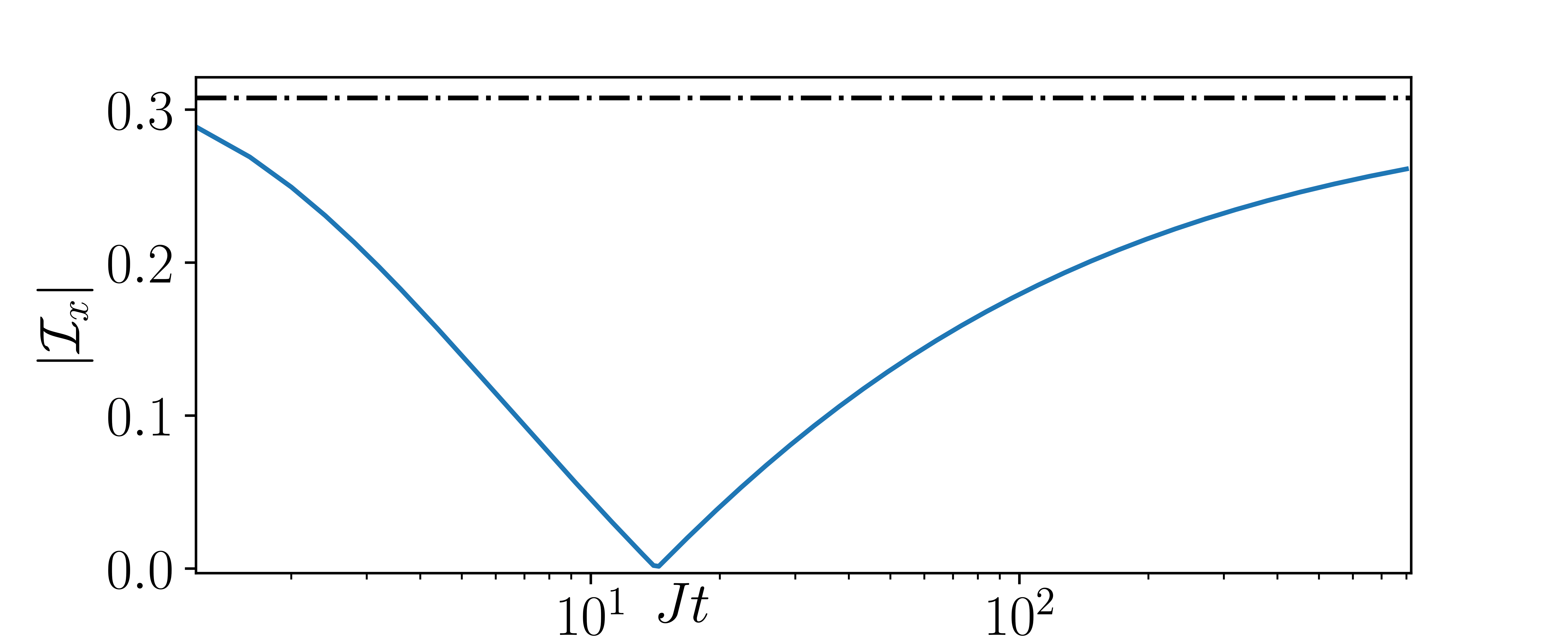}}
    \caption{\textbf{Analytic expectation value of the total net polarization} -- The expectation value of the total net polarization (solid blue) is evaluated using Eq.~\eqref{eq:x_of_t_gibbs}  together with Eq.~\eqref{eq:phi_n} and the (norm of the) asymptotic value (dashed-dotted black) using Eq.~\eqref{eq:time_asymptotic_value} with $\mathrm{Tr}\left[\mathcal{H}\rho_\mathrm{init}\right]=1/2 ~N_p p~ \delta\vartheta$. We used $\delta\vartheta=0.05\pi$ and $D=1$; the parameters ($N_p$, $p$, $J$) are the same as in fig.~\ref{fig:free_diffusion}.
    \textit{Summary.--} Using the local Gibbs approximation and the spatially inhomogenous potential gradient, we can analytically reproduce the (global) polarization inversion.
    }
    \label{fig:x_analytical}
\end{figure}

To analyze a site-dependent on-site potential $\potential_n$ we can immediately adapt almost all of the above analysis. In fact, apart from the last step, the derivation of the previous section is completely general. As outlined in Sec.~\ref{sec:effective_Hamiltonian}, the remaining effective on-site potential is expected to approximately scale as 
\begin{eqnarray}
    \potential_n \propto   \frac{1}{\vert n - n_{\mathrm{NV}} \vert^3} - \delta\vartheta,
\end{eqnarray}
where $n_{\mathrm{NV}}$ is the location of the NV-center. To emphasize and condense the most relevant physical implications of spatially dependent on-site fields with a constant asymptotic value, we use a simplified spatial dependence for the on-site field. To avoid dealing with divergences, we truncate the potential at $\potential_\mathrm{max}-\delta\vartheta$ as $n\rightarrow n_\mathrm{NV}$ (see fig.~\ref{fig:phi_of_r}):
\begin{eqnarray}
\label{eq:phi_n}
    J^{-1} \potential_n  = \begin{cases}
     \frac{1}{\vert n - n_{\mathrm{NV}} \vert^3} - \delta\vartheta, & \text{if $ \frac{1}{\vert n - n_{\mathrm{NV}} \vert^3} < J^{-1}\potential_\mathrm{max}$}.\\
    J^{-1}\potential_\mathrm{max}, & \text{otherwise}.
  \end{cases}
\end{eqnarray}
Plugging $\potential_n$ in Eq.~\eqref{eq:x_of_t_gibbs} and assuming $\sigma_E = \sqrt{D t}$, we obtain the absolute value of the total net polarization over time in the presence of an on-site potential of the form in Eq.~\eqref{eq:phi_n}. Figure~\ref{fig:x_analytical} displays the obtained time evolution curve for $D=1$. $\delta\vartheta$ is chosen positive so that far away from $n_{\mathrm{NV}}$, we obtain $\lim_{\vert n-n_{\mathrm{NV}}\vert \rightarrow \infty} \potential_n/J = -\delta\vartheta$. Initially, the energy density is localized in the regime where $\potential_n>0$; hence, we find a positive total net polarization. As time progresses an increasing amount of energy diffuses into the regime where $\potential_n<0$, leading to a decrease and, eventually, a sign inversion of the total net polarization. The timescale of the crossing depends sensitively on the initial state and the diffusion constant $D$. Notably, at late times when almost all energy is located in the region with $\potential_n<0$, the total net polarization reaches a steady value given by
\begin{eqnarray}
\label{eq:time_asymptotic_value}
       \lim_{t\rightarrow \infty} \mathcal{I}_x (t)
    = -\frac{\delta\vartheta}{\frac{3}{2}J+J \delta \vartheta^2}\mathrm{Tr}\left[\mathcal{H}_\pi\rho_\mathrm{init}\right].
\end{eqnarray}
For large values of the offset $\delta\vartheta$, the late-time steady value scales as $\sim 1/\delta\vartheta$. 

\begin{figure}
    \centering
   {\includegraphics[width=0.5\textwidth]{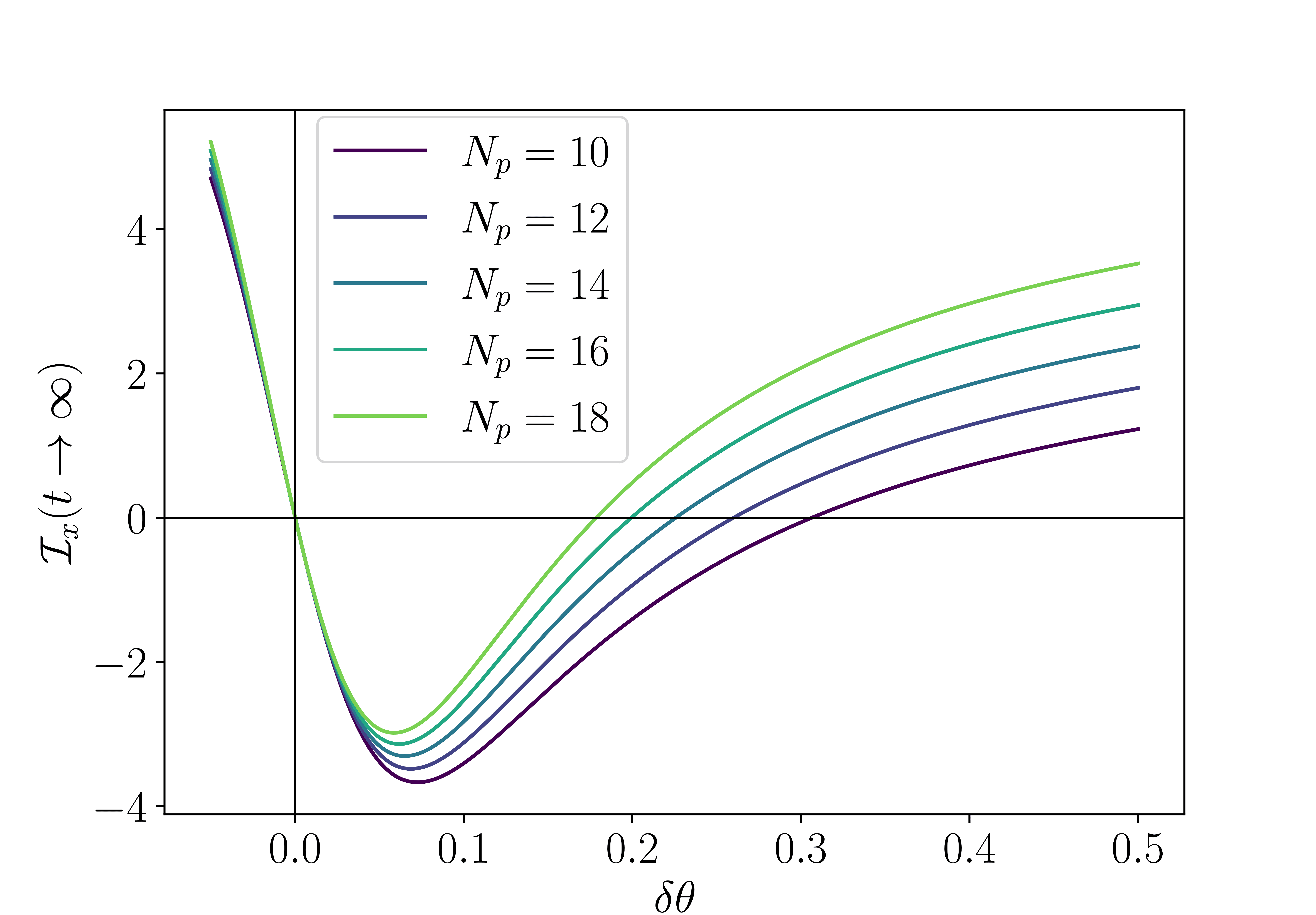}}
    \caption{\textbf{Steady value of the total $\xhat$-polarization} -- Analytic late time (steady) expectation value of the total $\xhat$-polarization with different initial states characterised by different values of $N_p$. Further parameters are as in fig.~\ref{fig:free_diffusion}. 
    \textit{Summary.--} Depending on the initial state, the sign inversion of the total $\xhat$-polarization only occurs for a finite interval of $\delta\vartheta$-values, see also fig.~\ref{fig:delta_phi_dep}. 
    }
    \label{fig:late_time_steady}
\end{figure}

Evidently, the steady value of the polarization is not only controlled by $\delta\vartheta$ but also by the total initial energy. Assuming a fixed initial polarization per spin $p$ as well as a fixed region of polarization, and given Eq.~\eqref{eq:phi_n}, the energy of the initial state is given by
\begin{eqnarray}
\label{eq:intial_energy_n}
    \mathrm{Tr}\left[\mathcal{H}_\pi\rho_\mathrm{init}\right] &=& p\!\sum_{ \substack{\mathrm{region~of}\\ \mathrm{polarization}}} \left(\frac{1}{\vert n - n_{\mathrm{NV}} \vert^3} - \delta\vartheta \right)  +\mathcal{O}(p^2)\nonumber\\
    &=&\frac{p}{2}( c - N_p\delta\vartheta)+\mathcal{O}(p^2),~~
\end{eqnarray}
where $c = \sum_{ \substack{\mathrm{region~of}\\ \mathrm{polarization}}} \frac{1}{\vert n - n_{\mathrm{NV}} \vert^3}$ is a positive constant, and $N_p$ is the total number of spins in the region of polarization. The late-time steady value thus has two zeros, cf.~Eq.~\eqref{eq:time_asymptotic_value}: $\delta\vartheta=0$ and $\delta\vartheta = c/N_p$. Clearly, the second zero depends strongly on the polarization profile in the initial state. The two zeros constrain the regime where we expect to find a diffusion-induced sign inversion (see fig.~\ref{fig:late_time_steady}). As $N_p$ increases, the $\delta\vartheta$ window, where the sign inversion can be observed, shrinks. 

In the experiment, $N_p$ is related to the hyperpolarization time: longer hyperpolarization times imply that more $\Cs$ spins further away from the NV-center are partially polarized. and the constant polarization $p$ becomes site dependent $p \rightarrow p(n)$.

\subsubsection{Polarization gradients}
\label{sec:mag_grad}

A very interesting aspect emerging from the above analysis is that the local polarization of a given spin at position $n$ (to leading order in $\potential_n$) is given by
\begin{eqnarray}
\label{eq:local_mag_0}
    \mathcal{I}_n^x(t) = \frac{\potential_n}{\frac{3}{2}J_n^2+\frac{1}{2}(\potential_{n}^2+\potential_{n+1}^2)}  \mathrm{Tr}\left[ \mathcal{H}_\pi \rho_\mathrm{init}\right] p_E(n,t)
\end{eqnarray}
In particular, in the limit of a small on-site potential $\potential_n$ we obtain a linear dependence 
\begin{eqnarray}
\label{eq:local_mag}
    \mathcal{I}_n^x (t)= \frac{2\potential_n}{3J_n^2+ \left(a\frac{\partial \potential_n}{\partial n}\right)^2} p_E(n,t)\mathrm{Tr}\left[ \mathcal{H}_\pi \rho_\mathrm{init}\right]+ \mathcal{O}(\potential_n^2),
\end{eqnarray}
where we used $\potential_{n+1}=\potential_n + a\frac{\partial \potential_n}{\partial n} +\mathcal{O}(a^2)$ with the lattice constant $a$. The local polarization at site $n$ is expected to be proportional to the on-site potential $\potential_n$. The latter is induced by the nearby presence of the NV center and is expected to vary on a length scale of nanometers which in turn implies a polarization gradient imposed on a nanometer lengthscale. 

\begin{figure}
    \centering
   {\includegraphics[width=0.55\textwidth]{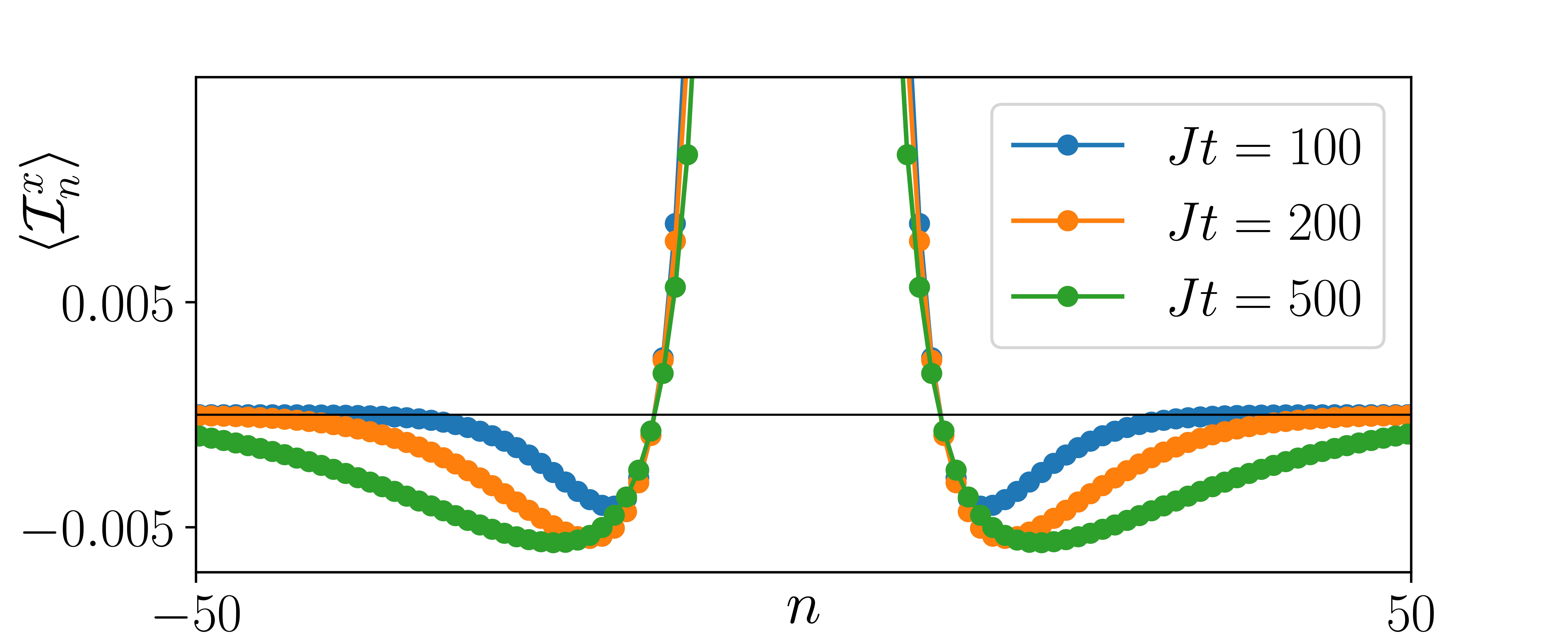}}
    \caption{\textbf{Spatially resolved $\xhat$-polarization } -- Local polarization according to Eq.~\eqref{eq:local_mag} for $\potential_n$ given by Eq.~\eqref{eq:phi_n} with lattice constant $a=0.2/\sqrt[3]{\pi}$. As  initial energy we use $\mathrm{Tr}\left[ \mathcal{H} \rho_\mathrm{init}\right]=1$. We assume $p_E(n,t)$ to be Gaussian with $\sigma_E= \sqrt{Dt}$ and $D=1$, $\potential_\mathrm{max}=\pi$ and $\delta\vartheta = 0.05 \pi$. Further parameters are as in fig.~\ref{fig:free_diffusion}. 
    \textit{Summary.--} With increasing time spins further away from the NV are getting more negatively polarized. However, the spatial zero crossing is fixed at $\potential_n=0$ for all times.
    }
    \label{fig:local_mag}
\end{figure}

In fig.~\ref{fig:local_mag} we show the distribution of the polarization for three different snapshots in time. 
The zero-crossing points (in space) that separate the positively polarized from the negatively polarized regime are found from Eq.~\eqref{eq:phi_n} setting $\potential_n=0$.
Note that this crossing point is time-independent and only determined by the applied on-site potential. As time progresses, spins further away from $n_{\mathrm{NV}}$ develop negative polarization, while simultaneously spins close to $n_{\mathrm{NV}}$ lose polarization (recall that the total net polarization is not conserved; thus these processes can happen in unequal proportions). Depending on the energy of the initial state (and more precisely, on the sign of the effective temperature in the prethermal plateau), a total negative polarization can develop over time and serve as a non-fine-tuned indicator for the presence of a nanoscale polarization gradient.

\subsubsection{Energy diffusion in inhomogeneous systems}

Let us first assume a constant $\potential_n= \potential$. As discussed in Sec.~\ref{sec:const_onsite}, in this case energy spreads diffusively in the system with a spatially uniform diffusion constant $D\neq 0$.
In the limit $\potential\rightarrow \infty$ the Hamiltonian $\mathcal{H}_\pi$ reduces to a single-particle problem with vanishing energy diffusion, $D \rightarrow 0$. For increasing values of $\potential$, we thus expect a decrease of $D$.

Let us now introduce the spatial dependence of $\potential_n$ given in Eq.~\eqref{eq:phi_n}. Far away from $n_{\mathrm{NV}}$ we expect to find a (spatially constant) diffusion constant $D_{\mathrm{asympt}}$. Likewise, close to $n_{\mathrm{NV}}$ (assuming the cut-off depicted in fig.~\ref{fig:phi_of_r}) we have $D_{n_{\mathrm{NV}}}$. Since the asymptotic value $\delta\vartheta\ll \potential_{n_{\mathrm{NV}}}/J$, we expect $D_{n_{\mathrm{NV}}}<D_{\mathrm{asympt}}$. The necessary interpolation between these two values implies a spatially dependent diffusion constant. Such a diffusion problem can be recast in a Fokker-Planck equation, which has no flat steady-state solution as $D\rightarrow D(n)$.

\begin{figure}
    \centering
  {\includegraphics[width=0.5\textwidth]{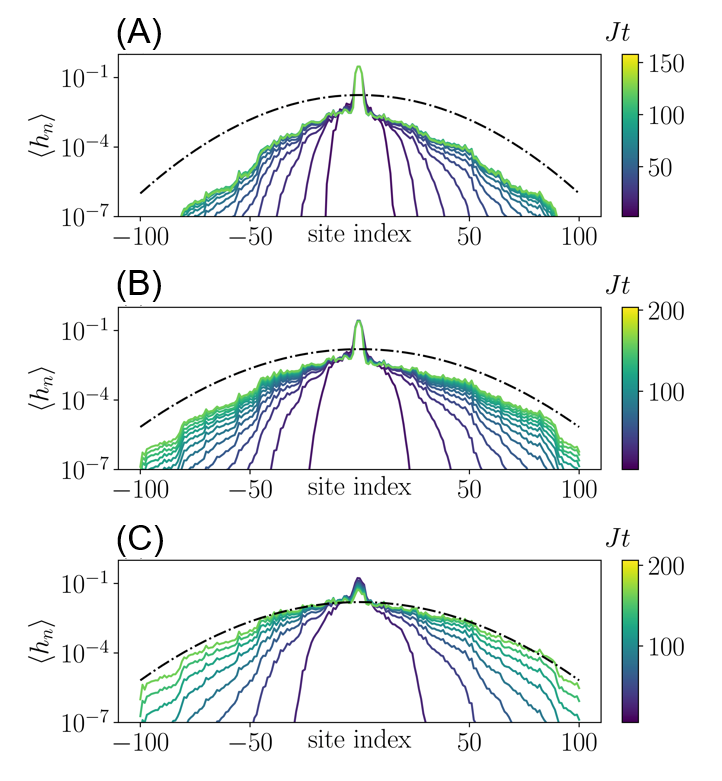}}
    \caption{\textbf{Time dependence of the energy distribution} -- Normalized distribution of energy when evolving a state of the form of Eq.~\eqref{eq:initial_state} under the Hamiltonian of Eq.~\eqref{eq:short_range_model} with the on-site potential schematically shown in fig.~\ref{fig:phi_of_r}, using $\delta\vartheta=0.04 \pi$ and three different truncation values (a) $\potential_\mathrm{max}=1.0\pi$, (b) $\potential_\mathrm{max}=0.5\pi$, and (c) $\potential_\mathrm{max}=0.25\pi$. The black dotted-dashed line corresponds to the Gaussian distribution of Eq.~\eqref{eq:gaussian} with $\sigma_E^2 = Dt$ where $D=1.35$. All remaining parameters are the same as inf fig.~\ref{fig:free_diffusion}.
    \textit{Summary.--} A strong spatial dependence of the on-site potential $\potential_n$ leads to a significant slowing down of energy transport. 
    }
    \label{fig:energy_distribution}
\end{figure}

Figure~\ref{fig:energy_distribution} shows the energy distribution as a function of time for three different truncation values of $\potential_\mathrm{max}$ at fixed $\delta \vartheta$. For $\potential_\mathrm{max}/\vert J_0\vert \gg 1$ (panel (a)), the distribution of energy is significantly distorted compared to the Gaussian distribution of Eq.~\eqref{eq:gaussian}. In particular, the spread of energy slows down significantly with increasing time.
When lowering $\potential_\mathrm{max}$, this effect is reduced and the spread of energy continues even at late times (panels (b) and (c)). Since our simulations can only reach finite times, we cannot rule out the possibility that the spread of energy slows down in (b) and (c) as well, and eventually comes to a halt with a space-dependent steady state. The slowing down observed in the closed one-dimensional short-range toy model resembles similar effects discussed in recent papers on Stark many-body localization \cite{Schulz2019,van2019bloch,Morong2021}. As shown in the next section, such effect is mitigated by virtue of dissipation.

Finally, let us notice that in the experimental three-dimensional long-range system, the (here) observed slowdown of diffusion is expected to be less pronounced as the relative importance of the pinning potential is drastically reduced. In that case, we expect that energy spreads diffusively and the late time steady value analysis of Sec.~\ref{sec:space_dep_phi} remains valid.

\subsubsection{Dissipation}
\label{sec:dissipation}

\begin{figure}[t!]
    \centering
   {\includegraphics[width=0.5\textwidth]{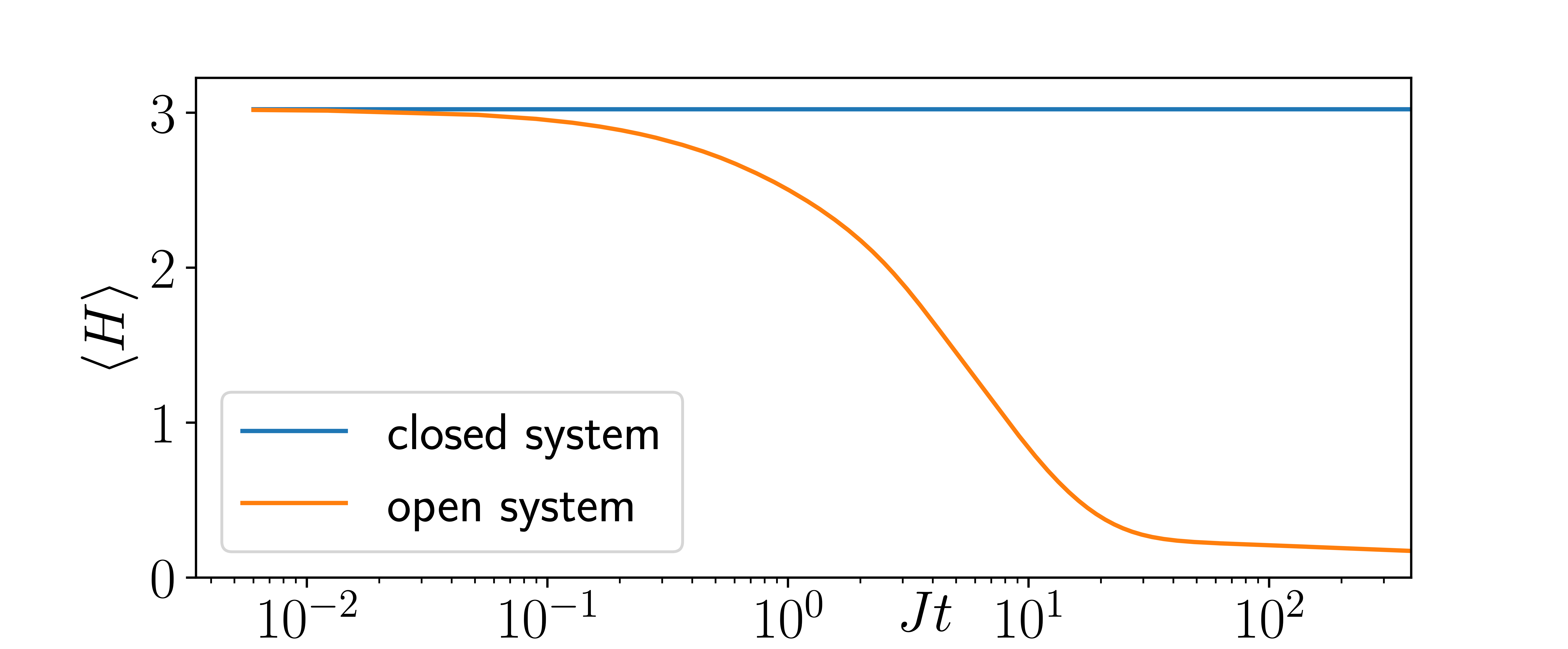}}
    \caption{\textbf{Time evolution of the total energy} --  Total energy as a function of time for the closed and open quantum dynamics with on-site dissipation at $n=n_{\mathrm{NV}}$ with coupling constants $\gamma_+=\gamma_-=\gamma_z=\vert J_0\vert$. Further parameters are the same as in fig.~\ref{fig:energy_distribution} (a), fig.~\ref{fig:free_diffusion}, respectively. 
    \textit{Summary.--} Coupling of the NV center to a phonon-bath acts as an energy sink on the spin system leading to energy leakage, in contrast to the energy-conserving close system dynamics.
    }
    \label{fig:energy_loss}
\end{figure}

In Section~\ref{sec:SI-open-system}, we discussed the impact of decoherence and dissipation on the $\Cs$ spins, which are influenced by the phonon bath generated by the diamond lattice and mediated by the NV center. Under the singular coupling limit, we derived the explicit form of the Lindblad master equation, which exhibits two crucial characteristics for our subsequent analysis. First, the Lindblad jump operators induce both dephasing and dissipation on the $\Cs$ spins when they are polarized along the $\xhat$ axis. Second, the coupling strength of the Lindblad jump operators decays more rapidly with distance from the NV center (${\sim}1/r^6$) compared to the dipole interaction (${\sim}1/r^3$). These properties enable us to simplify the effect of dissipation in our one-dimensional toy model used for the numerical simulations. Specifically, we only apply local jump operators for $n=n_{\mathrm{NV}}$, which represents the spin closest to the hypothetical NV center in our simulations. Moreover, in Eq.~\eqref{eq:lindblad} we employ the three jump operators 
\begin{eqnarray}
\label{eq:jump_ops}
 L_+ = \frac{1}{2}\left(\sigma_{n_{\mathrm{NV}}}^x + i \sigma_{n_{\mathrm{NV}}}^y\right), ~~ L_- = \frac{1}{2}\left(\sigma_{n_{\mathrm{NV}}}^x - i \sigma_{n_{\mathrm{NV}}}^y\right),~~ L_z = \sigma_{n_{\mathrm{NV}}}^z,
\end{eqnarray}
where $L_+$ and $L_-$ generate dissipation, while $L_z$ is responsible of dephasing. Note that in an open quantum system, energy is not a globally conserved quantity: the Lindblad jump operators in Eq.~\eqref{eq:jump_ops} with coupling constants $\gamma_+=\gamma_-=\gamma_z$, as implemented in the numerical simulations, act as an energy sink (see fig.~\ref{fig:energy_loss}).

\begin{figure}[t!]
    \centering
   {\includegraphics[width=0.5\textwidth]{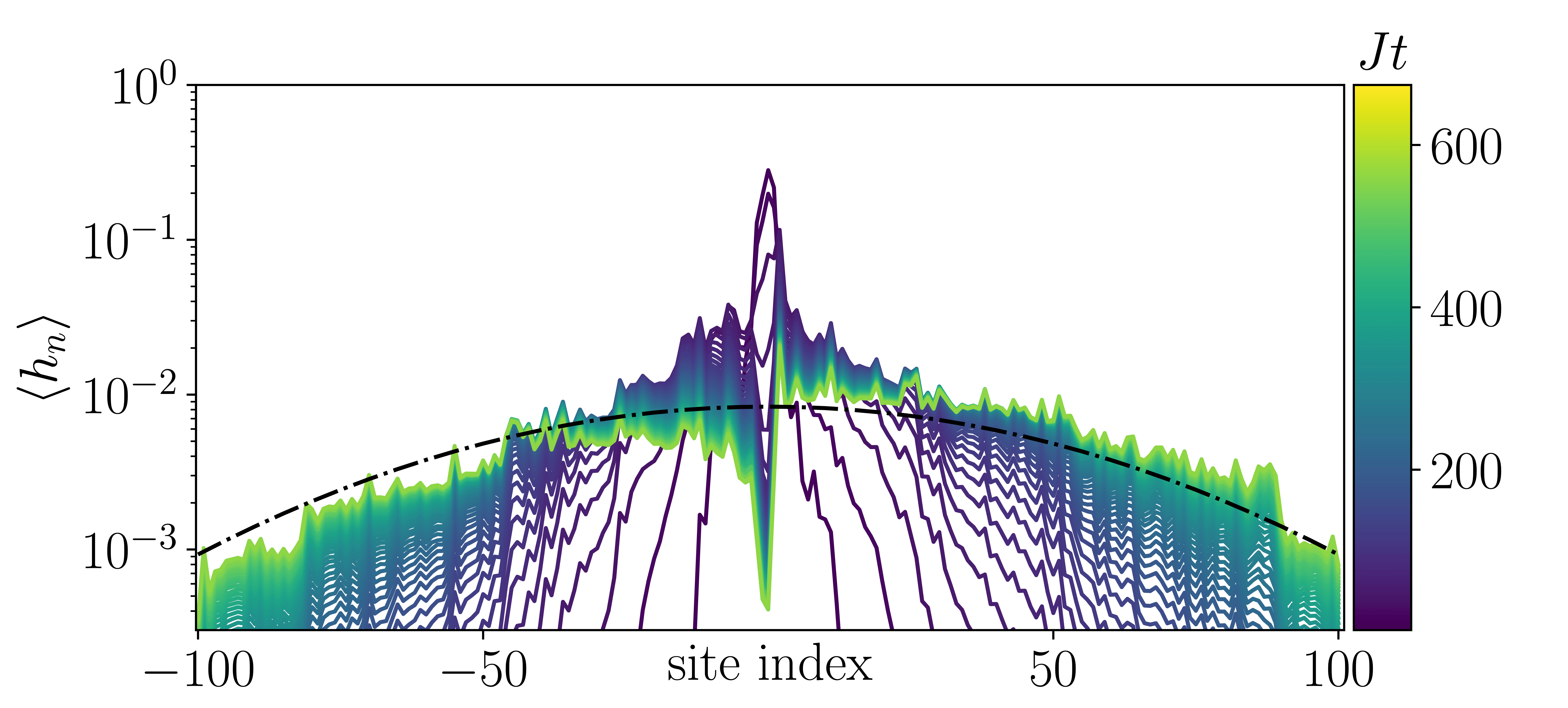}}
    \caption{\textbf{Time dependence of the energy distribution} -- Same plots as fig.~\ref{fig:energy_distribution} (a) with additional on-site dissipation at $n=n_{\mathrm{NV}}$ of coupling strength $\gamma_+=\gamma_-=\gamma_z=\vert J_0\vert$. Further parameters are the same as in fig.~\ref{fig:energy_distribution} (a), fig.~\ref{fig:free_diffusion}, respectively.
    \textit{Summary.--} In contrast to fig.~\ref{fig:energy_distribution} including dissipation leads to a Gaussian profile of the energy distribution at late times regardless of the cutoff $\potential_\mathrm{max}$.
    }
    \label{fig:energy_distribution_dissipation}
\end{figure}

Solving the respective subsystem Lindblad equation~\eqref{eq:lindblad} with the jump operators of Eq.~\eqref{eq:jump_ops} and the on-site potential sketched in fig.~\ref{fig:phi_of_r}, yields the time-dependent energy distribution shown in fig.~\ref{fig:energy_distribution_dissipation}. 

Apart from the local jump operators at $n=n_{\mathrm{NV}}$, in fig.~\ref{fig:energy_distribution} (a) and fig.~\ref{fig:energy_distribution_dissipation} we employ identical simulation parameters. Contrasting both figures, it is evident that dissipation changes the energy distribution, especially at late times: instead of having the bulk of the energy density accumulated around $n=n_{\mathrm{NV}}$ as in fig.~\ref{fig:energy_distribution} (a), the distribution in fig.~\ref{fig:energy_distribution_dissipation} resembles a Gaussian distribution with standard deviation $\sigma_E \sim \sqrt{Dt}$.

\begin{figure}[t!]
    \centering
   {\includegraphics[width=0.5\textwidth]{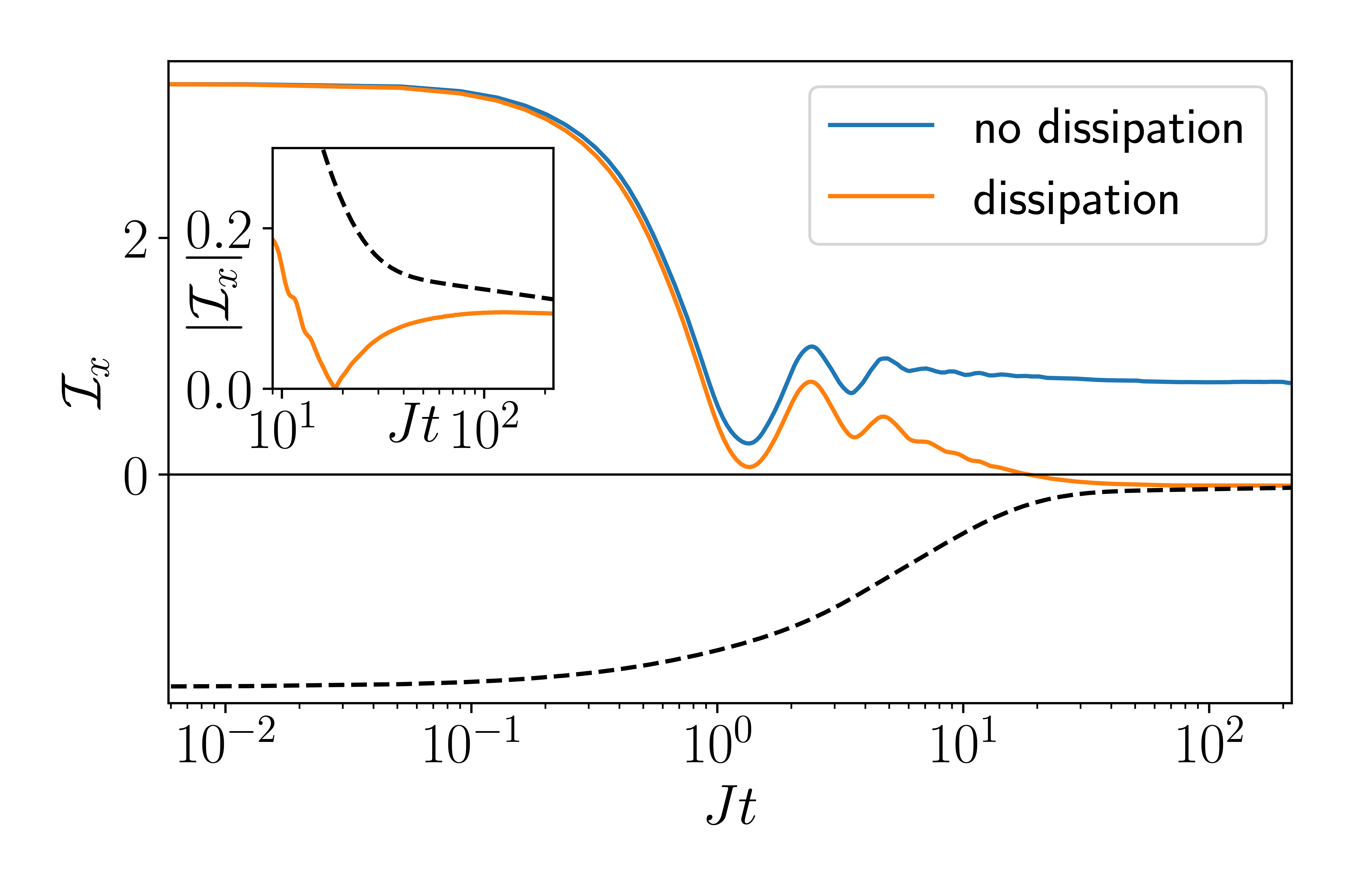}}
    \caption{\textbf{Time-evolution of the total net polarization} -- Time-evolution of the total net polarization with (orange) and without (blue) on-site dissipation at $n=n_{\mathrm{NV}}$. The black dashed line indicates the expected late time value in the presence of dissipation (at $n=n_{\mathrm{NV}}$ of coupling strength $\gamma_+=\gamma_-=\gamma_z=\vert J_0\vert$.) The inset shows a zoom-in. Further parameters are as in fig.~\ref{fig:energy_distribution} (a), fig.~\ref{fig:free_diffusion}, respectively.
    \textit{Summary.--} Additional dissipation can lead to sign inversion of the total $\xhat$-polarization as positive polarization close to the NV is absorbed.
    }   
    \label{fig:diffusion_dissipation_01}
\end{figure}

In fig.~\ref{fig:diffusion_dissipation_01} we show the time-evolution curves corresponding to fig.~\ref{fig:energy_distribution} (a) and fig.~\ref{fig:energy_distribution_dissipation}: notice that, while there is no sign-inversion of the total net polarization in case of a closed quantum system (where the majority of energy density remains localized around $n=n_{\mathrm{NV}}$), the open quantum system develops negative polarization whose value asymptotically approaches the expected late time steady-state (dashed black line in fig.~\ref{fig:diffusion_dissipation_01}):
\begin{eqnarray}
\label{eq:late_time_x_dissipation}
    \mathcal{I}_x (Jt \gg 1) =  -\frac{ J\delta\vartheta}{\frac{3}{2}J^2+ J^2\delta \vartheta^2}\mathrm{Tr}\left[\mathcal{H}_\pi\rho(t)\right].
\end{eqnarray}
Equation~\eqref{eq:late_time_x_dissipation} corresponds to Eq.~\eqref{eq:time_asymptotic_value} weighted with the fraction of energy left in the system at time $t$; this is brought in by using the time-dependent density matrix $\rho(t)$. The latter is justified whenever the diffusion timescales are much faster than the dissipation timescales, so that at every fixed time $t$ the system  can still be approximated by a local Gibbs state.

\begin{figure}
    \centering
   {\includegraphics[width=0.5\textwidth]{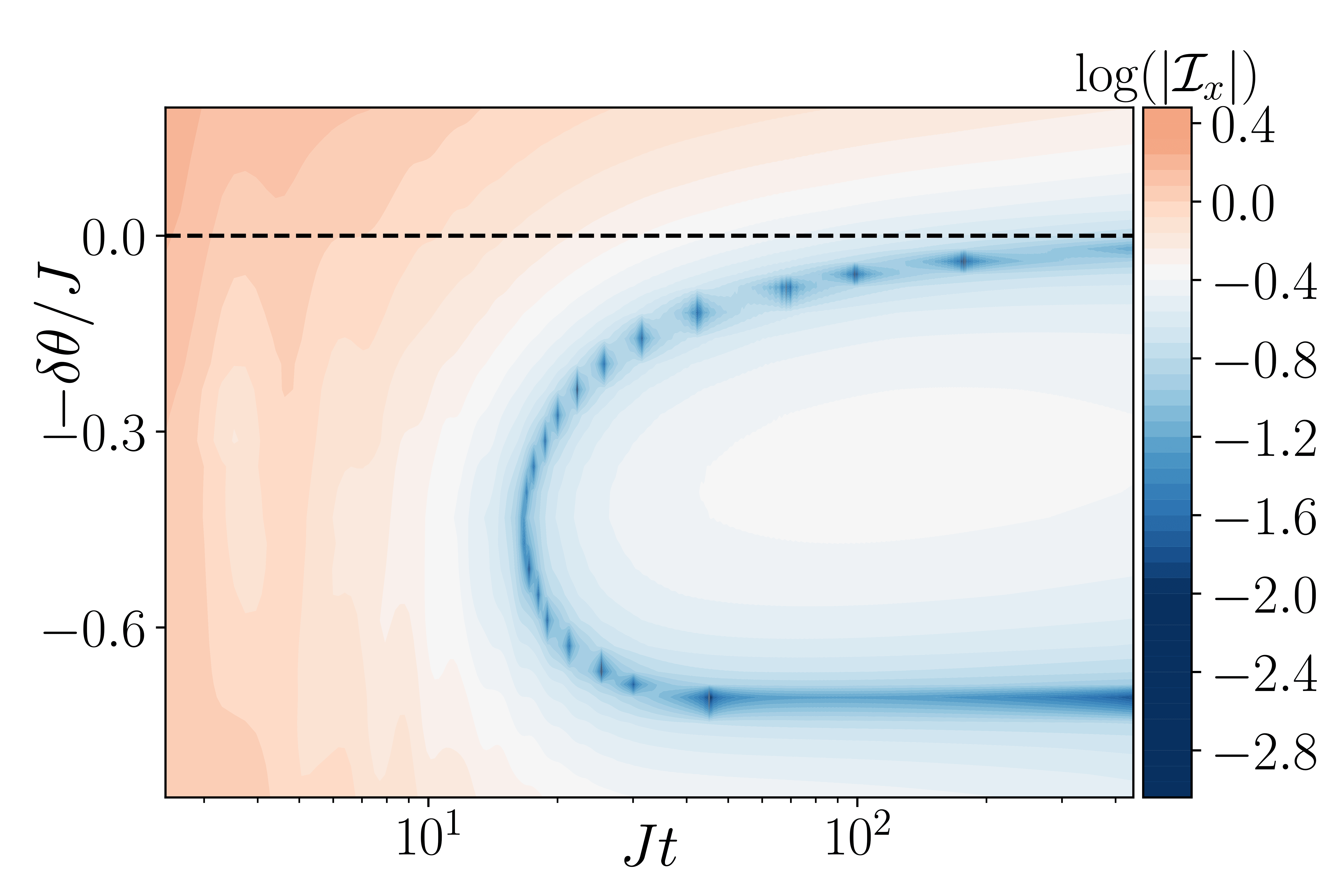}}
    \caption{\textbf{Time- and $\delta\vartheta$-dependence of the total net polarization} -- Time dependence of the total net polarization for different values of the asymptotic on-site potential $-\delta\vartheta$ for $\gamma_+=\gamma_-=\gamma_z=\vert J_0\vert $, $\potential_\mathrm{max}=0.5\pi$ and the lattice constant is $a=0.2 /\sqrt[3]{\pi}$. Further parameters are as in fig.~\ref{fig:free_diffusion}.
    \textit{Summary.--} As indicated in fig.~\ref{fig:late_time_steady} the sign inversion of the total $\xhat$-polarization only appears for a limited window of shifts $\delta\vartheta$, in agreement with the experimental results in \zfr{fig4}.
    }
    \label{fig:delta_phi_dep}
\end{figure}

With dissipation acting only at $n=n_\mathrm{NV}$ we find a single sign inversion of the total $\xhat$-polarization in a bounded region of $\delta\vartheta$ values, which indeed coincides with the analytical analysis carried out in Secs.~\ref{sec:energy_diff}-\ref{sec:space_dep_phi}. fig.~\ref{fig:delta_phi_dep} shows the corresponding results. Note that the shape of the sign-inversion arc can depend on the relative weight of the different parameters. However, for the applied potential (Eq.~\eqref{eq:phi_n}), we expect the upper asymptotic value to approach $\delta\vartheta=0$ for any set of parameters and different initial states. This offers an experimentally accessible way to calibrate the $\xhat$-kick angle at $\pi$ (corresponding to $\delta \vartheta =0$ in the toy model): an $\xhat$-kick angle of $\pi$ is expected to be insensitive to changes of the initial state. This holds true even in the presence of dissipation (decaying with the distance from the NV center). In contrast, the lower asymptotic value (here around $-\delta\vartheta \approx -0.6$) depends strongly on the initial state (see also the discussion of Sec.~\ref{sec:space_dep_phi}, in particular, fig.~\ref{fig:late_time_steady}). 

Note that, unlike Eq.~\eqref{eq:time_asymptotic_value}, Eq.~\eqref{eq:late_time_x_dissipation} is time-dependent. Therefore, in principle, depending on the initial state and the form of dissipation, $\mathrm{Tr}\left[\mathcal{H}_\pi\rho(t)\right]$ can change sign during the time evolution. It is thus possible to find several sign inversions of the total $\xhat$-polarization, provided that the timescales associated with dissipation are much slower than those related to diffusion. Notice that the sign inversion is only an indirect indicator for the presence of a nanoscale gradient; thus, having more than one sign-inversion process does not alter the general results.

\subsection{Polarization diffusion around \texorpdfstring{($\xt \approx \pi/2$)}{TEXT}: state engineering}
\label{sec:one_d_pi_half}

In contrast to the $\xhat$-kick angle close to $\pi$ where the only globally (quasi-)conserved quantity is energy, at kick angles around $\pi/2$ the corresponding effective Hamiltonian of Eq.~\eqref{eq:Heff_nopi_simple} additionally preserves the total $\xhat$-polarization. This significantly changes the relevant physics since, in this case, besides energy, also the $\xhat$-polarization (i.e., the spatially resolved $\xhat$-polarization) diffuses in the system. Thus, a uniformly polarized initial state cannot lead to a stable polarization gradient. Moreover, even for more complex initial states -- where polarization gradients might be present -- they cannot be detected by the experimentally available global measurements of the (quasi-)conserved energy and polarization.
The only way to induce a sign inversion of the total net polarization is to break conservation of the total $\xhat$-polarization by dissipation.

\subsubsection{Toy model Hamiltonian}
Along the lines of Sec.~\ref{sec:toy_model_pi}, we use a one-dimensional short-range toy model Hamiltonian akin to the effective Hamiltonian of Eq.~\eqref{eq:Heff_nopi_simple}

\begin{eqnarray}
\label{eq:short_range_model_pi_half}
    \mathcal{H}_{\pi/2} = \sum_k J_k \left(\frac{3}{2} \left(I^z_k I_{k+1}^z + I^y_k I_{k+1}^y \right)-\mathbf{I}_k \cdot \mathbf{I}_{k+1}\right).
\end{eqnarray}
Once again, to mimic the random spin position of $\Cs$ atoms in the original model, we use $J_k = J_0 + W_k$ where $W_k \in [-W,W]$ is a uniformly distributed random number.

\subsubsection{Initial states}

\begin{figure}
    \centering
  {\includegraphics[width=.5\textwidth]{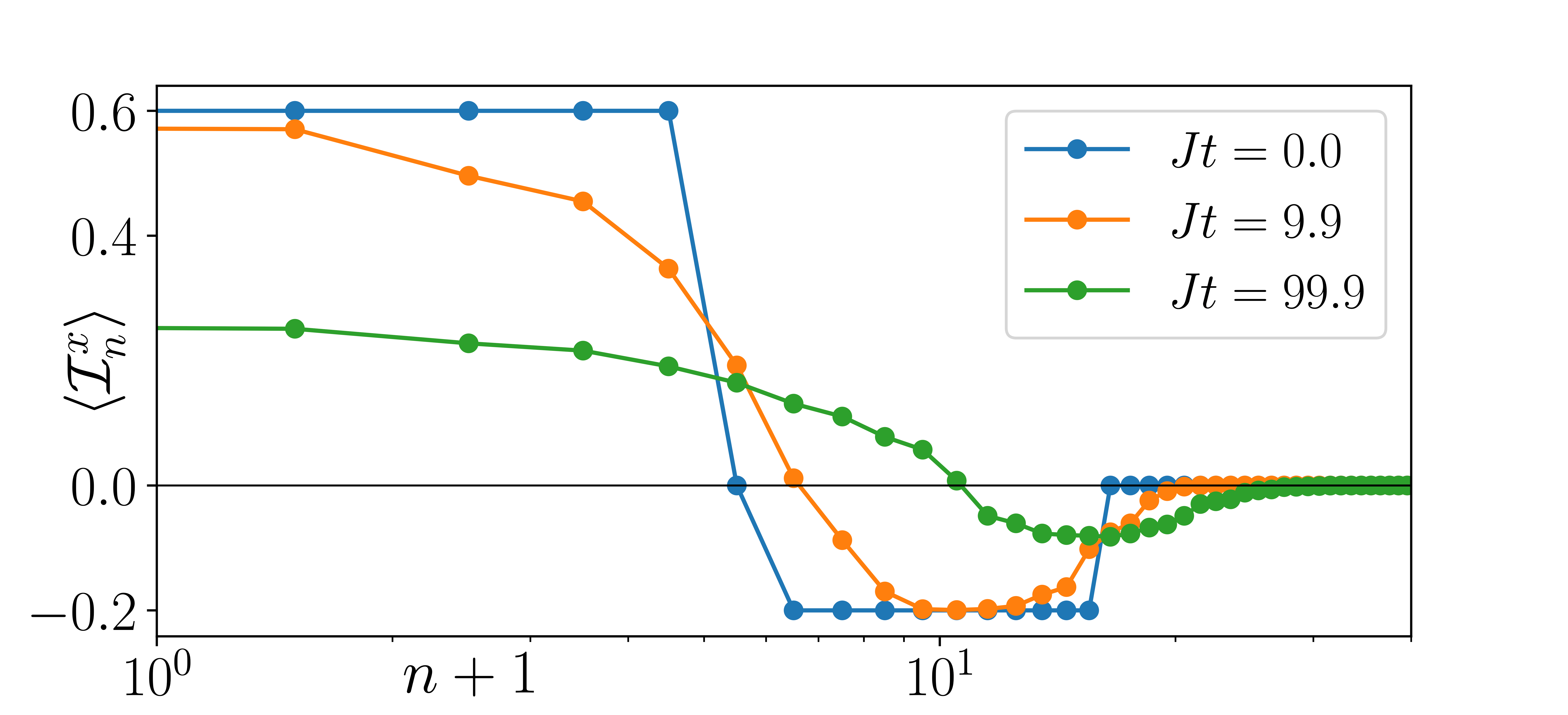}}
    \caption{\textbf{Polarization profile} -- polarization of individual spins of the initial state (blue) and the state at later snap-shots in time (see legend) for half of the investigated system evolved with LITE. 
    The initial domain wall melts and diffuses through the system. We plot the polarization as a function of $n+1$ to be able to show all the (half-chain) distribution on a log-scale. The region of polarization includes a total of $N_p=31$ spins where the $N_+=11$ spins closest to $n_\mathrm{NV}$ have $p_n=0.6$, while the remaining $N_-=20$ spins have $p_n=-0.2$. Further parameters are as in  fig.~\ref{fig:free_diffusion}.
    \textit{Summary.--} In the case of conserved $\xhat$-polarization an engineered inhomogenous spatial profile is flattened out over time with a time-dependent zero crossing.
    }
    \label{fig:polarization_profile}
\end{figure}

Similar to Sec.~\ref{sec:initial_states_pi}, we work with asymptotically translationally invariant states akin to Eq.~\eqref{eq:initial_state}; however, in the region of polarization, the local polarization now depends on the site index, i.e., $p\rightarrow p_n$. In practice, to mimic the experiment, we initialize a polarization profile that is positive close to $n_{NV}$ and negative further away from $n_\mathrm{NV}$. The profile is shown in fig.~\ref{fig:polarization_profile}. We evolve this state with the Hamiltonian of Eq.~\eqref{eq:short_range_model_pi_half} and perform numerical simulations using the LITE algorithm. Over time, the initial domain wall melts and diffuses through the system (see fig.~\ref{fig:polarization_profile}).

\subsubsection{Polarization diffusion}

\begin{figure}
    \centering
   {\includegraphics[width=0.5\textwidth]{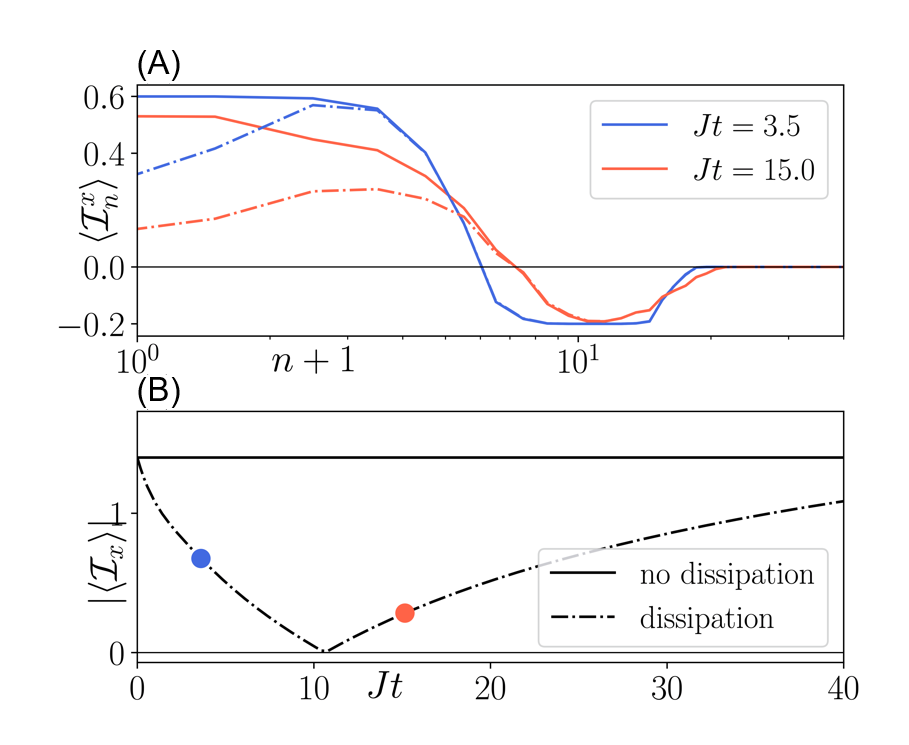}}
    \caption{\textbf{Time-evolution of the total net polarization} -- (a) Spatially resolved polarization of individual spins for two different snapshot times with (dashed-dotted) and without (solid) on-site dissipation at $n=n_\mathrm{NV}$ ($n_{\mathrm{NV}}=0$ here).  
    (b) Corresponding total net polarization as a function of time from time-evolution with the Hamiltonian of Eq.~\eqref{eq:short_range_model_pi_half}. 
    The black dashed-dotted line corresponds to a simulation including dissipation. The blue and red filled circles relate to the corresponding dashed-dotted line in (a) at the indicated time. We use the dipole initial state of fig.~\ref{fig:polarization_profile}. When dissipation is non-zero we set $\gamma_+=\gamma_-=\gamma_z=\vert J_0\vert$. Further parameters are as in fig.~\ref{fig:free_diffusion}.
    \textit{Summary.--} The total $\xhat$-polarization is a conserved quantity in the absence of dissipation, preventing the observation of a sign inversion. In the presence of dissipation the positive polarization close to the NV is dissipated faster than the negative polarization further out leading to a sign inversion in the total $\xhat$-polarization.
    }
    \label{fig:mag_inversion_pi_half}
\end{figure}

There is a striking difference between the dynamics generated by the $\xt=\pi/2$ kick Hamiltonian of Eq.~\eqref{eq:short_range_model_pi_half} and the $\xt=\pi$ kick Hamiltonian from Eq.~\eqref{eq:short_range_model}: with the total net polarization conserved, besides energy, also polarization diffuses through the system. As a consequence, even when starting from an initial state with a polarization dipole, there is no experimental way to resolve this spatial structure from the total net polarization (which is constant in time). Only if dissipation induced by the NV center is sufficiently strong to reduce the polarization close to $n_{\mathrm{NV}}$ (here the positively polarized part), can the total net polarization change significantly over time; under these conditions, a sign inversion of the total net polarization, similar to the one discussed on Sec.~\ref{sec:one_d_pi}, can be observed. These two scenarios are shown in fig.~\ref{fig:mag_inversion_pi_half}. The dissipation is modeled using on-site dissipators according to Eq.~\eqref{eq:jump_ops} with a non-zero dissipation coupling constant only for $n=n_\mathrm{NV}$. 

\subsection{Comparison between energy diffusion and polarization diffusion}
\label{subsec:comparison-energy-polarization}

The employed toy models of Sec.~\ref{sec:one_d_pi} and Sec.~\ref{sec:one_d_pi_half} can both be used to create the characteristic sign-inverting signature of the total net polarization for the respective parameter regimes and initial states. While seemingly similar, the underlying physics is fundamentally different in both cases: The toy model of Sec.~\ref{sec:one_d_pi} (which corresponds to the experimental system driven at with kick angles close to $\pi$) has no globally conserved quantity other than energy. In this case, energy diffuses and, over time, a polarization gradient builds up; in particular, the boundary between negatively and positively polarized spins $r_c$ remains constant in time. By contrast, the toy model investigated in Sec.~\ref{sec:one_d_pi_half} (which corresponds to the experimental system driven with $\xhat$-kick angles of $\pi/2$) has the total $\xhat$-polarization as an additional global (quasi-)conserved quantity. Thus, in this case, polarization diffuses in the system. This, in turn, implies that any artificially designed polarization domain wall is \textit{not} preserved in time, i.e., $r_c=r_c(t)$ changes over time. The exact functional form of $r_c(t)$, in this case, depends sensitively on the interplay of diffusion, dissipation, and the initial state. 

\begin{figure*}
    \centering
   {\includegraphics[width=\textwidth]{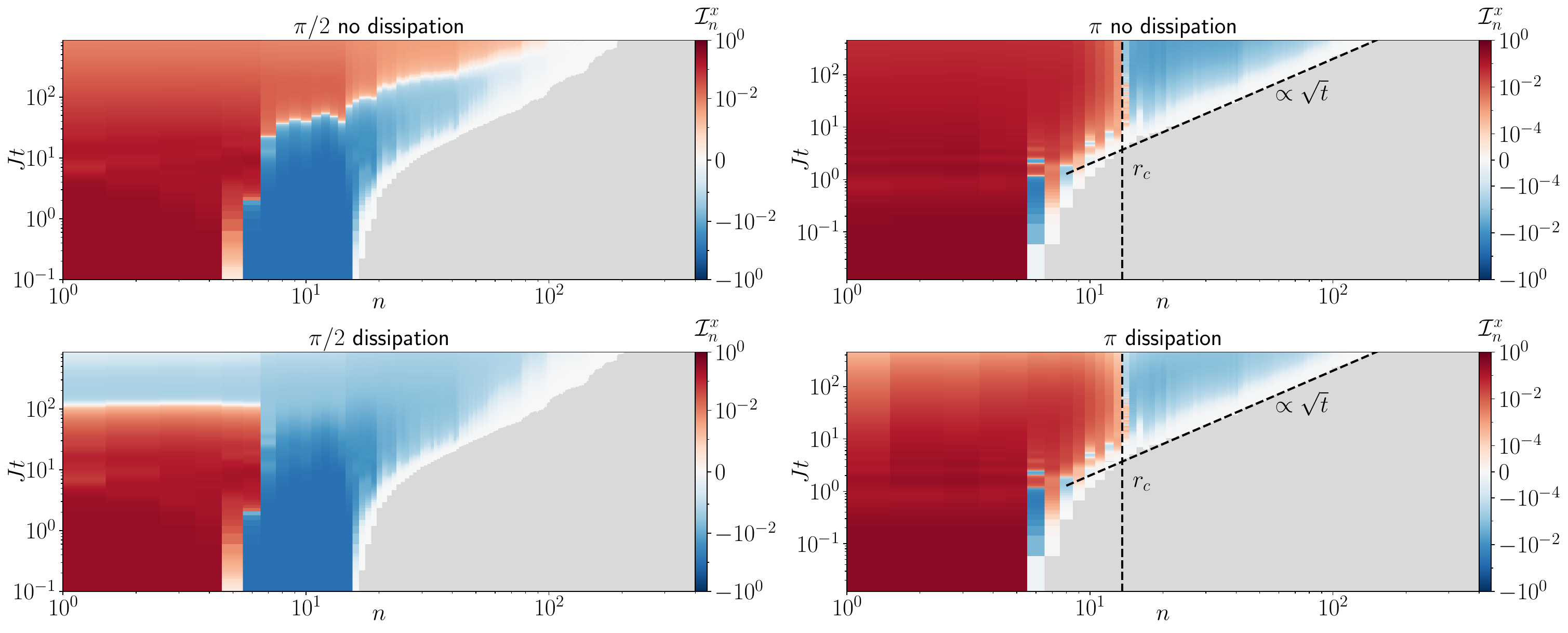}}
    \caption{\textbf{Comparison of the dynamics of the different toy models} -- 
    \textit{Left column:} Dynamics generated by the toy model of Sec.~\ref{sec:one_d_pi_half} using a domain wall initial state identical to the one used in fig.~\ref{fig:polarization_profile}.
    \textit{Right column:} Dynamics generated by the toy model of Sec.~\ref{sec:one_d_pi} with a uniformly polarized initial state with $N_p=11$ and $p=0.6$; the lattice constant is $a=0.2/\sqrt[3]{\pi}$. The vertical dashed line corresponds to $r_c=1/(a\sqrt[3]{\delta\vartheta})+n_\mathrm{NV}$ where $n_\mathrm{NV}=0$ and $\delta\vartheta = 0.05\pi$.
    Dissipation is applied in the form of on-site dissipators at $N= n_\mathrm{NV}$ with $\gamma_+ = \gamma_- = \gamma_z=\vert J_0\vert$. Moreover, we use $\potential_\mathrm{max}=0.2\pi$.
    The diagonal dashed line indicates the expected diffusive scaling with $\propto \sqrt{t}$, and serves as a guide to the eye. 
    Further parameters in all panels are the same as in fig.~\ref{fig:free_diffusion}. The colors are chosen such that values $>0$ appear in red colors and values $<0$ appear in blue, on a logarithmic scale. The grey area indicates the region where no polarization is present.
    \textit{Summary.--} Time evolution with respect to $\mathcal{H}_\pi$ qualitatively unaffected with a sign-inversion of polarization and a fixed zero crossing in both cases. For $\mathcal{H}_{\pi/2}$ the zero crossing is not fixed over time and is strongly affected by the presence of absence of dissipation.
    }
    \label{fig:one_d_comparison}
\end{figure*}

Figure~\ref{fig:one_d_comparison} shows a direct comparison between the dynamics of the different cases with and without dissipation. The right column depicts the evolution at an x-kick angle close to $\pi$ (toy model of Sec.~\ref{sec:one_d_pi}, Hamiltonian engineering approach): Using the applied potential given in Eq.~\eqref{eq:phi_n}, we can compute $r_c$ (dashed vertical line in the right column of fig.~\ref{fig:one_d_comparison}). As energy diffuses following a $\propto \sqrt{t}$ scaling, the polarization gradient builds up with $r_c$ being constant. Instead, in the left column panels we show the dynamics generated by an x-kick angle $\pi/2$ (toy model of Sec.~\ref{sec:one_d_pi_half}, State engineering approach). In both panels (i.e., with and without dissipation) the position of the domain wall separating the positively and negatively polarized regimes depends strongly on time, as a result of polarization diffusion.

\subsection{Summary}
\label{subsec:one-dimension-summary}

Using two numerically tractable one-dimensional short-range models akin to the three-dimensional long-range Hamiltonians present in the experiment, we have explored two different diffusion effects. While our toy model resembling the experimental regime of kick angles $\sim \pi$ only has energy as a globally conserved quantity, the corresponding toy model for kick angles $\sim \pi/2$ additionally has the total $\xhat$-polarization as a conserved charge. This additional constant of motion changes the diffusive behaviour: polarization diffuses through the system and domain-walls are not stable -- neither in space nor in time. In contrast, if energy is the only globally conserved quantity, polarization domain walls can build up in the system as energy diffuses. Such domain walls are then stable in space and time (within the duration of the prethermal plateau). We analyze this effect in detail by means of numerical and analytical methods: Assuming an initial state with a finite amount of energy whose density is located around the NV-center, which induces a spatially dependent on-site potential on $\Cs$ nuclei, we show that the total net polarization can undergo a sign inversion while energy diffuses through the system. Depending on the initial state and the asymptotic value of the on-site single particle potential, the late-time steady-state polarization can either take a negative or positive value. Importantly, for the single-particle potentials considered here, there is always a finite range of asymptotic single-particle potentials for which a sign-inversion of the total net polarization is obtained. 

Locally, the polarization of individual spins is proportional to the value of the single on-site potentials. For space-dependent on-site potentials, this implies polarization gradients, where the sign-inversion of the total net polarization might serve as a global indicator for the latter. To arrive at this result, we only assume diffusion of energy as well as a local Gibbs approximation of a given late-time state. Hence, it can straightforwardly be transferred to a three-dimensional model (as present in the experiment). 

Model-specifically, we find that in our theoretically tractable one-dimensional short-range toy model, energy remains bound in regions with strong on-site potentials and diffusion slows down significantly. Some aspects that underlie this behaviour can be transferred over to the actual three-dimensional system with long-range couplings -- such as a spatially-dependent diffusion constant; however, diffusive (or non-diffusive) properties of the experimental system cannot be deduced from this analysis, since localization effects are known to display a severe dependence on the dimensionality of the system and the range of interactions.  Including dissipation -- which is expected to appear in the experimental system as well -- enhances the diffusive behaviour of the theoretical one-dimensional short-range model. The expected late-time value of the total $\xhat$-polarization is then reduced by a factor accounting for the continuous loss of energy.

\section{\label{sec:dimensionality}Role of dimensionality and interaction range}

Numerically simulating long-range three-dimensional quantum systems with a non-vanishing linear extent is notoriously hard. However, the essential physics is often captured already by simpler models. In order to gain a qualitative understanding of the experimental system we therefore study simplified versions of the long-range three-dimensional system.
In particular, we restrict ourselves to exact quantum dynamics of long-range one-dimensional small systems~(Sec.~\ref{sec:effective_Hamiltonian}) and approximate quantum dynamics of a short-range one-dimensional quasi-infinite system~(Sec.~\ref{sec:approximate_dynamics}).

In general, one cannot expect that these simplifications lead to \textit{qualitatively} similar results. Therefore, in this section we summarize possible caveats of the simplifications and which behaviour is expected to generalize to three dimensions.

\subsection{Dynamics beyond the prethermal plateau}

Let us briefly comment on the Floquet prethermalization of the considered models. It was found experimentally before~\cite{Beatrez2022} that the three-dimensional long-range interacting model has a stable prethermal plateau which is algebraically long-lived in the driving period, i.e., the prethermal lifetime $T_\mathrm{prethermal}\propto 1/\omega^2$. This is in contrast to the exponential lifetime of a short-range interacting system, $T_\mathrm{prethermal}\propto \exp(-\omega/\text{const})$. However, the effective Hamiltonian analysis in Sec.~\ref{sec:effective_Hamiltonian} only relies on the existence of a prethermal plateau and not on its parametric scaling.

 \subsection{\label{subsec:approximate_3D}Equilibration and thermalization dynamics within the prethermal plateau}

In contrast to the discussion in Sec.~\ref{sec:approximate_dynamics}, the experimental system is described by a three-dimensional long-range Hamiltonian (instead of a short-range one-dimensional one). While the results derived in Sec.~\ref{sec:approximate_dynamics} are independent of the dimensionality of the problem Hamiltonian, they rely on the locality of the Hamiltonian. 

In the three-dimensional experimental system, the locality of the Hamiltonian may be violated as spin-spin interactions decay with a critical exponent. However, as far as expectation values are concerned, most of the energy is still stored on local scales. To see this, let us decompose the long-range Hamiltonian into two parts
\begin{eqnarray}
    H = H_{r\leq \ell_c} + H_{r > \ell_c},
\end{eqnarray}
where $H_{r\leq \ell_c}$ contain all dipolar coupling terms of spins with a distance up to $\ell_c$, while $H_{r> \ell_c}$ covers the rest. At late times, when the (diffusive) spread of information has reached the system size, the state of the system becomes thermal $\rho(t\rightarrow \infty)  = Z^{-1} \exp(-\beta H )$ with the partition function $Z = \mathrm{Tr}[ \exp(-\beta H )]$ and non-zero inverse temperature $\beta$. Using $\rho(t\rightarrow \infty)$, the energy found on scales larger than $\ell_c$ is given by

\begin{eqnarray}
    \langle H_{r > \ell_c}\rangle (t \rightarrow \infty) = \mathrm{Tr}\left[ H_{r > \ell_c} \rho(t\rightarrow \infty)\right] \approx \frac{ -\beta}{D}\mathrm{Tr}\left[ H_{r > \ell_c}^2\right]
\end{eqnarray}
with $D$ being the dimension of the Hilbert space. Note that, the mixed term $\mathrm{Tr}\left[ H_{r > \ell_c} H_{r \leq \ell_c}\right]=0$ vanishes as $H_{r > \ell_c} H_{r \leq \ell_c}$ contains strings with at least two non-trivial Pauli-operators and all non-trivial Pauli-strings are traceless, see Eq.~\eqref{eq:pauli_trace}. The remaining trace evaluates to
\begin{eqnarray}
    \mathrm{Tr}\left[ H_{r > \ell_c}^2\right] = D \sum_{\substack{ i<j \\ r_{ij}> \ell_c }} J_{\mathrm{exp}}^2 \frac{(3 \cos^2(\theta_{ij}) -1)^2}{r_{ij}^6}.
\end{eqnarray}
Assuming uniform $\Cs$ density and $\ell_c$ to be much larger than the average inter-spin distance we can replace the sum with an integral. 

This results in
\begin{equation}
\label{eq:H_ellc}
     \langle H_{r > \ell_c}\rangle (t \rightarrow \infty) \approx -2\pi \beta \frac{4 J_{\mathrm{exp}}^2}{5} \sum_i \int_{r>\ell_c}  \! \frac{\mathrm{dr}}{r^4} = \frac{-8 N \pi \beta J_{\mathrm{exp}}^2}{15 \ell_c^3}
\end{equation}
with $N$ the total number of spins. Up to corrections of order $\mathcal{O}(1/\ell_c^3)$ the energy is thus located on scales $<\ell_c$. 
\begin{eqnarray}
    \langle H \rangle (t\rightarrow \infty) = \langle H_{r\leq\ell_c} \rangle (t) + \mathcal{O}(1/\ell_c^3).
\end{eqnarray}
Similarly, at early times, for the initial state with finite polarization close to the NV-center (and featureless anywhere else), we have $\langle H_{r>\ell_c} \rangle (t) = 0$. Assuming diffusive transport the late time value of Eq.~\eqref{eq:H_ellc} poses an upper bound on the energy found on scales $>\ell_c$.
Thus, as far as we are not interested in correlation functions that exceed the scale $\ell_c$, we can neglect the energy stored at scales $>\ell_c$ and approximate the full Hamiltonian of the system using only terms up to $\ell_c$
\begin{eqnarray}
\label{eq:3D_hamiltonian_density}
    H \approx \sum_{ n }  h_{V_n^{\ell_c}},
\end{eqnarray}
where $ h_{V^{\ell_c}_n}$ is the $\ell_c$-local Hamiltonian density in the volume $V_n^{\ell_c}$ with center $n$ and diameter $\ell_c$. Similar arguments were derived in Ref.~\cite{Abanin2018_LongRange} in the context of Floquet heating. 

Equation~\eqref{eq:3D_hamiltonian_density} allows to repeat the steps outlined in Secs.~\ref{sec:energy_diff} -- \ref{sec:mag_grad}. In analogy to Eq.~\eqref{eq:beta_of_t} This yields local inverse temperatures of the form
\begin{eqnarray}
\label{eq:beta_of_t_3D}
    \beta_n(t) \approx \frac{\mathrm{Tr}\left[ H \rho_\mathrm{init}\right]}{\mathrm{Tr}\left[ h_{V_{n}^{\ell_c}}H\right]}\;
    p_E(n,t).~~~
\end{eqnarray}
where $\mathrm{Tr}\left[ h_{V_{n}^{\ell_c}}H\right]$ takes a non-zero positive value depending on $n$ and $\ell_c$. Following similar steps as derived in Secs.~\ref{sec:energy_diff} - \ref{sec:mag_grad}, we obtain the local polarization of a given spin $n$ 
\begin{eqnarray}
\label{eq:local_mag_3D}
    \mathcal{I}_n^x(t) = \frac{\potential_n}{4}\frac{\mathrm{Tr}\left[ H \rho_\mathrm{init}\right]}{\mathrm{Tr}\left[ h_{V_{n}^{\ell_c}}H\right]} p_E(n,t).
\end{eqnarray}
Finally, let us point out that interactions in the three-dimensional long-range system depend on the angle $\theta$ between the lattice position vector and the applied magnetic on-site potential. 
This leads to an angular dependence in the couplings and the spatially inhomogeneous on-site potential $\potential_n$ induced by the NV electron spin, which cannot be equivalently modeled in one dimension. 
Moreover, the radial dependence of the exact profile does not follow a simple $1/r^3$-dependence, see effective Hamiltonian analysis around Eq.~\eqref{eq:Heff_pi} and fig.~\ref{fig:Heff_summary} (b) for the exact form. 

The angular dependence in the spin-spin couplings has no qualitative impact on the observed dynamics. Nevertheless, it does change the details of the above analysis. 
For instance, the key results of diffusion and the fact that the effective potential profile will also be imprinted in the local polarization profile, Eq.~\eqref{eq:local_mag_3D}, remain unchanged.
However, angular-dependent spatially inhomogeneous on-site potentials lead to a quantitatively different local polarization profile following the local on-site potential, which is also found in the classical three-dimensional simulation, Sec.~\ref{sec:classical_simulation}.

\section{\label{sec:classical_simulation}Three-dimensional classical simulations on a diamond lattice}

In this section we complement the observations from the previous Secs.~\ref{sec:effective_Hamiltonian} and \ref{sec:approximate_dynamics} performed for few spin or infinite $1$D systems, with a classical simulation on a finite $3$D long-range system on a diamond lattice of $L=1000$ randomly placed spins.

Performing a classical simulation allows us to reach considerably larger system sizes of hundreds of spins, enabling the study of $3$D systems with a finite linear extent. At the same time, however, a classical simulation comes at the expense of neglecting all quantum correlations. 
The thermalization analysis in Sec.~\ref{subsec:ETH} and the approximate dynamics in Sec.~\eqref{sec:approximate_dynamics} indicate that the system starts in an high-temperature state and remains close to a high-temperature state. This suggests that the dynamics may be described well classically.

While the classical simulation scales only linearly in the number of spins $L$, the number of spins in a three-dimensional system scales cubic $L\propto r^3$ in the linear dimension $r$. Thus, studying quasi infinite systems, i.e., systems with a linear extension far exceeding the support of the initial state, is beyond the scope even for the classical simulation. Therefore, we cannot study free diffusion in these systems. Hence, we restrict the study of three-dimensional systems to test the predictions from the ETH-analysis in Sec.~\ref{subsec:ETH} about the formation of spatially inhomogeneous local polarization starting from a homogenous polarized state in three dimensions.

As discussed before, it is sufficient to consider the Hamiltonian time evolution generated by the effective Hamiltonian~\eqref{eq:Heff_nopi_simple} for intermediate times, since we are only interested in the prethermal properties.

\subsection{\label{subsec:classical_algorithm} Classical simulation algorithm}

\textit{Equations of Motion. -- }
Notice that the effective Hamiltonian analysis in Sec.~\ref{sec:effective_Hamiltonian} was performed for a quantum system. However, the same analysis and arguments of the Floquet prethermal plateau are expected to remain valid also in the classical limit~\cite{Pizzi21, ye2021floquet}.
The starting point for the classical equations of motion is the quantum Heisenberg equation of motion for the expectation values of the spin operators with respect to the three-dimensional Hamiltonian, Eq.~\eqref{eq:Heff_pi_simple},
\begin{equation}
    \label{eq:Heisenberg_EOM}
    \begin{aligned}
        \frac{\mathrm{d}}{\mathrm{d}t} \left<\mathbf{I}_k\right>(t) &= \left< i \left[ \Heffpi ,\, \mathbf{I}_k \right] \right>(t) 
            = \left< \mathbf{I}_k \times \boldsymbol{\nabla}_{\mathbf{I}_k} \Heffpi \right>(t)
    \end{aligned}
\end{equation}
where $\mathbf{I}_k = (I_k^x, I_k^y, I_k^z)^T$ and spin index $k$ and $\times$ denotes the vector (outer) product in $3$D.
In the second line of Eq.~\eqref{eq:Heisenberg_EOM} we used the spin-algebra $[I^\alpha_k ,\, I^\beta_l] = i\delta_{kl} \sum_\gamma \epsilon_{\alpha\beta\gamma} I^\gamma_k$, with the fully anti-symmetric Levi-Civita symbol $\epsilon_{\alpha\beta\gamma}$.
The classical approximation amounts to a mean potential approximation $\left<I^\alpha_k I^\beta_l\right> {\approx} {\left< I^\alpha_k\right>} {\left< I^\beta_l\right>}$. Therefore, the classical equations of motion for the spin expectation values $\mathcal{I}_k^\alpha {=}\left<I_k^\alpha\right>$ read
\begin{equation}
    \label{eq:Classical_EOM}
        \frac{\mathrm{d}}{\mathrm{d}t} \boldsymbol{\mathcal{I}}_k(t) =  \boldsymbol{\mathcal{I}}_k(t) \times \boldsymbol{\nabla}_{\boldsymbol{\mathcal{I}}_k} \mathcal{H}(\boldsymbol{\mathcal{I}}_k(t)) \, ,
\end{equation}
where $\mathcal{H}{=}\Heffpi(\boldsymbol{\mathcal{I}}_k)$ is a quadratic function in the spin expectation values and is obtained from $\Heffpi$ by replacing $\mathbf{I}_k {\to} \boldsymbol{\mathcal{I}}_k$.

Notice, that Eq.~\eqref{eq:Classical_EOM} only contains $2L$ independent degrees of freedom instead of $2^L$ for a quantum system of $L$ spins, thus, leading to an exponential reduction of degrees of freedom. Therefore, the classical simulation enables us to study much larger systems compared to the quantum simulation~($L{\approx} 1000$). Further notice that Eq.~\eqref{eq:Classical_EOM} is a non-linear equation of motion as the right-hand-side is quadratic in $\boldsymbol{\mathcal{I}}_k$ and must thus be solved iteratively. The classical chaotic dynamics renders time evolution up to very long times harder compared to the linear quantum equations.

\textit{Initial State. -- } 
The initial state in the experiment is assumed to be well described by the density matrix $\rho_0 \approx (\identity + \sum_k \mu I^x_k)/Z$, where $\mu_k$ describes the local profile of the initial state and $\mu_n\ll1$ is sufficiently small such that $\rho_0$ is a positive definite density matrix. 
Therefore, we may write $\rho_0 \approx \exp(\sum_k \mu_k I_k^x)/Z \overset{\mu\ll1}{\approx} (\identity + \sum_k \mu_k I_k^x)/Z$, to have a well defined density matrix for all values of $\mu_n$. 
Since this density matrix does not have any correlations between spins, i.e., $\rho_0 = \bigotimes_{k=1}^L \left[ \exp(\mu_k I_k^x)/Z_k \right]$, it can also be used as an initial state for the classical simulation.
In particular, we use the heatbath Monte-Carlo algorithm~\cite{Loison2004} to sample individual classical spin configurations and average over a large number of the configurations to obtain a thermal average. 
The algorithm works as follows. We first draw, for each individual spin $k=1,\,\dots,\,L$, the $\mathcal{I}_k^x$ component with $\mathcal{I}_k^x= \log(1 + u [e^{2\mu_k} - 1])/\mu_k$, where $u$ is a random variable uniformly distributed on the unit interval. Then, the  $\mathcal{I}_k^{y,z}$ components are drawn uniformly from a circle of radius $[{1 - (\mathcal{I}_k^x)^2}]^{1/2}$. 
Finally, we compute the relevant (time-dependent) expectation values for each state~(trajectory) and eventually average over many initial spin configurations.
Averaging over many spin configurations the $\mathcal{I}_k^{y,z}$ components average to zero, while the $\mathcal{I}_k^x$-component is biased towards positive values, i.e. leading to a $\xhat$-polarized ensemble.

In the experiment the initial state is generated via hyperpolarization, i.e. the NV centers are polarized via a resonant drive and due to dipole-dipole interactions this polarization diffusive through the surrounding system of $\Cs$atoms. Therefore, the initial state likely possesses a spatial profile with parts of the system closer to the NV being polarized more strongly. We mimic this in the classical simulation by choosing a domain-wall like state
\begin{equation}
\label{eq:chemical_potential}
    \mu_n = \begin{cases}
        \mu & r \leq r_\mathrm{pol} \,,\\
        0   & r > r_\mathrm{pol} \, ,
    \end{cases}
\end{equation}
where spins within a given radius $r_\mathrm{pol}$ are polarized and spins beyond this radius are not polarized, see also initial state in fig.~\ref{fig:classical_simulation}A(iii).

\textit{System geometry. -- }
The system of interest are the nuclear spins of $\Cs$-atoms which are randomly distributed on a diamond lattice with a lattice spacing $a=0.357\,\mathrm{nm}$ and a density of roughly $n=0.5\,\%$ of lattice sites being occupied by $\Cs$-atoms. 

However, the $\Cs$-spins within a radius of $r<r_\mathrm{c}=1.7\,\mathrm{nm}$, the so-called frozen core, are far-detuned from $\Cs$-spins at larger radii and cannot be detected by the measurement procedure implemented in the experiment. Therefore, we consider only spins that have a minimal distance $r_\mathrm{min}>r_\mathrm{c}$ from the electron. Since we cannot simulate all ${\sim 10^4}$ spins surrounding a single NV-center, we further restrict ourselves to a region close to the chosen crossing radius, which we choose to be $r_c\approx 6.5\,\mathrm{nm}\,\cdot\,\sqrt[3]{|3\cos(\theta)^2-1|}$ and we set $r_\mathrm{min}=3\,\mathrm{nm}$.

Drawing the sites occupied by $\Cs$-atoms on the diamond lattice randomly may lead to two or more $\Cs$-atoms being in close proximity of each other, leading to a strong interaction two orders of magnitude stronger than the median coupling. Such configurations can also occur in the experiment and this strong coupling will dominate the short-time dynamics. However, the long-time dynamics are hardly impacted by such rare configurations. Since we are only interested in the long-time dynamics of the system and simulating systems which possess a separation of time scales demands expensive high precision simulations, we neglect lattice configurations where $\Cs$-atoms are closer than some minimal distance $d_\mathrm{min}$ apart. We choose ${d_\mathrm{min}=2\,a}$ leading to the interaction energy scales $E_\mathrm{interaction}$ being comparable to the median coupling $J$, i.e., $E_\mathrm{interaction} \leq 5\,J$. This entire procedure corresponds to an effective coarse-graining of time, neglecting short-time dynamics $tJ \ll 1$.

In summary, we draw $L$ many spin positions on a diamond lattice of size $L/n$ with inter-spin distance $r_{ij}\geq 2a$ and distance from the NV-center $r \geq r_\mathrm{min}=3\,\mathrm{nm}>r_\mathrm{c}$. This ensures that we avoid the aforementioned large energy scales and populate a mean ratio of $n$ $\Cs$-spins. In the experiment the observed data corresponds to an average over the environment of many NV-centers. Therefore, we also average our results over many lattice configurations.

\subsection{\label{eq:classical_results} Spin inversion in classical simulations}

\begin{figure}[t]
    \centering
  {\includegraphics[width=.5\textwidth]{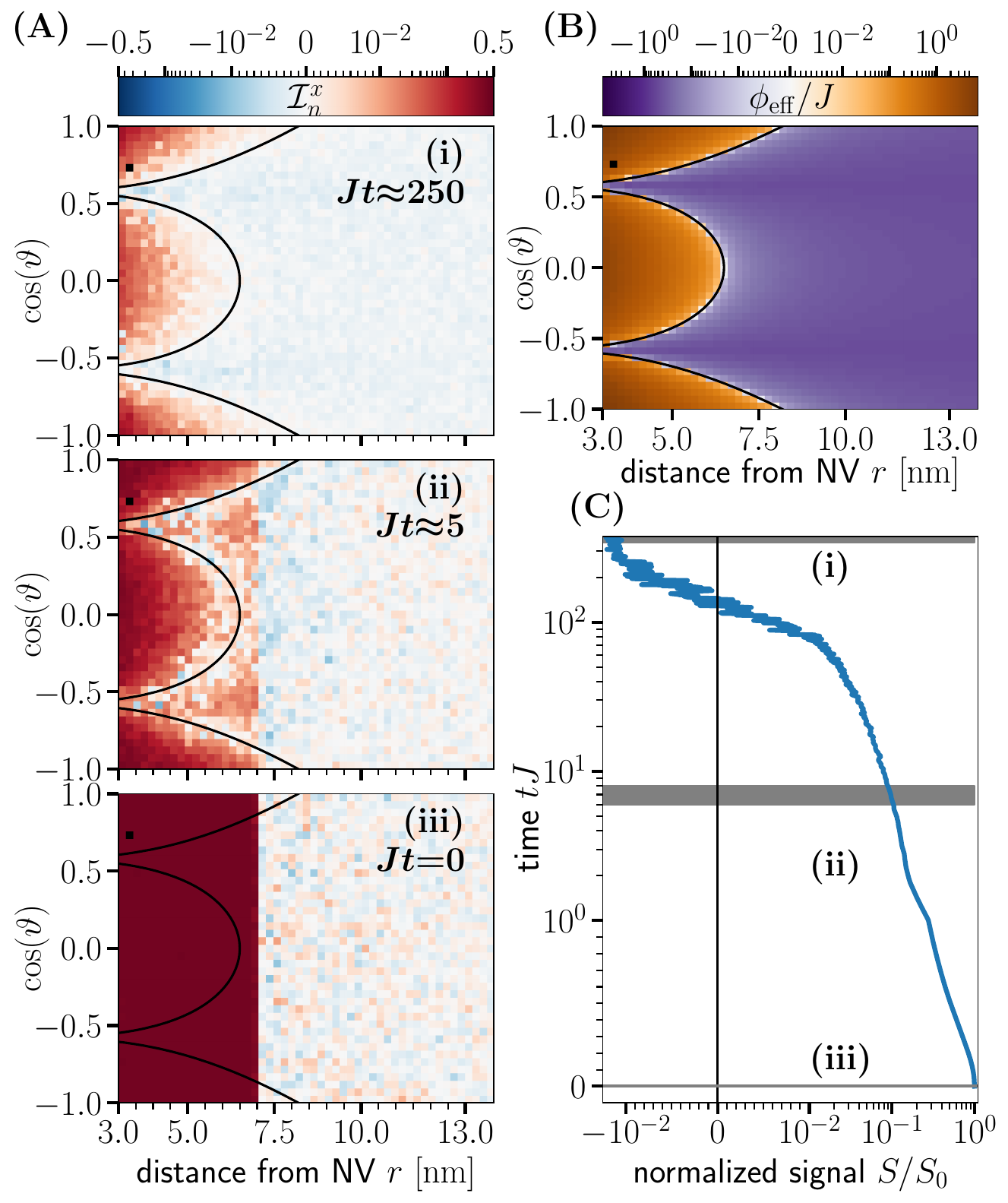}}
    \caption{
    \textbf{Classical Simulation Results:}
    Classical simulation of spin locking around ${\vartheta\approx\pi}$, Eq.~\eqref{eq:Classical_EOM}, of an ensemble of $1000$ spins, averaged over $100$ lattice configurations and $150$ initial states.
    \textbf{(A)} Coarse-grained polarization profile taken at different time cuts (time flowing from bottom to top) from early time (iii) to late time (i) as indicated in (C). Black squares indicate that no data exists for that parameter value. Black line indicates the $\theta$-dependent crossing radius $r_\mathrm{c}$.
    \textbf{(B)} Applied effective on-site Potential. 
    \textbf{(C)} Time evolution of signal $S$ normalized with respect to initial value $S_0$.
    \textit{Summary.--}In agreement with analytical predictions and the quantum simulation results the local polarization at late times follows the applied local potential up to an overall constant (the inverse temperature $\beta$).
    }
    \label{fig:classical_simulation}
\end{figure}

In the following, we study the spin locking regime near ${\vartheta\approx \pi}$ using the classical simulation introduced above. As already mentioned, this enables us to simulate the full three-dimensional long-range interacting system for a large, yet limited number of spins. The results are shown in fig.~\ref{fig:classical_simulation}.

\textit{Local polarization profile.--} The initial state for the time-evolution is a homogeneously polarized state with a polarization of $\mathcal{I}_n^x\approx 0.4$ per site, see fig.~\ref{fig:classical_simulation}~(A)(iii). As time progresses the state develops a spatially inhomogeneous local polarization distribution, fig.~\ref{fig:classical_simulation}~(A)(ii). Eventually at late times, fig.~\ref{fig:classical_simulation}~(A)(iii), the polarization is equal to the applied effective on-site potential, fig.~\ref{fig:classical_simulation}~(B) up to a proportionality constant given by the inverse temperature $\beta$; this is expected from the ETH analysis in Sec.~\ref{subsec:ETH}. Therefore, the classical simulation also affirms the conjecture that the experimental observations are caused by the formation of a robust spatially inhomogeneous local polarization profile. In this case the polarization gradient around each NV encloses $1000$-spins and a macroscopic distance of $>10\,\mathrm{nm}$ with $\approx 16\,\%$ negatively polarized and $\approx 84\,\%$ positively polarized spins. We found that the precise profile is strongly influenced by the choice of parameters, in particular the pulse duration~$\tkick$, Rabi frequency~$\Rabi$, and polarization of the electron $P$.

\textit{Integrated polarization.--} 
As expected (see also Sec.~\ref{subsec:ETH}) the signal $S=\sum_{n=1}^L \mathcal{I}_n^x$ is not conserved by the dynamics but decays slowly in time, see fig.~\ref{fig:classical_simulation}(C). In agreement with other simulation methods the net polarization eventually inverts at late times as the energy spreads outwards and more parts of the outer region are getting negatively polarized. 
Notice, that the negative signal is much weaker in magnitude than the initial positive signal than observed in other simulations or the experiment. One possible explanation is the lack of dissipation such that at all times the strong positive contribution at $r{<}r_\mathrm{c}(\theta)$ does not decay and thus decreases the magnitude of the negative signal.

In summary, the classical simulation confirms in a three-dimensional long-range model the analytical and numerical results obtained for one-dimensional toy models in Secs.~\ref{sec:approximate_dynamics} and \ref{sec:effective_Hamiltonian}. 
In particular, the macroscopic, stable, spatially inhomogeneous profile formed at late times reflects the profile of the effective on-site potential induced on the nuclear spins by the interaction with the NV.



   
    
    